\documentclass[12pt]{article}
\usepackage{amsfonts,amssymb,graphicx}
\usepackage{fullpage}
\usepackage{subfig}
\usepackage{rotating}
\usepackage{sidecap}
\pdfoutput=1
\usepackage{ifpdf}
\ifpdf
  \usepackage{color}
  \definecolor{darkblue}{rgb}{0.3,0.3,0.6}
    \definecolor{darkgreen}{rgb}{0,0.6,0}
  \usepackage[pdftitle={Deformations on Tilted Tori and Moduli Stabilisation at the Orbifold Point},pdfauthor={Michael Blaszczyk, Gabriele Honecker, Isabel Koltermann},pdfsubject={String theory},pdfkeywords={string theory},colorlinks=true,linkcolor=black,urlcolor=darkblue,filecolor=darkblue,citecolor=darkblue,menucolor=darkblue,breaklinks=true,bookmarks=true,anchorcolor=black,unicode=true]{hyperref}
\else
  
\fi
\usepackage[numbers,compress]{natbib}
\usepackage{a4wide}
\usepackage{longtable}
\usepackage[table]{xcolor}
\usepackage{graphicx}
\usepackage[centertags]{amsmath}
\usepackage{multirow}
\usepackage{slashed}
\usepackage{pdflscape}
\usepackage{caption}  
\usepackage{MnSymbol} 

\usepackage{mathtools}
\usepackage{bbm}
\usepackage{dsfont}
\usepackage{hhline}

\newcommand{\bCentering}{\centering}
\newcommand{\bCaption}{\caption}

\def\muc{\multicolumn}

\def\Z{\mathbb{Z}}
\def\R{\mathbb{R}}

\def\1{{\bf 1}}
\def\2{{\bf 2}}
\def\3{{\bf 3}}
\def\4{{\bf 4}}
\def\6{{\bf 6}}
\def\8{{\bf 8}}
\def\OR{\Omega\mathcal{R}}

\def\targ#1#2{\genfrac{[}{]}{0pt}{}{#1}{#2}}

\def\targ2#1#2{\genfrac{}{}{0pt}{}{#1}{#2}}

\definecolor{blus}{rgb}{0.1,0.1,0.8}
\definecolor{GreenYellow}{cmyk}{0.15,0,0.69,0}
\definecolor{Yellow}{cmyk}{0,0,1,0}
\definecolor{Goldenrod}{cmyk}{0,0.10,0.84,0}
\definecolor{Dandelion}{cmyk}{0,0.29,0.84,0}
\definecolor{Apricot}{cmyk}{0,0.32,0.52,0}
\definecolor{Peach}{cmyk}{0,0.50,0.70,0}
\definecolor{Melon}{cmyk}{0,0.46,0.50,0}
\definecolor{YellowOrange}{cmyk}{0,0.42,1,0}
\definecolor{Orange}{cmyk}{0,0.61,0.87,0}
\definecolor{BurntOrange}{cmyk}{0,0.51,1,0}
\definecolor{Bittersweet}{cmyk}{0,0.75,1,0.24}
\definecolor{RedOrange}{cmyk}{0,0.77,0.87,0}
\definecolor{Mahogany}{cmyk}{0,0.85,0.87,0.35}
\definecolor{Maroon}{cmyk}{0,0.87,0.68,0.32}
\definecolor{BrickRed}{cmyk}{0,0.89,0.94,0.28}
\definecolor{Red}{cmyk}{0,1,1,0}
\definecolor{OrangeRed}{cmyk}{0,1,0.50,0}
\definecolor{RubineRed}{cmyk}{0,1,0.13,0}
\definecolor{WildStrawberry}{cmyk}{0,0.96,0.39,0}
\definecolor{Salmon}{cmyk}{0,0.53,0.38,0}
\definecolor{CarnationPink}{cmyk}{0,0.63,0,0}
\definecolor{Magenta}{cmyk}{0,1,0,0}
\definecolor{VioletRed}{cmyk}{0,0.81,0,0}
\definecolor{Rhodamine}{cmyk}{0,0.82,0,0}
\definecolor{Mulberry}{cmyk}{0.34,0.90,0,0.02}
\definecolor{RedViolet}{cmyk}{0.07,0.90,0,0.34}
\definecolor{Fuchsia}{cmyk}{0.47,0.91,0,0.08}
\definecolor{Lavender}{cmyk}{0,0.48,0,0}
\definecolor{Thistle}{cmyk}{0.12,0.59,0,0}
\definecolor{Orchid}{cmyk}{0.32,0.64,0,0}
\definecolor{DarkOrchid}{cmyk}{0.40,0.80,0.20,0}
\definecolor{Purple}{cmyk}{0.45,0.86,0,0}
\definecolor{Plum}{cmyk}{0.50,1,0,0}
\definecolor{Violet}{cmyk}{0.79,0.88,0,0}
\definecolor{RoyalPurple}{cmyk}{0.75,0.90,0,0}
\definecolor{BlueViolet}{cmyk}{0.86,0.91,0,0.04}
\definecolor{Periwinkle}{cmyk}{0.57,0.55,0,0}
\definecolor{CadetBlue}{cmyk}{0.62,0.57,0.23,0}
\definecolor{CornflowerBlue}{cmyk}{0.65,0.13,0,0}
\definecolor{MidnightBlue}{cmyk}{0.98,0.13,0,0.43}
\definecolor{NavyBlue}{cmyk}{0.94,0.54,0,0}
\definecolor{RoyalBlue}{cmyk}{1,0.50,0,0}
\definecolor{Blue}{cmyk}{1,1,0,0}
\definecolor{Cerulean}{cmyk}{0.94,0.11,0,0}
\definecolor{Cyan}{cmyk}{1,0,0,0}
\definecolor{ProcessBlue}{cmyk}{0.96,0,0,0}
\definecolor{SkyBlue}{cmyk}{0.62,0,0.12,0}
\definecolor{Turquoise}{cmyk}{0.85,0,0.20,0}
\definecolor{TealBlue}{cmyk}{0.86,0,0.34,0.02}
\definecolor{Aquamarine}{cmyk}{0.82,0,0.30,0}
\definecolor{BlueGreen}{cmyk}{0.85,0,0.33,0}
\definecolor{Emerald}{cmyk}{1,0,0.50,0}
\definecolor{JungleGreen}{cmyk}{0.99,0,0.52,0}
\definecolor{SeaGreen}{cmyk}{0.69,0,0.50,0}
\definecolor{Green}{cmyk}{1,0,1,0}
\definecolor{ForestGreen}{cmyk}{0.91,0,0.88,0.12}
\definecolor{PineGreen}{cmyk}{0.92,0,0.59,0.25}
\definecolor{LimeGreen}{cmyk}{0.50,0,1,0}
\definecolor{YellowGreen}{cmyk}{0.44,0,0.74,0}
\definecolor{SpringGreen}{cmyk}{0.26,0,0.76,0}
\definecolor{OliveGreen}{cmyk}{0.64,0,0.95,0.40}
\definecolor{RawSienna}{cmyk}{0,0.72,1,0.45}
\definecolor{Sepia}{cmyk}{0,0.83,1,0.70}
\definecolor{Brown}{cmyk}{0,0.81,1,0.60}
\definecolor{Tan}{cmyk}{0.14,0.42,0.56,0}
\definecolor{Gray}{cmyk}{0,0,0,0.50}
\definecolor{Black}{cmyk}{0,0,0,1}
\definecolor{White}{cmyk}{0,0,0,0}
\definecolor{LightGray}{gray}{0.8}

\definecolor{mygr}{rgb}{0,0.6,0}
\definecolor{mygrey}{rgb}{0,0.1,0.2}
\definecolor{myblue}{rgb}{0,0.5,0.9}
\definecolor{myblue2}{rgb}{0,0.5,0.5}
\definecolor{myorange}{rgb}{1,0.5,0}
\definecolor{mypurple}{rgb}{0.6,0,1}
\definecolor{mygolden}{rgb}{1,0.8,0.2}

\newcommand{\bCaptionfonts}{\small}
\makeatletter
\long\def\@makecaption#1#2{%
  \vskip\abovecaptionskip
  \sbox\@tempboxa{{\bCaptionfonts #1: #2}}%
  \ifdim \wd\@tempboxa >\hsize
    {\bCaptionfonts #1: #2\par}
  \else
    \hbox to\hsize{\hfil\box\@tempboxa\hfil}%
  \fi
  \vskip\belowcaptionskip}
\makeatother

\makeatletter
\let\ORIGINALlatex@openbib@code=\@openbib@code
\renewcommand{\@openbib@code}{\ORIGINALlatex@openbib@code\setlength{\itemsep}{1ex plus.5ex minus.5ex}\setlength{\parsep}{0pt}}
\makeatother

\def\mathtabfix#1#2#3{\begin{table}[th]\bCentering\resizebox{\linewidth}{!}{$#1$}\bCaption{#3}\label{tab:#2}\end{table}}

\renewcommand{\arraystretch}{1.3}

\allowdisplaybreaks[4]

\begin{document}
\begin{center}
\begin{flushright}
{\small MITP/15-055\\ 
\today}

\end{flushright}

\vspace{25mm}
{\Large\bf Deformations on Tilted Tori and Moduli Stabilisation at the Orbifold Point}
\vspace{12mm}

{\large Michael Blaszczyk${}^{\clubsuit}$, Gabriele Honecker${}^{\heartsuit}$ and Isabel Koltermann${}^{\spadesuit}$
}

\vspace{8mm}
{
\it PRISMA Cluster of Excellence \& Institut f\"ur Physik  (WA THEP), \\Johannes-Gutenberg-Universit\"at, D-55099 Mainz, Germany
\;$^{\clubsuit}${\tt blaszczyk@uni-mainz.de},~$^{\heartsuit}${\tt Gabriele.Honecker@uni-mainz.de},~$^{\spadesuit}${\tt kolterma@uni-mainz.de}}

\vspace{15mm}{\bf Abstract}\\[2ex]\parbox{140mm}{
We discuss deformations of orbifold singularities on tilted tori in the context of Type IIA orientifold model building with D6-branes on special Lagrangian cycles.
Starting from $T^6/(\mathbb{Z}_2 \times \mathbb{Z}_2)$, we mod out an additional $\mathbb{Z}_3$ symmetry to describe phenomenologically appealing backgrounds and reduce to $\mathbb{Z}_3$ and $\Omega\mathcal{R}$ invariant orbits of deformations. While D6-branes carrying SO$(2N)$ or USp$(2N)$ gauge groups do not constrain deformations, D6-branes with U$(N)$ gauge groups develop non-vanishing D-terms if they couple to previously singular, now deformed cycles. We present examples for both types of D6-branes, and in a three-generation Pati-Salam model on $T^6/(\mathbb{Z}_2 \times \mathbb{Z}_6')$ we find that ten out of 15 twisted complex structure moduli are indeed stabilised at the orbifold point by the existence of the brane stacks.
}
\end{center}

\thispagestyle{empty}
\clearpage 

\tableofcontents
\setlength{\parskip}{1em plus1ex minus.5ex}
%

\section{Introduction}\label{S:intro}

String model building is since its beginnings plagued by the conflict of interest between generic Calabi-Yau threefolds as compact backgrounds and a small number of geometrically accessible  singular orbifold limits thereof. While the former class of compactifications basically restricts to the supergravity regime, the latter admits the use of powerful Conformal Field Theory (CFT) techniques.

In both cases, however, the dilaton and geometric moduli form flat directions. On the one hand, this allows to tune parameters to achieve phenomenologically acceptable values of the strengths of gauge and gravitational interactions, but on the other hand, the flat directions impede the predictive power of string theory. 
Over the years, various mechanisms to achieve (partial) moduli stabilisation have been proposed, most notably closed string background fluxes in Type II orientifolds~\cite{Grana:2005jc}, which
back-react on the geometry and change it away from Calabi-Yau manifolds or singular limits thereof.
This change in geometry further exacerbates the construction of explicit stringy vacua for particle physics and cosmology, see e.g. the discussion for the six-torus, potentially endowed with some $\Z_2 \times \Z_2$ symmetry in~\cite{Camara:2005dc,Aldazabal:2006up}.\footnote{
At this point, it is worth noting  that model building in Type IIA orientifolds has to date only provided {\it gobally} defined models on the six-torus and its orbifolds, see e.g.~\cite{Blumenhagen:2002gw,Honecker:2004kb,Honecker:2004np,Blumenhagen:2005tn,Bailin:2006zf,Bailin:2007va,Gmeiner:2007we,Gmeiner:2007zz,Gmeiner:2008xq,Bailin:2008xx,Forste:2008ex,Honecker:2012jd,Honecker:2012qr,Bailin:2013sya,Ecker:2014hma} for GUT and MSSM-like models with fractional D6-branes stuck at some $\Z_2$ singularities and~\cite{Cvetic:2001tj,Cvetic:2001nr,Honecker:2003vq,Gmeiner:2005vz,Blumenhagen:2005mu,Blumenhagen:2006ci,Ibanez:2012zz} for more comprehensive lists of references, while {\it semi-local} models on  (smooth) Calabi-Yau manifolds exist to our best knowledge only in terms of hypersurfaces in weighted in projective spaces~\cite{Palti:2009bt}, which might possibly be extended to Complete Intersection Calabi-Yau manifolds using the recently reported special Lagrangian cycles in~\cite{Apruzzi:2014dza}. {\it Semi-local} refers here to the uncertainty in the contribution of the O6-planes to the RR tadpole cancellation conditions and the K-theory constraints, which would require a classification of USp- vs. SO-type probe branes. For toroidal orbifolds, such a classification can be done via CFT techniques, see e.g.~\cite{Gmeiner:2009fb,Forste:2010gw,Honecker:2011sm}, while for properties of special Lagrangians on generic Calabi-Yau manifolds, results in the mathematics literature are scarce~\cite{Joyce:2001xt,Joyce:2001nm}.
}

Within heterotic $E_8 \times E_8$ model building, it was noticed a few years ago that some orbifold singularities cannot be resolved without breaking supersymmetry. 
While there the identification of moduli is complicated due to the fact that all fields group into charged representations under some gauge group, it was shown that blow-ups of orbifold singularities~\cite{Nibbelink:2009sp, Blaszczyk:2010db, Nibbelink:2007rd} lead to non-vanishing Fayet-Iliopoulos terms of Abelian gauge factors and thus supersymmetry breaking at one-loop \cite{Atick:1987gy,Blumenhagen:2005ga}.

In~\cite{Blumenhagen:2005pm,Blumenhagen:2005zg,Blumenhagen:2005zh}, the analogous argument was formulated first for the $SO(32)$ heterotic string and then S-dualised to the Type I string with fluxed D9/D5-branes, where the
Fayet-Iliopoulos terms arise as $\alpha^{\prime}$ corrections. 
Employing mirror symmetry to arrive at Type IIA/$\OR$ orientifolds, we expect that now complex structure deformations away from the singular orbifold point lead to supersymmetry breaking. 
Contrary to heterotic orbifold models, Type II orientifolds allow for a unique identification of the geometric moduli from the closed string sector, and indeed in~\cite{Blaszczyk:2014xla} we found first evidence that deformations in the presence of D-branes with U(1) symmetries break supersymmetry. In this article, we generalise the simplified discussion of~\cite{Blaszczyk:2014xla} to backgrounds of interest for D-brane model building, in particular the $T^6/(\Z_2 \times \Z_6^{(\prime)} \times \OR)$ orientifolds, which provide three particle generations due to the inherent $\Z_3$ subgroup~\cite{Honecker:2012qr,Ecker:2014hma,Ecker:2015vea}.

Our new findings in this article show that a large number of geometric moduli can be stabilised even without invoking closed string background fluxes and severe back-reactions on the geometry. In addition, our computation of periods over (special) Lagrangian cycles provide the tree-level value of the gauge couplings~\cite{Aldazabal:2000cn}, and upon deformation previously identical couplings can either be enhanced or diminished. This property is particularly interesting in the context of the hierarchy between strong and weak interactions in some explicit Pati-Salam model presented here.

This article is organised as follows: in section~\ref{S:HypersurfaceFormalism} we briefly review the hypersurface formalism and apply it for the first time to hexagonal lattices, as required for orbifold backgrounds with an underlying $\Z_3$ symmetry. The main focus lies on the $\Z_2 \times \Z_6^{\prime}$ orbifold with discrete torsion, and extensions to  $\Z_2 \times \Z_6$  with a different $\Z_6$ action are briefly addressed in section~\ref{Sss:Z2Z6}.
In section~\ref{S:ConcreteModels}, we discuss two different classes of examples on the $\Z_2 \times \Z_6^{\prime}$ orbifold with discrete torsion and address the question of (twisted) moduli stabilisation: in the first class, all D-branes are orientifold invariant with SO or USp gauge factors, while D-branes in the second class carry unitary gauge groups whose D-terms account for a non-trivial (twisted) moduli potential. 
Our conclusions are given in section~\ref{S:Conclusions}.
Last but not least, in appendix~\ref{A:1} we collect technical details for the computations on the $\Z_2 \times \Z_6^{\prime}$ orbifold, and in appendix~\ref{A:2} we provide a generic overview of Lagrangian lines on tori of untilted and tilted shape beyond the hexagonal ones in the main text.

\section{Deformations of Orbifold Singularities and Hypersurface Formalism}\label{S:HypersurfaceFormalism}

In this section we discuss three-cycles, on which D6-branes in Type IIA/$\OR$ orientifold models can be wrapped. 
Our focus lies on supersymmetric D-brane configurations and the phenomenologically appealing $T^6/ (\Z_2 \times \Z_6^{\prime})$ orbifold background, for which explicit examples will be discussed in section~\ref{S:ConcreteModels}.

\subsection{The geometric setup}\label{Ss:Geometry}

This section briefly describes the geometric setting we are dealing with. For a more detailed discussion see e.g.\ \cite{Blaszczyk:2014xla}. We start with the geometry of toroidal orbifolds and the fractional three-cycles therein. Then the hypersurface formalism is reviewed, first for two-tori and then for the $T^6/ (\Z_2 \times \Z_{2N})$ orbifold with discrete torsion including its deformations. We focus in particular on the behaviour of {\it (special) Lagrangian} three-cycles under deformations.

\paragraph{Three-cycles on orbifolds:}
We focus on orbifolds of the type $T^6 / (\Z_2 \times \Z_{2N})$ with discrete torsion where the six-torus $T^6$ is factorisable, i.e.\ it can be written as a product of three mutually orthogonal two-tori, $T^6 = T^2_{(1)} \times T^2_{(2)} \times T^2_{(3)}$. The orbifold group contains a $\Z_2 \times \Z_2$ subgroup which allows to switch on a global discrete torsion phase
$\eta=\pm 1$~\cite{Vafa:1994rv}. Consequently, the $\Z_2 \times \Z_2$ singularities can be deformed $(\eta=-1)$ rather than blown-up $(\eta=+1)$, which results in exceptional three-cycles at the former fixed loci that can be used for D6-brane model building purposes~\cite{Blumenhagen:2005tn,Forste:2010gw}. Thus, we describe the orbifold as $\bigl( T^6 / (\Z_2 \times \Z_2) \bigr) / \Z_N$ where the $\Z_N$ ideally does not lead to exceptional three-cycles, but rather restricts the way how the $\Z_2 \times \Z_2$  singularities can be deformed, cf.\ table~\ref{tab:Hodge-numbers} for the corresponding Hodge numbers per twist sector. 
\mathtabfix{
\renewcommand{\arraystretch}{1.3}
\begin{array}{|c||c|c|c|c|c|c|c|c||c|}\hline
\muc{10}{|c|}{\text{\bf Hodge numbers $\boldsymbol{(h_{11},h_{21})}$ of factorisable $\boldsymbol{T^6/(\Z_2 \times \Z_{2N})}$ orbifolds with discrete torsion}}
\\\hline\hline
T^6/ & \multicolumn{8}{|c||}{\text{Lattice: } \text{SU}(2)^6} &  \left(\begin{array}{c}  h_{11} \\ h_{21}  \end{array}\right)
 \\\hline
 \text{Sector} & \text{Bulk}
 & (0,\frac{1}{2},\frac{-1}{2}) &  (\frac{1}{2},\frac{-1}{2},0) &  (\frac{1}{2},0,\frac{-1}{2}) & \muc{4}{|c||}{} & \text{Total}
\\\hline
 \begin{array}{c}  \mathbb{Z}_2 \times \mathbb{Z}_2 \\ \eta=-1 \end{array}&
   \left(\begin{array}{c} 3 \\   3 \end{array}\right) 
 & \left(\begin{array}{c} 0 \\ 16 \end{array}\right)
 & \left(\begin{array}{c} 0 \\ 16 \end{array}\right)
  & \left(\begin{array}{c} 0 \\ 16 \end{array}\right)
  & \muc{4}{|c||}{} &   \left(\begin{array}{c} 3 \\ 51 \end{array}\right)
\\\hline\hline
T^6/ & \multicolumn{8}{|c||}{\text{Lattice: }  \text{SU}(2)^2 \times \text{SU}(3)^2} &
\\\hline
 \text{Sector} & \text{Bulk}
 & (0,\frac{1}{2},\frac{-1}{2}) 
& (\frac{1}{2},\frac{-1}{2},0) 
 &   (\frac{1}{2},0,\frac{-1}{2})
 & (0,\frac{1}{3},\frac{-1}{3}) 
 & (0,\frac{1}{6},\frac{-1}{6}) 
 & (\frac{-1}{2},\frac{1}{3},\frac{1}{6}) 
 & (\frac{-1}{2},\frac{1}{6},\frac{1}{3}) 
 &  \text{Total}
\\\hline
 \begin{array}{c}  \mathbb{Z}_2 \times \mathbb{Z}_6 \\ \eta=-1 \end{array}
 &  \left(\begin{array}{c} 3 \\   1 \end{array}\right) 
 & \left(\begin{array}{c} 0 \\ 6 \end{array}\right)
 & \left(\begin{array}{c} 0 \\  4 \end{array}\right) 
 & \left(\begin{array}{c} 0 \\  4 \end{array}\right)
 & \left(\begin{array}{c} 8 \\ 2 \end{array}\right)
 & \left(\begin{array}{c} 0 \\ 2 \end{array}\right)
 & \left(\begin{array}{c} 4 \\ 0 \end{array}\right)
 & \left(\begin{array}{c} 4 \\ 0 \end{array}\right)
 & \left(\begin{array}{c} 19  \\ 15+4  \end{array}\right)
 \\\hline\hline
T^6/ &  \multicolumn{8}{|c||}{\text{Lattice: }  \text{SU}(3)^3} & 
 \\\hline
 \text{Sector} & \text{Bulk}
& (0,\frac{1}{2},\frac{-1}{2}) 
& (\frac{1}{2},\frac{-1}{2},0) 
&  (\frac{1}{2},0,\frac{-1}{2})
& (\frac{-2}{3},\frac{1}{3},\frac{1}{3}) 
&  (\frac{-1}{3},\frac{1}{6},\frac{1}{6}) 
& (\frac{1}{6},\frac{-1}{3},\frac{1}{6}) 
& (\frac{1}{6},\frac{1}{6},\frac{-1}{3})  
 & \text{Total}
\\\hline
\begin{array}{c}    \mathbb{Z}_2 \times \mathbb{Z}_6'  \\ \eta =-1  \end{array} 
& \left(\begin{array}{c} 3 \\   0 \end{array}\right)
& \left(\begin{array}{c} 0 \\  5 \end{array}\right)
& \left(\begin{array}{c} 0 \\  5 \end{array}\right)
&  \left(\begin{array}{c} 0 \\  5  \end{array}\right)
& \left(\begin{array}{c} 9 \\ 0 \end{array}\right)
& \left(\begin{array}{c} 1 \\  0 \end{array}\right)
& \left(\begin{array}{c} 1 \\ 0 \end{array}\right)
& \left(\begin{array}{c} 1 \\ 0 \end{array}\right)
& \left(\begin{array}{c} 15  \\  15 \end{array}\right)
\\\hline
\end{array}
}{Hodge-numbers}{Hodge numbers of factorisable $T^6/(\Z_2 \times \Z_{2N})$ orbifolds with discrete torsion.
For both choices $\Z_2 \times \Z_6^{(\prime)}$, 16 independent three-cycles in $\Z_2$-twisted sectors are reduced to 4, 5 or 6 independent ones due to the additional $\Z_3$ symmetry.
For $\Z_2 \times \Z_6$, two out of the four additional $\Z_3$- and $\Z_6$-twisted sectors contribute $h_{21}^{\Z_3+\Z_6}=4$, while all other three-cycles - on which this article focuses - stem from the bulk and $\Z_2$-twisted sectors.}
Orbifolds containing a $\Z_3$ factor are of particular phenomenological interest~\cite{Honecker:2012qr,Honecker:2013kda,Ecker:2014hma}.

In such geometries there are three types of three-cycles:
\begin{itemize}
 \item A one-cycle on a two-torus $T^2_{(k)}$ is specified by its two integer-valued coprime wrapping numbers $(n_k, m_k)$. Three-cycles on the six-torus are products of three one-cycles, one on each two-torus. Adding up all orbifold images of such torus cycles, one obtains a bulk cycle on the orbifold,
 \begin{align}
  \Pi^{\rm bulk} = \sum_{\Z_2 \times \Z_{2N} \atop \text{images}} \left( n_1 \pi_1 + m_1 \pi_2 \right) \otimes \left( n_2 \pi_3 + m_2 \pi_4 \right) \otimes \left( n_3 \pi_5 + m_3 \pi_6 \right) \,, \label{Eqn:bulkcycle}
 \end{align}
with basis one-cycles $\pi_k$ oriented as in figure~\ref{fig:Square_tilted_lattice}.
 \item Locally, a resolved $\Z_2^{(k)}$ singularity along $T^4_{(k)} \equiv T^2_{(i)} \times T^2_{(j)}$, labelled by $(\alpha\beta)$, results in a two-cycle $e_{\alpha\beta}^{(k)}$ which topologically is a two-sphere. In the case of discrete torsion, the remaining $\Z_2$ factor acts with a minus sign on that two-cycle, so in order to obtain an invariant exceptional cycle one has to take the product with a one-cycle in the $k^{\rm th}$ two-torus:
 \begin{align}
  {\boldsymbol \varepsilon}_{\alpha\beta}^{(k)}  = \sum_{\Z_N \atop \text{images}}  2 \, e_{\alpha\beta}^{(k)} \otimes \pi_{2k-1} \,, \qquad 
  \tilde{\boldsymbol \varepsilon}_{\alpha\beta}^{(k)}  =  \sum_{\Z_N \atop \text{images}}  2 \, e_{\alpha\beta}^{(k)} \otimes \pi_{2k} \, ,
\label{Eqn:excycle}
 \end{align}
 where the sum over $\Z_N$ orbifold images for $T^6/(\Z_2 \times \Z_{2N})$ with discrete torsion has to be taken.\footnote{These exceptional three-cycles exist only for $N$ odd, i.e.\ $N=1,3$, since the discrete torsion phase for $N=2$ acts trivially on the $\Z_2$ twisted sectors, see~\cite{Forste:2010gw} for details.}
\item If a bulk cycle happens to pass through the orbifold singularities, it becomes its own $\Z_2$-orbifold image. Thus the sum in equation~\eqref{Eqn:bulkcycle} becomes redundant (up to the additional $\Z_N$ factor within $\Z_{2N}$, which also appears on the right hand sides of equation~\eqref{Eqn:excycle}). However, such a fractional cycle is closed only if one adds an appropriate contribution from the exceptional cycles:
\begin{align}
 \Pi^{\rm frac} = \frac14 \Pi^{\rm bulk} + \frac14 \sum_{k=1}^{3} \sum_{\alpha_k, \beta_k} \left(  x^{(k)}  \boldsymbol{\varepsilon}_{\alpha_k\beta_k}^{(k)} + y^{(k)} \boldsymbol{\tilde{\varepsilon}}_{\alpha_k\beta_k}^{(k)} \right) \,,
\end{align}
with the integer-valued exceptional wrapping numbers $(x^{(k)},y^{(k)})$ composed of a limited choice of linear combinations of $(n_k,m_k)$ weighted with sign factors due to $\Z_2$ eigenvalues $(\pm)$ and discrete Wilson lines $(\vec{\tau})$ with $\tau^i \in \{0,1\}$, as detailed e.g.\ in table~7 of~\cite{Ecker:2014hma}.
Here the sum over the fixed points denoted by $\alpha_k \in T^2_{(i)}$, $\beta_k \in T^2_{(j)}$ has to be chosen appropriately to match the position of the bulk two-cycle within $T^4_{(k)}$, 
which we  parametrise below by $(\vec{\sigma})$ with $\sigma^i \in \{0,1\}$.
\end{itemize}

\paragraph{Supersymmetry conditions:}

In orientifold models, the condition for $\mathcal{N}=1$ supersymmetry in four dimensions requires that D6-branes may only wrap three-cycles $\Pi$ which satisfy the {\it special Lagrangian (sLag)} 
 conditions,
\begin{align}
 \mathcal{J}_{1,1} \big|_{\Pi} = 0 \,, \label{Eqn:LagCond} \\
 \Im\left(\Omega_{3}\right) \big|_{\Pi} = 0 \,, \quad  \Re\left(\Omega_{3}\right) \big|_{\Pi} > 0 \,, \label{Eqn:sLagCond} 
\end{align}
where $ \mathcal{J}_{1,1}$ denotes the K\"ahler form and $\Omega_{3}$ the holomorphic volume form.
If a cycle only satisfies \eqref{Eqn:LagCond}, it is called {\it Lagrangian}, and the three-cycles we constructed in the previous paragraph automatically have this property. The {\it sLag} conditions \eqref{Eqn:sLagCond} then impose constraints on the wrapping numbers and the complex structure moduli. For example, it can happen that the bulk part of a fractional cycle is {\it sLag}, but its exceptional part is not. If a D6-brane wraps such a cycle, the model is only supersymmetric if the volume of the exceptional part is zero, i.e.\ at the orbifold point.

\subsubsection{Lagrangian cycles in the hypersurface formalism}
In order to describe the geometry including the deformations of all singularities, the orbifold of the factorised six-torus
has to be reformulated as a hypersurface in an ambient toric space. To this end we start with elliptic curves in $\mathbbm{P}^2_{112}$ and Lagrangian lines therein. Then we write down the $T^6/(\Z_2 \times \Z_2)$ orbifold and arbitrary deformations thereof as a hypersurface and show how the Lagrangian lines from the elliptic curves are inherited. By modding out another $\Z_3$ symmetry we can achieve descriptions of the deformed \mbox{$T^6/(\Z_2 \times \Z_{6}^{(\prime)})$} orbifolds with discrete torsion.

\begin{figure}[th]
\includegraphics[width=0.9\textwidth]{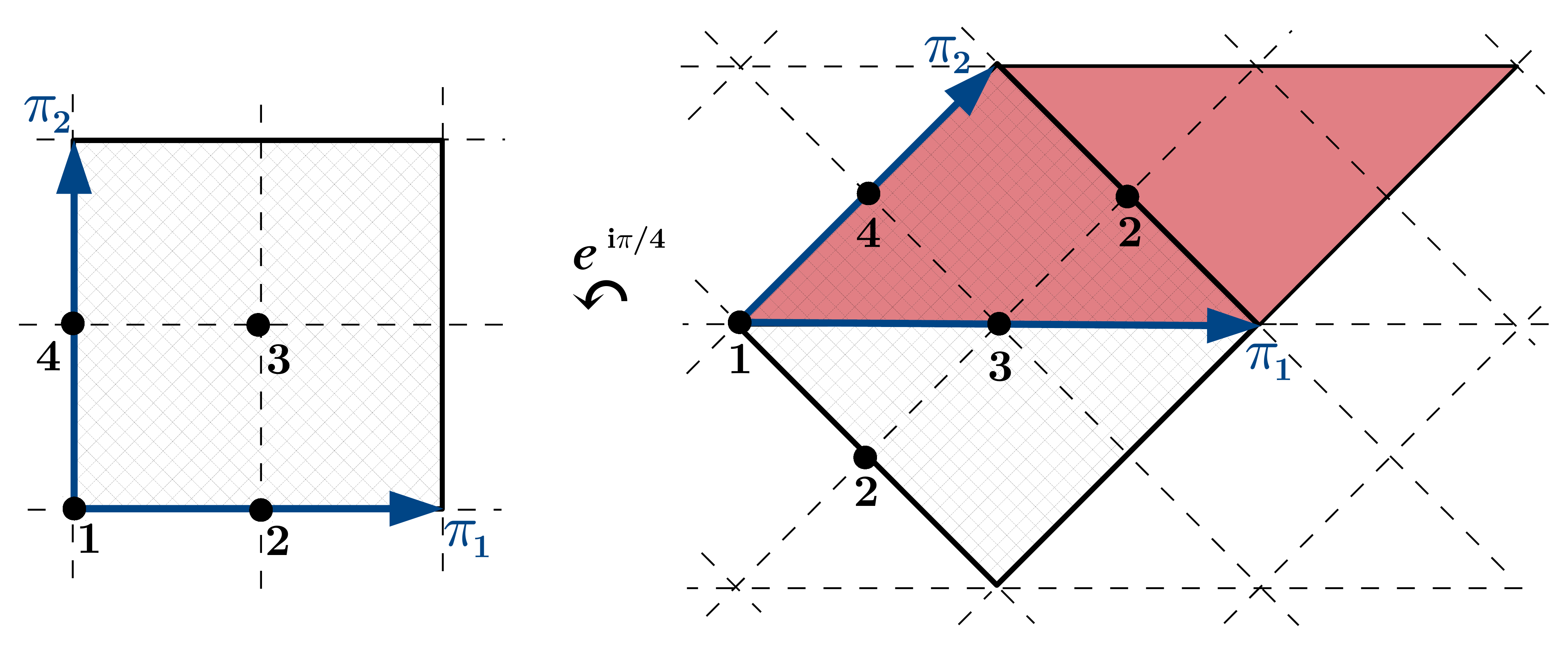}
\caption{Square torus with underlying untilted ({\bf a}-type) lattice at the left hand side and tilted ({\bf b}-type) lattice at the right hand side. The fixed points are labelled in such a way that rotating the square torus on the tilted ({\bf b}-type) lattice by $e^{\pi i/4}$ gives the same index notation as for the square torus with underlying untilted ({\bf a}-type) lattice.
The anti-holomorphic orientifold involution ${\cal R}$ acts by reflection along the horizontal axis passing through the $\Z_2$ fixed points 1 and 2 for the {\bf a}-type and 1 and 3 for the {\bf b}-type lattice. Throughout the article, we use analogous fixed point labels on generic tilted tori, in particular on hexagonal ones.
}
\label{fig:Square_tilted_lattice}
\end{figure}

\paragraph{Elliptic curves:}
For later purposes it is most convenient to transcribe the two-torus as a hypersurface in a complex weighted projective space $\mathbbm{P}^2_{112}$ with homogeneous coordinates $x$, $v$ and $y$ of weights 1, 1 and 2, respectively. An elliptic curve is the zero locus of a polynomial $f$ of degree four, which can generally be chosen to be in Weierstrass form,
\begin{align}
f := -y^2 + F(x,v) \stackrel{!}{=} 0 \,, \qquad F(x,v):= 4 v x^3 - g_2 v^3 x - g_3 v^4 \,, \label{Eqn:EllipticCurveEquation}
\end{align}
with real coefficients $g_2$, $g_3$. The anti-holomorphic orientifold involution $\sigma_{\cal R}$ acts as $(y,x,v) \mapsto (\overline y, \overline x, \overline v)$. The $\Z_2$ reflection acts by $y \mapsto -y$, so its fixed points are exactly the roots of $F(x,v)$, see figure~\ref{fig:Square_tilted_lattice} for our choice of labels 1,2,3,4.
Thus, it is convenient to write the polynomial in factorised form,
\begin{align}
F(x,v) = 4 \, v \, ( x-\epsilon_2 v) \cdot ( x- \epsilon_3 v ) \cdot (x - \epsilon_4 v) \,,
\end{align}
with $g_2 = - 4 \sum_{i<j} \epsilon_i \epsilon_j $, $g_3 = 4 \epsilon_2 \epsilon_3 \epsilon_4$ and $ \epsilon_2 + \epsilon_3 + \epsilon_4 = 0$. To make the $g_i$ real there are two possibilities:
\begin{itemize}
\item $\epsilon_4 < \epsilon_3 < \epsilon_2$ with all $\epsilon_i$ real corresponds to untilted tori or {\bf a}-type lattices,
\item $\epsilon_2 = \overline\epsilon_4 =: \epsilon$ and $\epsilon_3$ real corresponds to tilted tori or {\bf b}-type lattices.
\end{itemize}
If $g_2=0$, the elliptic curve exhibits an additional $\Z_3$ symmetry acting by $z \mapsto e^{2 \pi i /3} z$, i.e.\ it is a torus whose defining lattice is hexagonal as a special case of a tilted torus. This restriction is required whenever a $\Z_3$ subgroup acts on a given two-torus.

On the two-torus $T^2_{(k)}$, there is an infinite set of one-cycles specified by the coprime wrapping numbers $(n_k,m_k)$ and in addition by a continuous displacement parameter. The set of three-cycles which can be used for model building is, however, finite, since most cycles overshoot the RR tadpole cancellation conditions. The one-cycles which are linearly realised in terms of the homogeneous coordinates of the elliptic curve are just a finite set which depends on the complex structure. We focus on such cycles since they are rather easy to parametrise and since they serve our purposes sufficiently well. Basically, they are the ``horizontal''  and ``vertical'' cycles with wrapping numbers $(n,m)$ = (1,0) and (0,1) for untilted tori or $(n,m)$ = (1,0) and (-1,2) for tilted tori, respectively, which pass through the $\Z_2$ fixed points. The results are summarised in table~\ref{tab:T2LagLines_Untilted_Tilted}.

\begin{table}[th]
\centering
\resizebox{\linewidth}{!}{
\subfloat[]{
\begin{tabular}{|>{\centering}c|c|c|} \hline
\muc{3}{|c|}{\text{\bf \textit{Lag} lines of a-type lattice}}
\\\hline\hline
Label & Condition & Picture \\
\hline 
	 ${\bf aI}$
	& $x/v \geq \epsilon_2 $
	&  \raisebox{-5pt}{\includegraphics[scale=0.2]{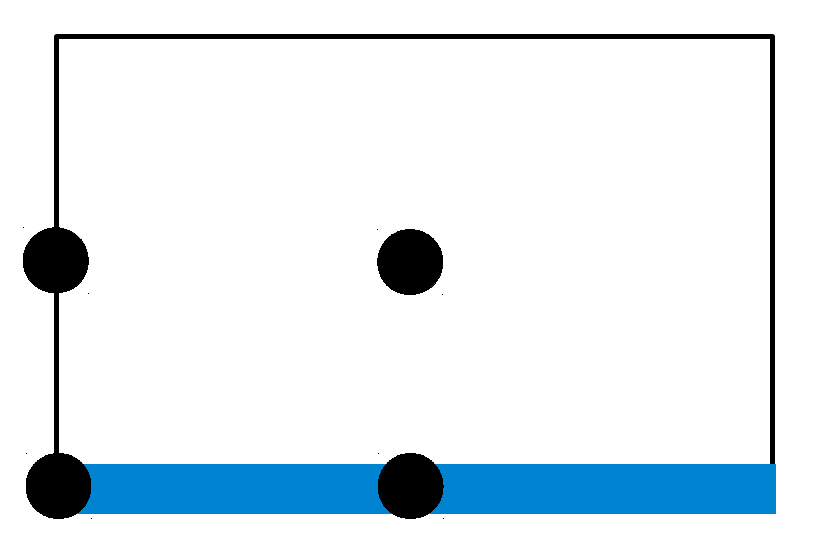}} \\
\hline
	 ${\bf aII}$
	& $\epsilon_2 \geq x/v \geq \epsilon_3 $
	&  \raisebox{-5pt}{\includegraphics[scale=0.2]{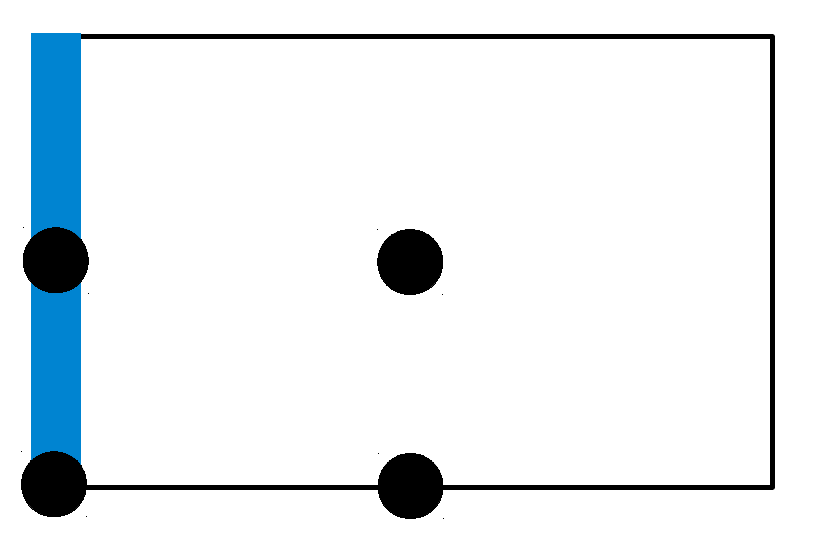}} \\
\hline
	 ${\bf aIII}$
	& $\epsilon_3 \geq x/v \geq \epsilon_4 $
	& \raisebox{-5pt}{\includegraphics[scale=0.2]{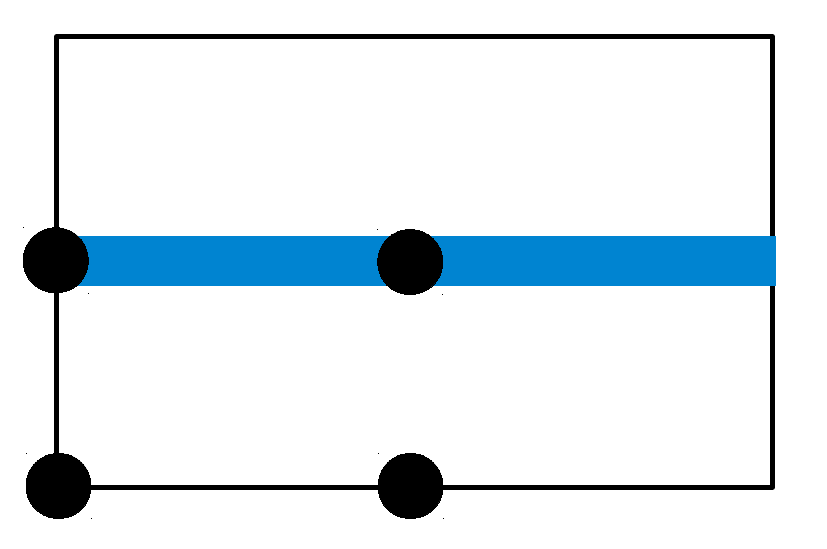}} \\
\hline
	 ${\bf aIV}$
	& $\epsilon_4 \geq x/v $
	& \raisebox{-5pt}{\includegraphics[scale=0.2]{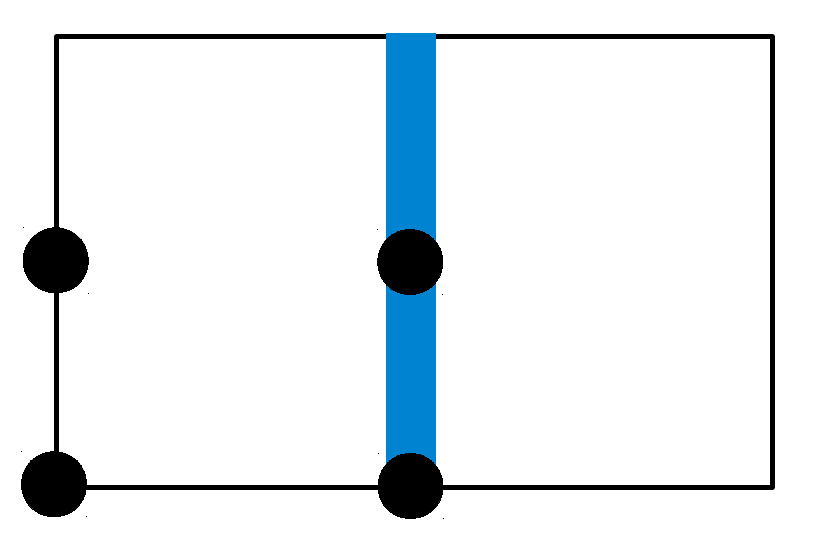}} \\
\hline
\end{tabular}
}
\subfloat[]{
\begin{tabular}{|c|c|c|} \hline
\muc{3}{|c|}{\text{\bf \textit{Lag} lines of b-type lattice}}
\\\hline\hline
Label & Condition & Picture \\
\hline
	 ${\bf bI}$
	& $x/v \geq - 2 \Re (\epsilon) $
	&  \includegraphics[scale=0.2]{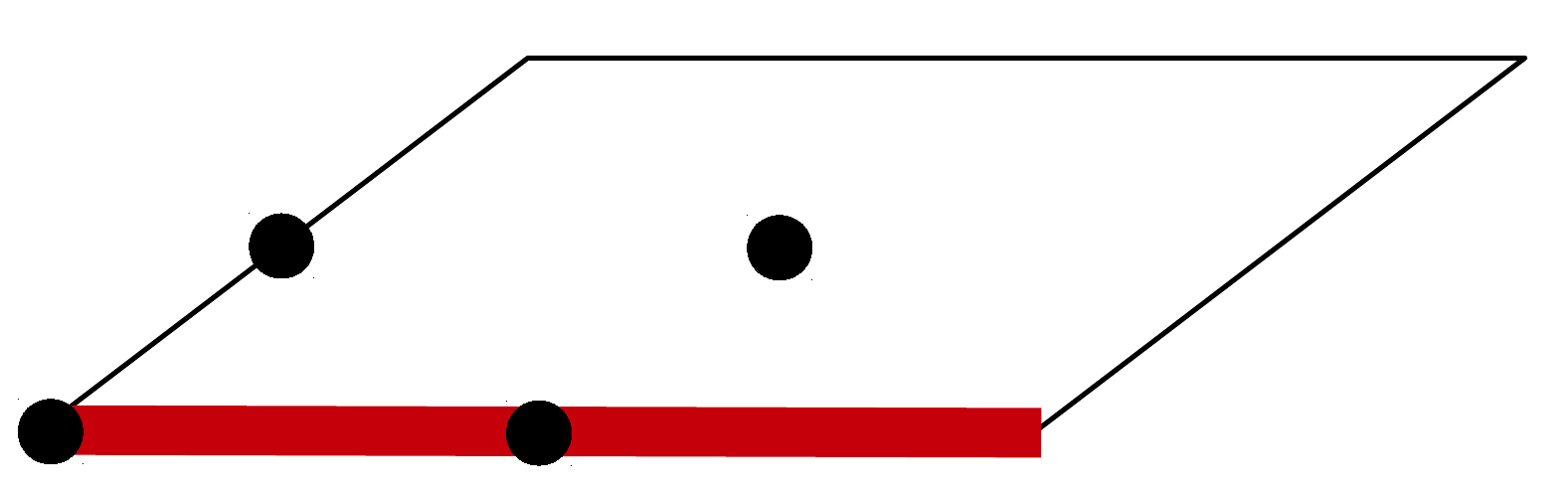} \\
\hline
	 ${\bf bII}$
	&  $x/v \leq - 2 \Re (\epsilon) $
	&  \includegraphics[scale=0.2]{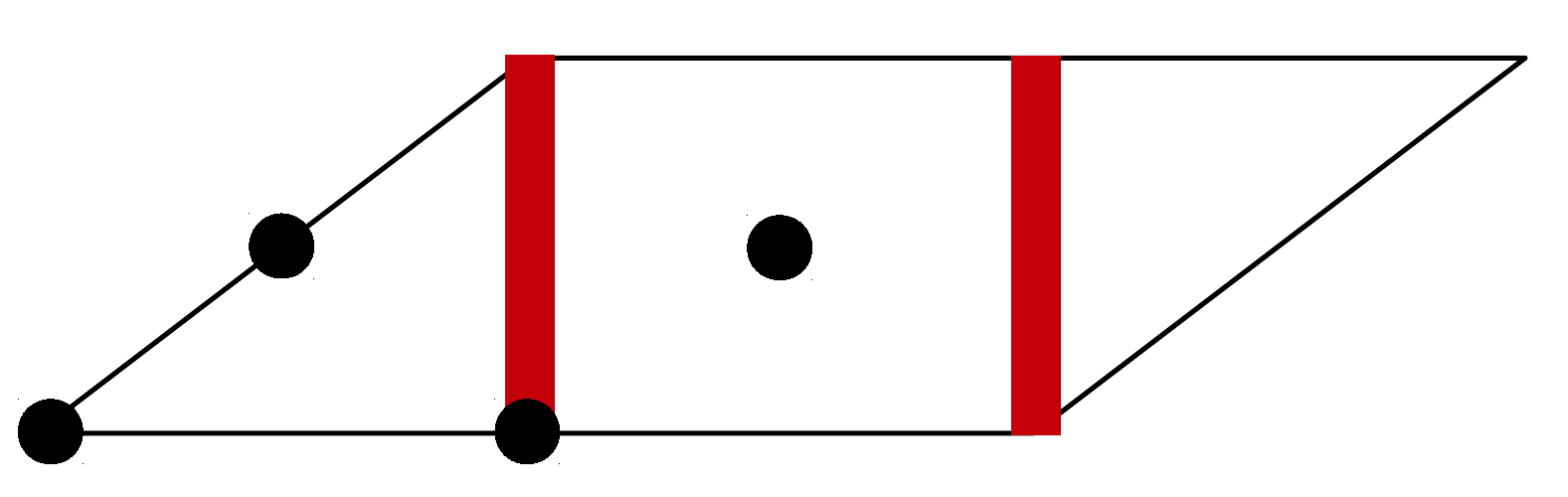} \\
\hline
	 ${\bf bIII}$
	& $
	\begin{array}{c}
	\left| x/v - 2 \Re (\epsilon) \right|^2 = |\epsilon|^2 + 8 \Re (\epsilon)^2 \\
	\Re(x/v) \geq \Re(\epsilon) 
	\end{array}	$
	&  \includegraphics[scale=0.2]{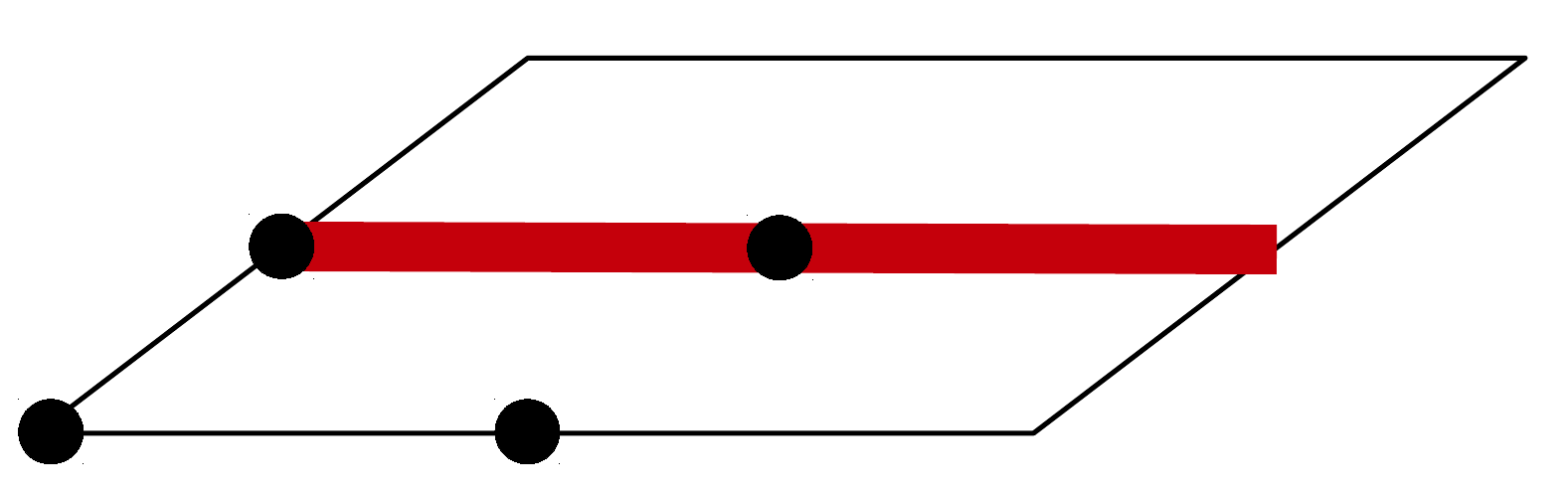} \\
\hline
	 ${\bf bIV}$
	& $
	\begin{array}{c}
	\left| x/v - 2 \Re (\epsilon) \right|^2 = |\epsilon|^2 + 8 \Re (\epsilon)^2 \\
	\qquad \Re(x/v) \leq \Re(\epsilon)
	\end{array} $
	&  \includegraphics[scale=0.2]{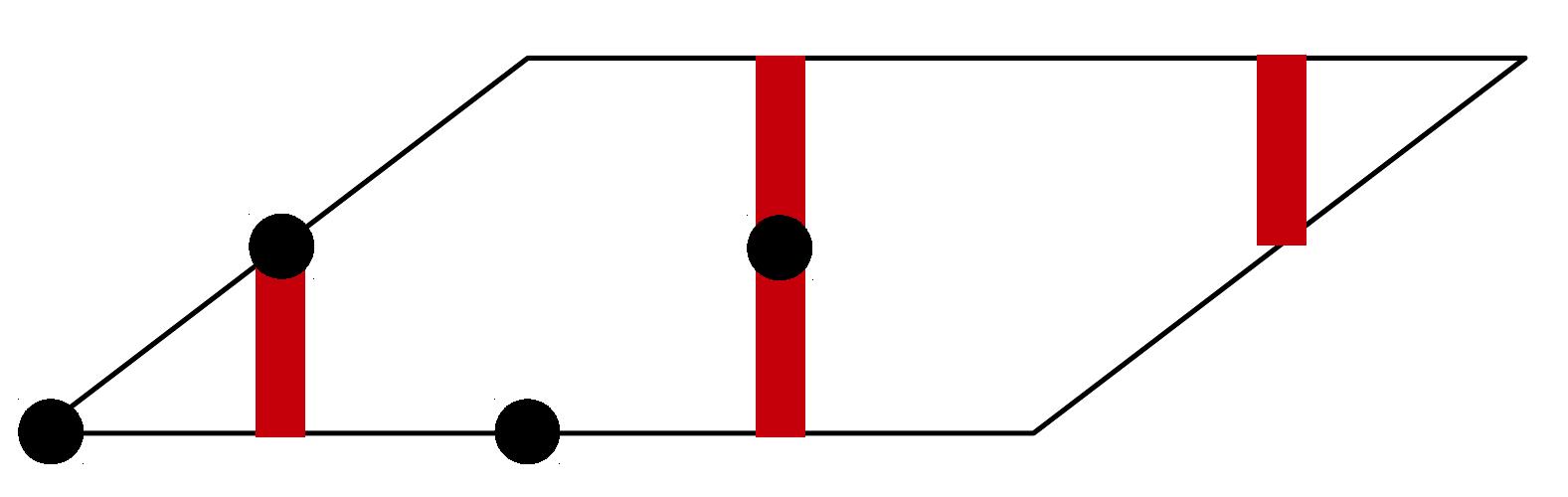} \\
\hline
\end{tabular}
}}
\caption{Linearly realised {\it Lag} lines on the (a) untilted and (b) tilted elliptic curve. }
\label{tab:T2LagLines_Untilted_Tilted}
\end{table}

There are two special cases for the complex structure in which the number of linear one-cycles enhances:
\begin{itemize}
\item If $g_3=0$, the elliptic curve has an underlying square lattice, which is at the same time tilted and untilted. Thus, both types of one-cycles ({\bf aX} and {\bf bX}, {\bf X} = {\bf I},\ldots,{\bf IV} 
in table~\ref{tab:T2LagLines_Untilted_Tilted}(a) and (b), respectively) can be constructed in the untilted lattice with the {\bf bX} rotated by 45 degrees (and analogously for the tilted lattice), see figure \ref{fig:Square_tilted_lattice} and appendix \ref{A:2}.
\item If $g_2=0$, the underlying lattice is hexagonal, which is a special tilted torus. In this case, the $\Z_3$ symmetry can be used to triple the amount of one-cycles. More precisely, we recover not only all {\bf bX} cycles in table~\ref{tab:T2LagLines_Untilted_Tilted}(b), but also their by $\pm \frac{2\pi}{3}$ rotated images.
\end{itemize}

\subsubsection{$\boldsymbol{T^6/(\Z_2 \times \Z_{2N})}$ as a hypersurface}

Now we construct a deformable version of $T^6/(\Z_2 \times \Z_{2N})$ with discrete torsion by starting with $T^6/(\Z_2 \times \Z_{2})$ with discrete torsion and modding out the remaining $\Z_N$ symmetry in section~\ref{Ss:Orbs-with-Z3}. The construction of toroidal orbifolds which allow for a blow-up of the singularities, e.g.\ for models without discrete torsion, was explained in~\cite{Blaszczyk:2011hs, Lust:2006zh}. The $\Z_2 \times \Z_2$ orbifold can be embedded in the toric space with weights $q_i$ shown in the following table:
\begin{center}
\begin{tabular}{|c|c|c|c|c|c|} 
\hline 
Coordinate & $x_1,v_1$ & $x_2,v_2$ & $x_3,v_3$ & $y$ & $f(y,x_i,v_i)$ \\ 
\hline 
$q_1$ & 1 & 0 & 0 & 2 & 4 \\
\hline 
$q_2$ & 0 & 1 & 0 & 2 & 4 \\
\hline 
$q_3$ & 0 & 0 & 1 & 2 & 4 \\
\hline
\end{tabular}
\end{center}
Here, $f(y,x_i,v_i)$ denotes the equation whose zero locus is the orbifold or its deformation. The most general form of $f$ is 
\begin{align}
\begin{split}
f &= -y^2 + F_1(x_1,v_1) F_2(x_2,v_2) F_3(x_3,v_3) \\
&- \sum_{i\neq j \neq k \neq i} \sum_{\alpha,\beta=1}^4 \varepsilon^{(i)}_{\alpha\beta} F_i(x_i,v_i) \delta F_j^{(\alpha)} (x_j,v_j) \delta F_k^{(\beta)} (x_k,v_k) \\
&+ \sum_{\alpha,\beta,\gamma =1}^4 \varepsilon_{\alpha\beta\gamma} \delta F_1^{(\alpha)} (x_1,v_1) \delta F_2^{(\beta)} (x_2,v_2) \delta F_3^{(\gamma)} (x_3,v_3) 
\, , 
\label{Eqn:T6Z2Z2HypersurfaceEquation}
 \end{split}
\end{align}
where we use the symbol $\varepsilon^{(i)}_{\alpha\beta}$ for the deformation parameter associated to the exceptional three-cycles $\boldsymbol{\varepsilon}^{(i)}_{\alpha\beta}$ in equation \eqref{Eqn:excycle}.

We explain the terms which appear in \eqref{Eqn:T6Z2Z2HypersurfaceEquation}:
\begin{itemize}
\item The $F_i$ are homogeneous polynomials of degree four as in \eqref{Eqn:EllipticCurveEquation}. They encode the information on the complex structure of the $i^{\rm th}$ two-torus. If we set $\varepsilon^{(i)}_{\alpha\beta} \equiv \varepsilon_{\alpha\beta\gamma} \equiv 0$, we see that the $x_i,v_i$ dependent part factorises into three pieces, each of them describing one two-torus $T^2_{(i)}$. However, the fact that there is only one $y$ coordinate indicates that a $\Z_2 \times \Z_2$ symmetry has been modded out. 
\item The $\delta F_i^{(\alpha)}$ are also homogeneous polynomials of degree four such that together with $F_i$ they constitute a basis of the symmetric polynomials $S^4(x_i,v_i)$ of fourth order. Furthermore, we choose them such that $\delta F_i^{(\alpha)}$ and $F_i$ have the same zeros up to the $\alpha^{\rm th}$ one. In this way $\delta F_i^{(\alpha)}$ corresponds to deforming the $\alpha^{\rm th}$ $\Z_2$-fixed point, see equation~\eqref{Eq:Def-Polynomials_Z2Z6} below for details in the case with additional $\Z_3$ symmetry.
\item The parameter $\varepsilon^{(i)}_{\alpha\beta}$ is responsible for deforming the $\Z_2^{(i)}$ singularity with label $(\alpha\beta)$ on $T^4_{(i)} \equiv T^2_{(j)} \times T^2_{(k)}$ 
with $\{ijk\}$ some permutation of \{123\}. Altogether there are $3 \times 4 \times 4$ such parameters, one for each $\Z_2$ singularity.
\item Finally, there is the possibility for 64 terms with coefficients $\varepsilon_{\alpha\beta\gamma}$. In a string model these are not free parameters, but are determined by $\varepsilon^{(i)}_{\alpha\beta}$ such that 64 conifold singularities remain at the points where the $\Z_2 \times \Z_2$ singularities used to intersect, as argued in \cite{Vafa:1994rv}.
\end{itemize}

One last ingredient in the hypersurface formalism is the expression for the holomorphic three-form. It is chosen such that on the one hand it reproduces the familiar expression $dz_1 \wedge dz_2 \wedge dz_3$ in the orbifold limit, and on the other hand contains the complex structure moduli, and in particular the deformation moduli $\varepsilon^{(i)}_{\alpha\beta}$ as parameters. In the coordinate patch $v_i \equiv 1$ the holomorphic three-form takes the form:
\begin{align}
\Omega_3 = \frac{dx_1 \wedge dx_2 \wedge dx_3}{y(x_i)} \,, \label{eqn:Omega3}
\end{align} 
up to a normalisation constant and a possible phase. Here the function $y(x_i)$ is obtained by imposing $f=0$ in equation \eqref{Eqn:T6Z2Z2HypersurfaceEquation} and fixing one branch of the square root.

More details on the untilted square torus can be found in~\cite{Blaszczyk:2014xla}, and in appendix~\ref{A:2} we give a brief account on the generic case with tilted and untilted tori, introducing for the first time the pictorial view of {\it Lag} lines used in section~\ref{Ss:Orbs-with-Z3} below also for the square torus.
\\
From the following section~\ref{Ss:Orbs-with-Z3} onwards, our discussion in the main text focuses on hexagonal lattices and the phenomenologically appealing $T^6/(\Z_2 \times \Z_6^{\prime} \times \OR)$ orientifold with discrete torsion.

\subsection[Orbifolds with additional $\Z_3$ actions: $\Z_2 \times \Z_6'$]{Orbifolds with additional $\boldsymbol{\Z_3}$ actions: $\boldsymbol{\Z_2 \times \Z_6'}$ }\label{Ss:Orbs-with-Z3}

Since this work focuses on models of orbifolds of type $\Z_2 \times \Z_6'$, we want to explain their geometry in some more detail. The orbifold action is given by 
\begin{align}
\theta^k \omega^l : z_i \longmapsto e^{2 \pi i ( k v^i + l w^i )} z_i\,, \qquad \vec v = \frac12( 1, -1, 0) \,, \quad \vec w = \frac16(-2,1,1) \,,
\end{align}
and the orbifold symmetry fixes the underlying lattice to be the root lattice of SU$(3)^3$, up to normalisations for each two-torus, i.e.\ in particular all (untwisted) complex structure moduli are frozen such that all two-tori are hexagonal. We find the following twisted sectors:
\begin{itemize}
\item Three $\Z_2$ twisted sectors: $\theta, \omega^3, \theta\omega^3$. Each of them has 16 fixed planes, which are labelled by two fixed points $(\alpha\beta)$ with $\alpha,\beta = 1,2,3,4$. The fixed point $(11)$ on $T^4_{(k)}$ in the $\Z_2^{(k)}$ twisted sector stays invariant under the remaining $\Z_2$ orbifold action, whereas the other 15 fixed points get mapped onto each other by the $\omega$ action as $2 \rightarrow 3 \rightarrow 4 \rightarrow 2$, which acts as a $\Z_3$ on the $T^6/(\Z_2 \times \Z_2)$ geometry. In the case with discrete torsion, we read off from table~\ref{tab:Hodge-numbers} that each of the five triplets of fixed points possesses one complex structure modulus or, in other words, (the $\Z_3$ orbit of) each fixed point on $T^4_{(k)}$ tensored with a one-cycle on $T^2_{(k)}$ contributes to the geometry.
\item One $\Z_3$ twisted sector: $\omega^2$. It has 27 fixed points $(\alpha,\beta,\gamma)$, labelled by $\alpha,\beta,\gamma = 1,\tilde{2},\tilde{3}$ in each two-torus, which are subject to $\Z_2 \times \Z_2$ identifications, thus leaving nine fixed points with one K\"ahler modulus each (cf.\ table~\ref{tab:Hodge-numbers}). For example, if $\alpha=1$ corresponds to the fixed point at the origin, (111) is $\Z_2 \times \Z_2$ invariant, $(11\tilde{2})+(11\tilde{3})$ forms a pair under $\Z_2^{(1)}$, and $(\tilde{2}\tilde{2}\tilde{2})+(\tilde{2}\tilde{3}\tilde{3})+(\tilde{3}\tilde{3}\tilde{2})+(\tilde{3}\tilde{2}\tilde{3})$ forms a quadruplet under the full $\Z_2 \times \Z_2$.
Since these lead to two- and four-cycles after blow-up, they are not interesting for wrapping D6-branes.
\item Three $\Z_6$ twisted sectors: $\omega, \theta\omega, \theta\omega^4$. After $\Z_2$ identifications, they have two fixed points: one at the origin without moduli and one coinciding with some orbit of $\Z_3$ fixed points with a K\"ahler modulus.
\end{itemize}

We would like to reformulate the geometry with $\Z_3$ symmetry in the hypersurface formalism. To do so, we specify the polynomial $F_i$ and its deformations\footnote{In order to include the additional $\Z_3$ action with a simple deformation behaviour as in equation~\eqref{Eqn:Z3DeltaF}, the polynomials in equation \eqref{Eq:Def-Polynomials_Z2Z6} had to be slightly modified compared to the polynomials specified in equation (39) of \cite{Blaszczyk:2014xla}, which are given in a more general form.} $\delta F_i^{(\alpha)}$ and study the $\Z_3$ action on them:
\begin{align} \label{Eq:Def-Polynomials_Z2Z6}
\begin{split}
F_i(x_i,v_i) &= 4 \left( v_i x_i^3 -  v_i^4 \right) \ =\  4 v_i \left( x_i - v_i \right)\left( x_i - \xi v_i \right)\left( x_i - \xi^2 v_i \right) \,, \\
\delta F_i^{(1)} &= 4 \left( x_i^4 -  x_i v_i^3 \right)\,, \\
\delta F_i^{(2)} &= 4 v_i^2 \left( v_i - x_i \right)\left( v_i - \xi x_i \right) \,, \\
\delta F_i^{(3)} &= 4 v_i^2 \left( v_i - \xi x_i \right)\left( v_i - \xi^2 x_i \right) \,, \\
\delta F_i^{(4)} &= 4 v_i^2 \left( v_i - \xi^2 x_i \right)\left( v_i - x_i \right) \,, \\
\end{split}
\end{align}
where here and from now on we abbreviate $\xi \equiv e^{2 \pi i / 3}$. The $\Z_3$ action on the homogeneous coordinates is $\omega^2 : (x_1,x_2,x_3) \mapsto \xi (x_1,x_2,x_3)$. As required, it leaves $F_i$ invariant, and the deformation polynomials $\delta F_i^{(\alpha)}$ transform as follows:
\begin{align}
\delta F_i^{(1)} \mapsto \xi \delta F_i^{(1)} \,, \qquad \delta F_i^{(2)} \mapsto \delta F_i^{(3)} \mapsto \delta F_i^{(4)} \mapsto \delta F_i^{(2)} \,. \label{Eqn:Z3DeltaF}
\end{align}
Note that the only invariant deformation polynomial is $\delta F^{(2)}_i + \delta F^{(3)}_i + \delta F^{(4)}_i$. However, such a transformation must also leave the torus complex structure invariant and thus can be absorbed in rescalings of the coordinates. In order to deform the $\Z_2$ singularities we must find invariant polynomials of the form $F_i \delta F_j^{(\alpha)} \delta F_k^{(\beta)}$, cf.\ equation~\eqref{Eqn:T6Z2Z2HypersurfaceEquation}, which we can interpret as restrictions on the coefficients $\varepsilon^{(i)}_{\alpha\beta}$. We find:
\begin{itemize}
\item $\varepsilon^{(i)}_{11} = 0 $ in accordance with the fact that the fixed point at the origin has no complex structure modulus,
\item $\varepsilon^{(i)}_{31}=\xi\varepsilon^{(i)}_{21}=\xi^2\varepsilon^{(i)}_{41} =: \varepsilon^{(i)}_{1}$,
\item $\varepsilon^{(i)}_{13}=\xi\varepsilon^{(i)}_{12}=\xi^2\varepsilon^{(i)}_{14}=:\varepsilon^{(i)}_{2}$,
\item $\varepsilon^{(i)}_{22}=\varepsilon^{(i)}_{33}=\varepsilon^{(i)}_{44}=:\varepsilon^{(i)}_{3}$,
\item $\varepsilon^{(i)}_{23}=\varepsilon^{(i)}_{34}=\varepsilon^{(i)}_{42}=:\varepsilon^{(i)}_{4}$,
\item $\varepsilon^{(i)}_{24}=\varepsilon^{(i)}_{32}=\varepsilon^{(i)}_{43}=:\varepsilon^{(i)}_{5}$,
\end{itemize}
and for later convenience we define $\varepsilon^{(i)}_{4\pm5}$ via $\varepsilon^{(i)}_{4/5} = \left( \varepsilon^{(i)}_{4+5} \pm i \varepsilon^{(i)}_{4-5} \right) /2$. Note that the same numbering is used for ($\Z_3$ orbits of) deformed $\Z_2$-singularities, i.e.\ for $e^{(i)}_{\rho}$ with $\rho = 1,\ldots,5$.

Thus, as expected from the Hodge numbers in table~\ref{tab:Hodge-numbers}, we find five independent deformation parameters in each $\Z_2$ twisted sector. Although there was some ambiguity in the definition of the deformation polynomials $\delta F_i^{(\alpha)}$, any other $\Z_3$ invariant definition would lead to the same result as can be seen by explicitly counting invariant polynomials of degree $(4,4,4)$ in the coordinates $x_i, v_i$. Note that the number of fixed points is actually larger than the number of allowed deformations. Therefore it is worth mentioning that the counting of deformations, which is totally independent from any string theory, agrees with the number of complex structure deformations obtained from conformal field theory.

In addition, we also count the number of invariant ``triple-deformation" terms (cf.\ equation~\eqref{Eqn:T6Z2Z2HypersurfaceEquation}) by checking the restrictions on $\varepsilon_{\alpha\beta\gamma}$. The invariant combinations are:
\begin{itemize}
\item $\varepsilon_{111}$,
\item $\varepsilon_{112} = \xi \varepsilon_{113} = \xi^2 \varepsilon_{114}$\ \  plus two combinations with permuted indices,
\item $\varepsilon_{1,\alpha,\beta} = \xi^2 \varepsilon_{1,\alpha+1,\beta+1} = \xi \varepsilon_{1,\alpha+2,\beta+2}$\ \  and permutations where $\alpha,\beta = 2,3,4$, and addition is modulo three,
\item $\varepsilon_{\alpha,\beta,\gamma} = \varepsilon_{\alpha+1,\beta+1,\gamma+1} = \varepsilon_{\alpha+2,\beta+2,\gamma+2}$, with conventions for $\alpha,\beta,\gamma$ as before.
\end{itemize}
This gives a total amount of $1 + 3 + 9 + 9 = 22$ allowed parameters, compared to $64$ parameters on the $\Z_2 \times \Z_2$ orbifold. The $64$ corresponded to the number of fixed points under the whole $\Z_2 \times \Z_2$ orbifold group. The remaining $\Z_3$ action leaves the one $\Z_2 \times \Z_2$ fixed point at the origin invariant and identifies the other 63 fixed points in triplets, thus we find $22$ such enhanced singularities on $\Z_2 \times \Z_6'$, in agreement with the number of allowed parameters. Therefore a proper tuning of the parameters $\varepsilon_{\alpha\beta\gamma}$ can indeed ensure that the deformation space has 22 conifold (or worse) singularities.

\subsubsection{Orientifold involution}

The orientifold involution acts on the homogeneous coordinates by complex conjugation, $\sigma_\mathcal{R}: (y,x_i,v_i) \mapsto (\overline{y},\overline{x}_i,\overline{v}_i)$. For this to be a symmetry of the hypersurface we require $\sigma_\mathcal{R}(f) = \overline{f}$, which implies that all parameters of the polynomial $f$ must be real. In the case of the undeformed orbifold, this means that each two-torus must be of rectangular or tilted shape, in accordance with the original argument that the anti-holomorphic orientifold involution acts as automorphism on the compactification lattice~\cite{Blumenhagen:1999md,Blumenhagen:1999ev,Forste:2000hx,Blumenhagen:2000ea,Forste:2001gb,Blumenhagen:2002wn}. Furthermore, we find restrictions on the deformation parameters which we discuss explicitly for the hexagonal case. The orientifold involution acts on the deformation polynomials given in equation~\eqref{Eq:Def-Polynomials_Z2Z6} as follows,
\begin{align}
 \delta F_i^{(1)} \mapsto \overline{\delta F_i}^{(1)}\,,  \quad
 \delta F_i^{(2)} \mapsto \overline{\delta F_i}^{(4)}\,,  \quad
 \delta F_i^{(3)} \mapsto \overline{\delta F_i}^{(3)}\,,  \quad
 \delta F_i^{(4)} \mapsto \overline{\delta F_i}^{(2)}\,,
\end{align}
which reflects the fact that the $\Z_2$ fixed points 1 and 3 are mapped onto themselves whereas the $\Z_2$ fixed points 2 and 4 get mapped onto each other, see the right hand side of figure~\ref{fig:Square_tilted_lattice} for the fixed point labels used for the depicted tilted quadratic as well as for the hexagonal {\bf A}-type lattice,\footnote{Analogously to the square tori of {\bf a}- and {\bf b}-type, there exist two orientations of hexagonal lattices, specified as {\bf A}- and {\bf B}-type, with the one-cycle $\pi_2$ at angle $\pi/3$ and $\pi/6$, respectively, with respect to the ${\cal R}$-invariant direction spanned by $\pi_1$.
} 
which has an angle of $\pi/3$ instead of $\pi/4$ among the basis one-cycles $\pi_1$ and $\pi_2$ shown in figure~\ref{fig:Square_tilted_lattice}.
In case of the $T^6/(\Z_2 \times \Z_6')$ orbifold, the $\Z_3 \subset \Z_6'$ and orientifold symmetries lead to the restrictions on the deformation parameters as summarised in table~\ref{tab:Z3Restrictions}.

\begin{table}[t]
\resizebox{\linewidth}{!}{
\begin{tabular}{|c|c|c|c|} \hline
\muc{4}{|c|}{\text{\bf Restrictions on the deformation parameters of $\boldsymbol{T^6/(\Z_2 \times \Z_6')}$}}
\\\hline\hline
$\rho$ & Parameter identifications & Parameter range & Exceptional wrapping numbers \\
\hline 1 & $\xi \varepsilon^{(i)}_{21}=\varepsilon^{(i)}_{31}=\xi^2 \varepsilon^{(i)}_{41}$ & $\mathbb{R}$ & $x^{(i)}_1, y^{(i)}_1 $ \\
\hline 2 & $\xi \varepsilon^{(i)}_{12}=\varepsilon^{(i)}_{13}=\xi^2 \varepsilon^{(i)}_{14}$ & $\mathbb{R}$ & $x^{(i)}_2, y^{(i)}_2 $ \\
\hline 3 & $\varepsilon^{(i)}_{22}=\varepsilon^{(i)}_{33}=\varepsilon^{(i)}_{44}$ & $\mathbb{R}$ & $x^{(i)}_3, y^{(i)}_3 $ \\
\hline 4,\,5 & $\varepsilon^{(i)}_{23}=\varepsilon^{(i)}_{34}=\varepsilon^{(i)}_{42} =  \overline\varepsilon^{(i)}_{24}=\overline\varepsilon^{(i)}_{32}=\overline\varepsilon^{(i)}_{43}$ & $\mathbb{C}$ & $x^{(i)}_4, y^{(i)}_4, x^{(i)}_5, y^{(i)}_5 $ \\
\hline
\end{tabular}}
\caption{$\mathbb{Z}_3$ and $\Omega \mathcal{R}$ restrictions on deformation parameters and wrapping numbers for the associated exceptional three-cycles.}
\label{tab:Z3Restrictions}
\end{table}

One finds that, in each twisted sector, there are three $\Z_3$-triplets of $\Z_2$ fixed points which are preserved by the orientifold involution and thus are restricted to have one {\it real} deformation parameter each, while the remaining two triplets of $\Z_2$ fixed points are exchanged by the orientifold involution, leading to only one {\it complex} deformation parameter. To summarise, the orientifold involution reduces five complex deformation parameters to three real ones plus a complex one. The resulting hypersurface equation for deformations restricted to just one $\Z_2^{(k)}$ sector is shown in equation \eqref{Eqn:Z6HypersurfaceEqn} in appendix \ref{A:1}.

\begin{SCfigure}
\centering
\caption{The blue regions represent the areas of positive values of $y^2(x_i)$. These are the positions of the four mutually non-adjacent octants in the real $x_1$-$x_2$-$x_3$-plane, which coincide with the position of the O6-planes. As detailed in table~\ref{tab:LagLinesHex}, the intersections of the octants are at the coordinates $x_i = 1, \infty$.} 
\includegraphics[width=0.34\textwidth]{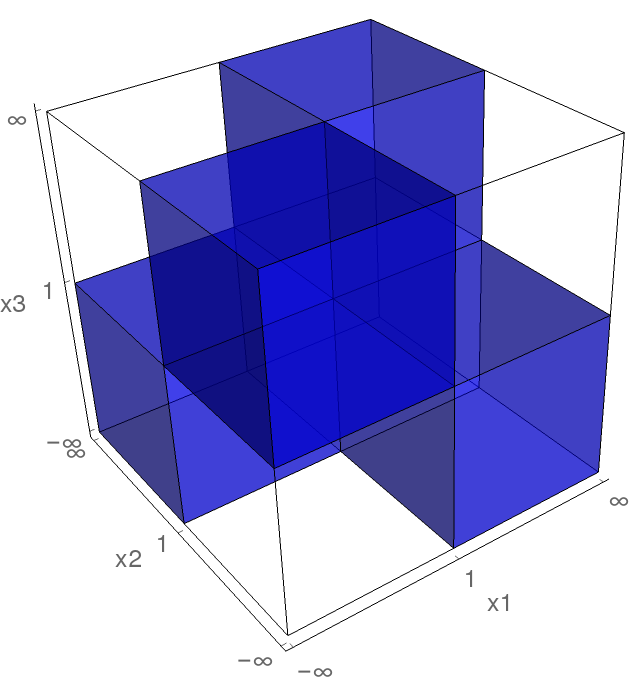} \hspace{20pt}
\label{fig:FourOctants}
\end{SCfigure}

\paragraph{Orientifold plane and \textbf{\textit{sLag}} cycles:}
Having fixed the orientifold action and its restrictions on the deformation parameters $\varepsilon^{(i)}_{\alpha\beta}$, we now determine its fixed set, i.e.\ the set of O6-planes. Since the $\Z_2 \times \Z_2$ action is automatically built in the hypersurface formalism, we have to consider the fixed sets of $\sigma_\mathcal{R}$ times all $\Z_3$ elements.\footnote{Together, the orientifold and the $\Z_3$ groups combine to the symmetric group $S_3$.} 
However, since the $\Z_3$ element $\omega^2$ maps the O6-planes from the sectors $\sigma_\mathcal{R}$, $\sigma_\mathcal{R}\omega^{2}$ and  $\sigma_\mathcal{R}\omega^4$ onto each other, it is sufficient to consider only $\sigma_\mathcal{R}$ O6-planes. 
It is then convenient to go to a coordinate patch where $v_i \neq 0$ and use the $\left(\mathbb{C}^*\right)^3$ action to set $v_i \equiv 1$. Therefore, the O6-planes lie in the three-plane spanned by the real $x_1,x_2,x_3$ coordinates and, more precisely, they are the subset of that plane where $y^2$, written as a function of the $x_i$ by equation \eqref{Eqn:T6Z2Z2HypersurfaceEquation}, is positive. At the orbifold point, this area consists of four mutually non-adjacent octants of the three-plane, separated by the coordinate planes, see the blue coloured regions in figure~\ref{fig:FourOctants}. This is a three-dimensional generalisation of figure~3 in \cite{Blaszczyk:2014xla}. In terms of toroidal wrapping numbers $(n_i,m_i)$, the four ($\Z_3$ orbits of) O6-planes for the so-called {\bf AAA}-orientation\footnote{In~\cite{Honecker:2012qr} it was shown that the {\it a priori} four different choices of background lattice orientations boil down to two physically distinct sets ${\bf AAA} \leftrightarrow {\bf ABB}$ and ${\bf AAB} \leftrightarrow {\bf BBB}$. The first set turns out to be favoured by model building with a minimal (or vanishing) amount of exotic matter charged under the QCD-stack, and the choice of the $\OR$-orbit as exotic O6-plane orbit allows for the maximal possible rank 16 of the overall gauge group.
The explicit models in section~\ref{S:ConcreteModels} refer to this particular choice of background.} of the hexagonal lattice are given in the following table:
\begin{center}
\begin{tabular}{|c|c|c|}
\hline
Orientifold sector & O6-plane charge & Wrapping numbers \\
\hline
$\OR$ & $\eta_{\OR} $ & (1,0;1,0;1,0) \\
\hline
$\OR\Z_2^{(1)}$ & $\eta_{\OR\Z_2^{(1)}} $ & (1,0;-1,2;1,-2)\\
\hline
$\OR\Z_2^{(2)}$ & $\eta_{\OR\Z_2^{(2)}} $ &  (1,-2;1,0;-1,2) \\
\hline
$\OR\Z_2^{(3)}$ & $\eta_{\OR\Z_2^{(3)}} $ & (-1,2;1,-2;1,0) \\
\hline
\end{tabular}
\end{center}
Each of these cycles is undisplaced so that it passes only through the fixed points 1 and 3 in each two-torus, either ``horizontally'' or ``vertically''. Furthermore, we introduced the O6-plane charges in the second column which can take the values $+1$ (normal O6-plane) or $-1$ (exotic O6-plane). These charges are related to the discrete torsion as $\eta =\eta_{\OR} \eta_{\OR\Z_2^{(1)}} \eta_{\OR\Z_2^{(2)}} \eta_{\OR\Z_2^{(3)}} $, so in our case (i.e.\ with discrete torsion) we must choose one of the O6-planes to be exotic\footnote{The choice of three exotic O6-planes does not allow for supersymmetric D6-brane solutions to the bulk RR tadpole cancellation conditions, see~\cite{Blumenhagen:2005tn,Forste:2010gw,Honecker:2012qr,Ecker:2014hma} for details.}.
Since the O6-planes are their own $\Z_2 \times \Z_2$ images, they seem to be fractional cycles which contain contributions from exceptional three-cycles. However, CFT calculations show that the Klein bottle and M\"obius strip amplitudes do not have any contribution from twisted sectors in the tree channel (see e.g.~\cite{Blumenhagen:2002gw,Honecker:2004kb}), thus the sum of all O6-planes is topologically a (fraction of a) pure bulk cycle. 

\begin{table}[ht!]
\bCentering
\resizebox{\linewidth}{!}{
\begin{tabular}{|c|c|c|c|c|c|}\hline
\muc{5}{|c|}{\textbf{Properties of all linearly realised \textit{Lag} lines on the hexagonal torus}}
\\\hline\hline
Wrapping numbers & Displacement  & Equation in $x$ & Label & \\
\hline\hline 
 	& \raisebox{-12pt}{0}
	& \raisebox{-12pt}{$x \ge 1$}
	& \raisebox{-12pt}{$\bf bI^0$}
	& \raisebox{2ex - .95\height}{\includegraphics[scale=0.16]{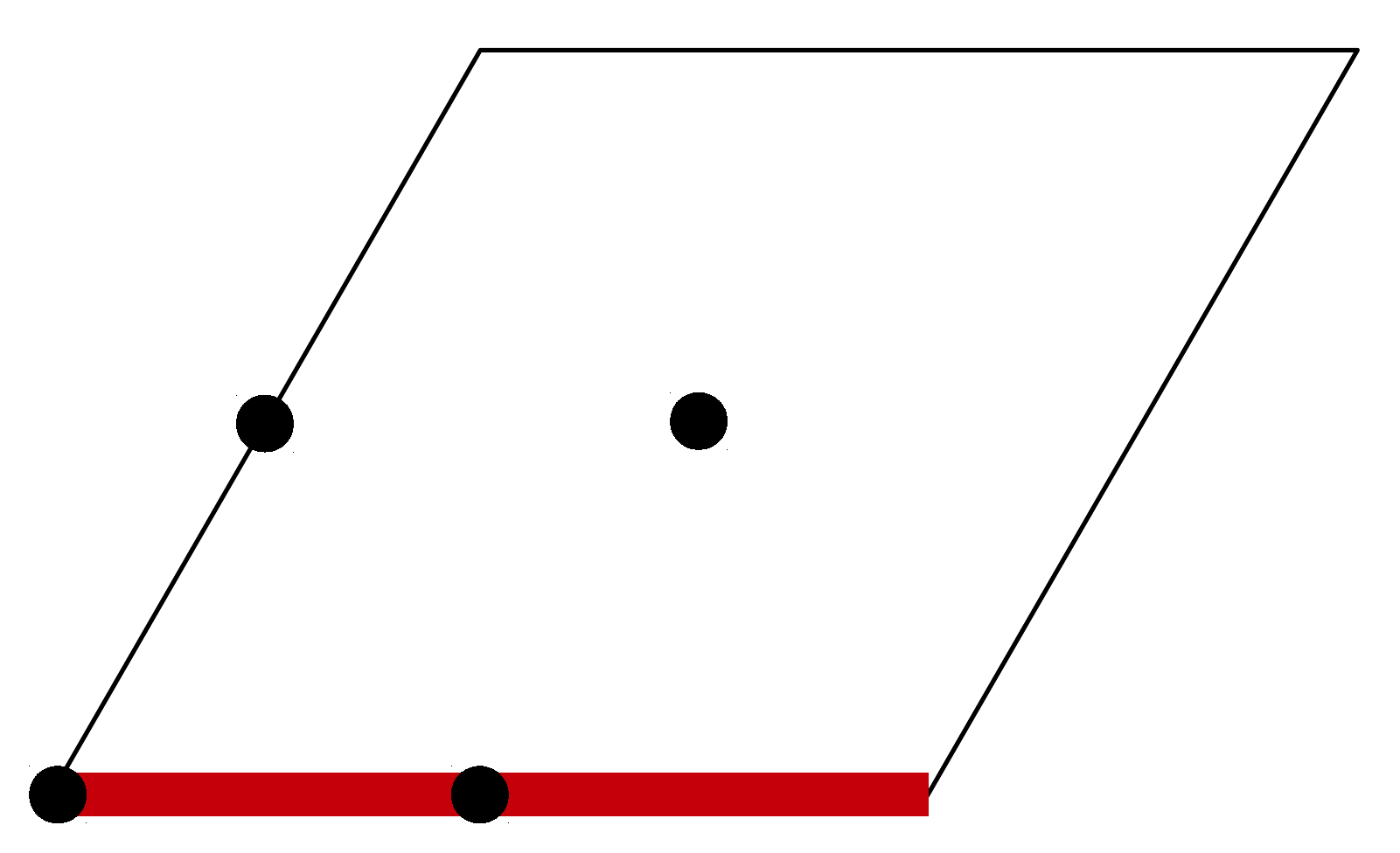}} \\
\cline{2-5} 
 \multirow{-2}{*}{$\pm(1,0)$}
	& \raisebox{-12pt}{1}
	& \raisebox{-12pt}{$| x - 1 |^2 = 3 \,, \Re(x) \le -1/2   $}
   & \raisebox{-12pt}{$\bf bIII^0$} 
   & \raisebox{2ex - .95\height}{\includegraphics[scale=0.16]{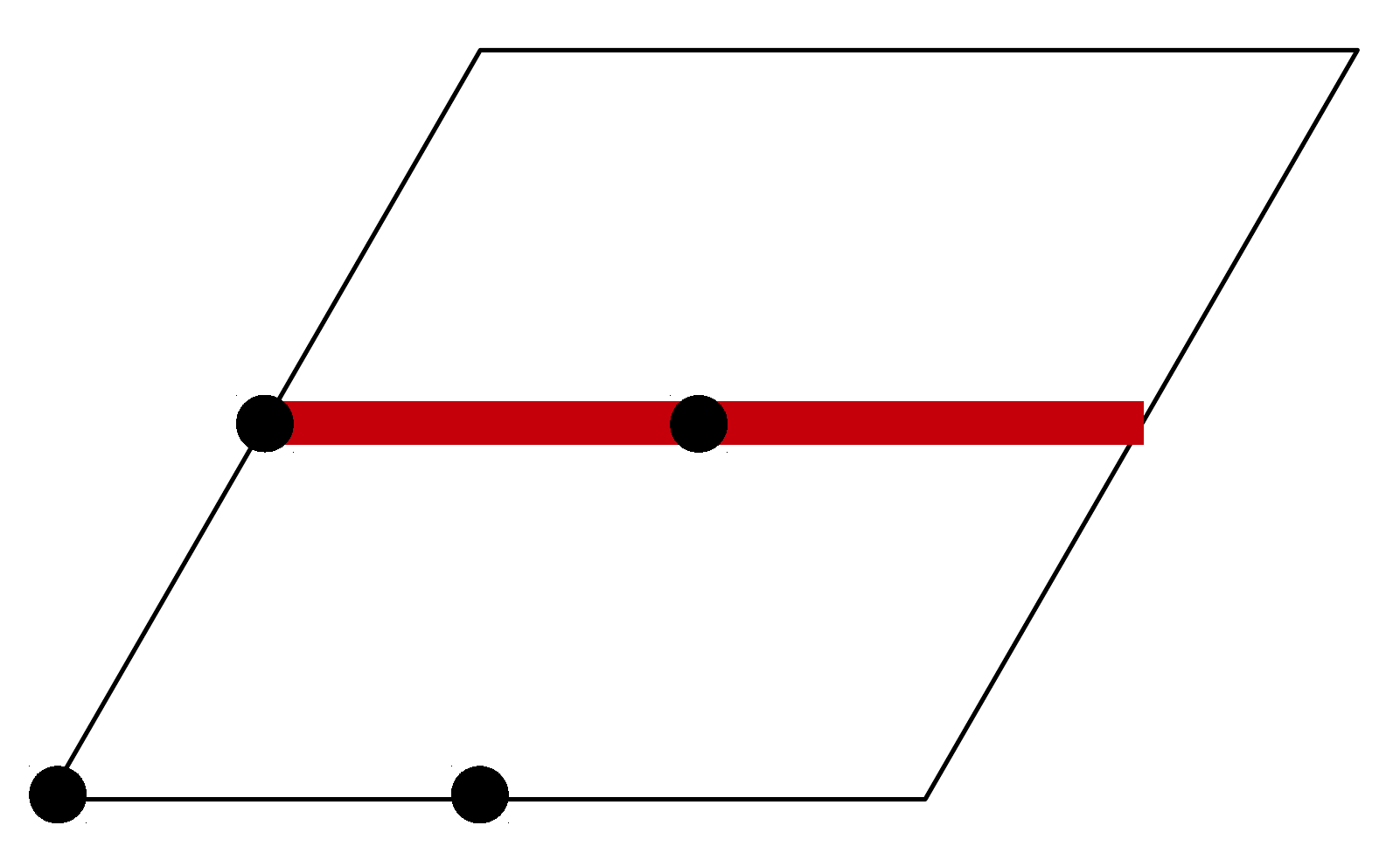}} \\
\cline{1-5} 
	&  \raisebox{-12pt}{$0$}
	& \raisebox{-12pt}{$x \le 1$}
	&  \raisebox{-12pt}{$\bf bII^0$} 
	&  \raisebox{2ex - .95\height}{\includegraphics[scale=0.16]{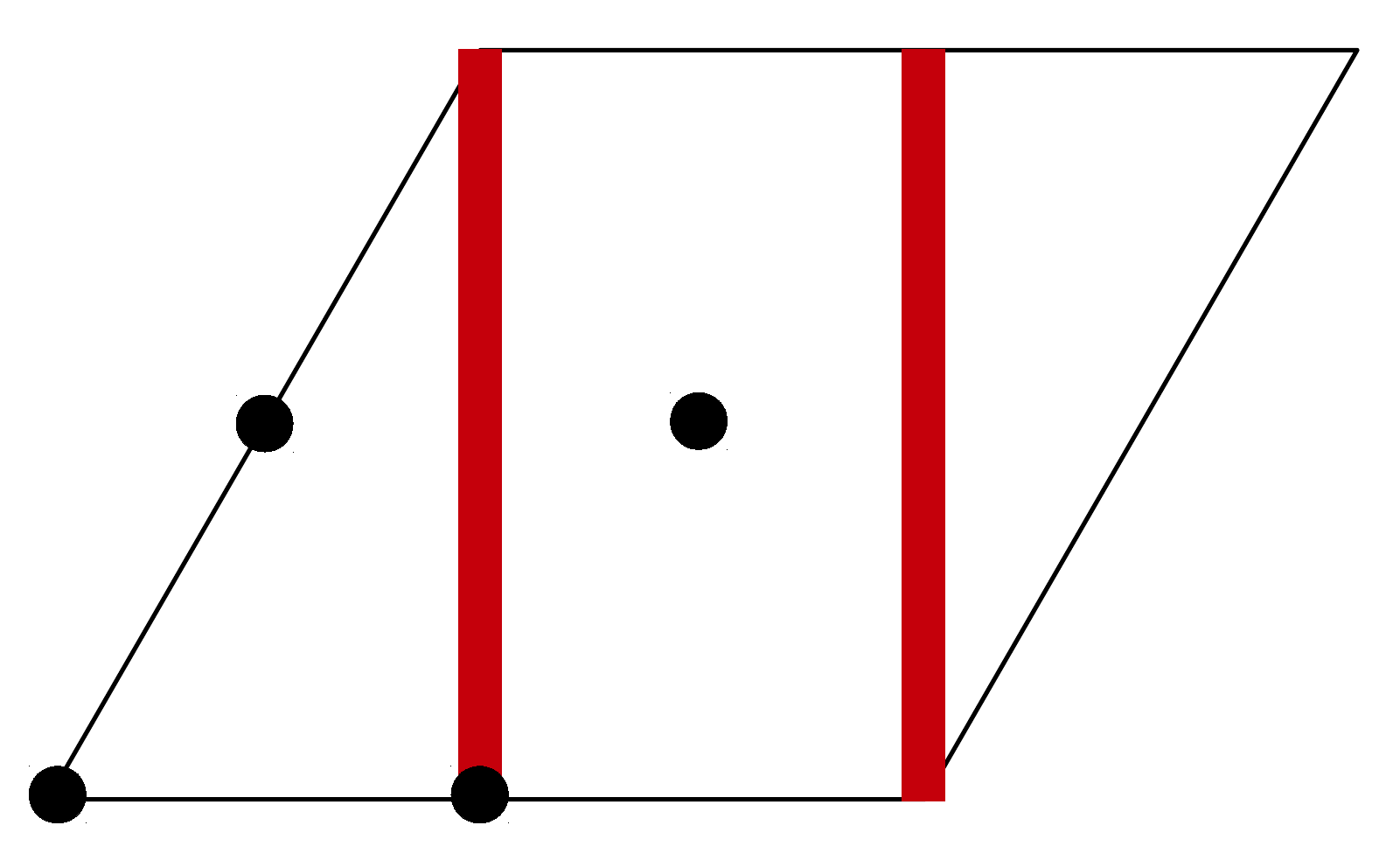}} \\
\cline{2-5} 
\multirow{-2}{*}{$\pm(-1,2)$}
	& \raisebox{-12pt}{$1$}
	& \raisebox{-12pt}{$| x - 1 |^2 = 3 \,, \Re(x) \ge -1/2   $}
 	& \raisebox{-12pt}{$\bf bIV^0$} 
 	& \raisebox{2ex - .95\height}{\includegraphics[scale=0.16]{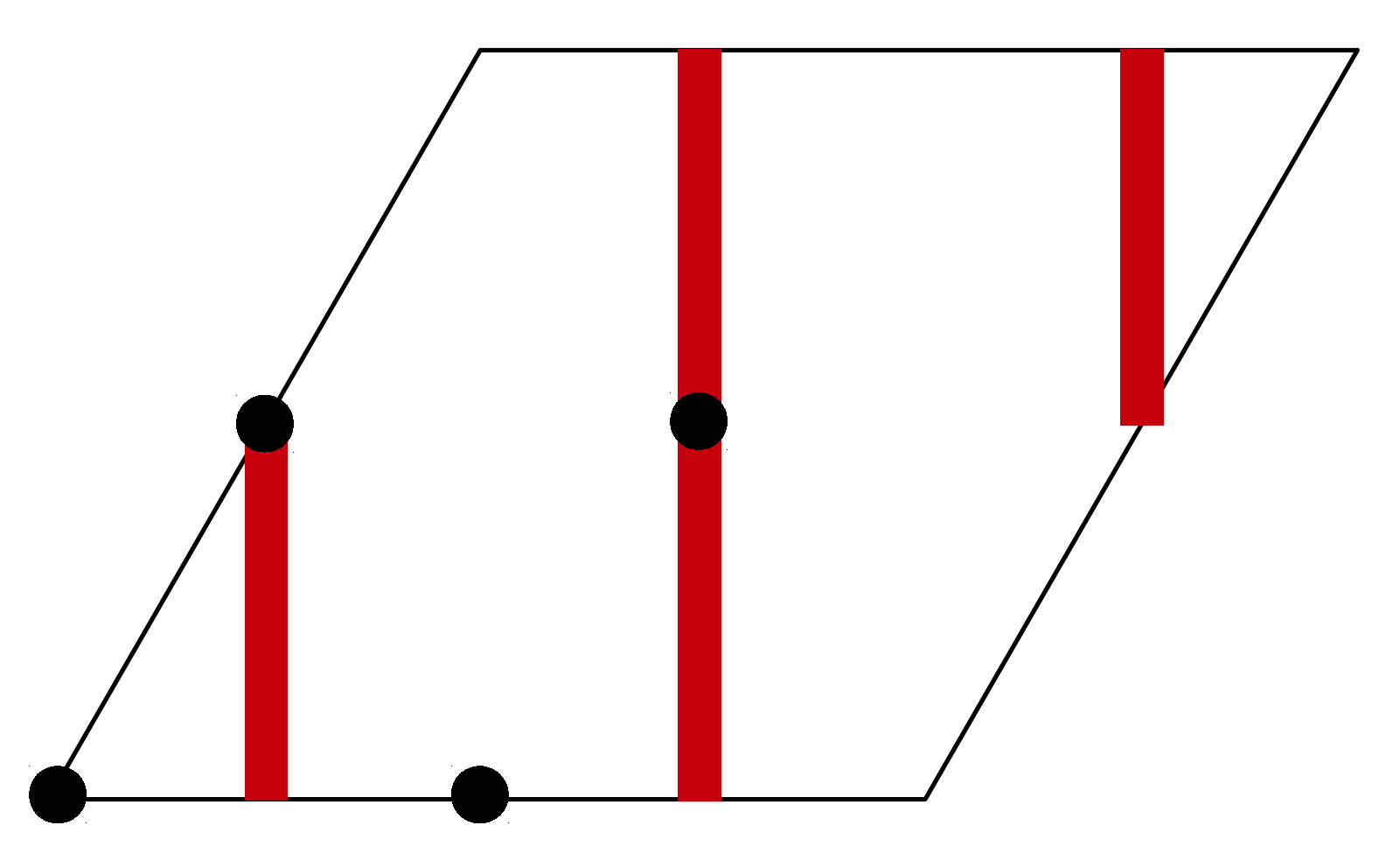}} \\
\hline\hline 
 	& \raisebox{-12pt}{0}
	& \raisebox{-12pt}{$\xi^2 x \ge 1$}
	& \raisebox{-12pt}{$\bf bI^-$}
	& \raisebox{2ex - .95\height}{\includegraphics[scale=0.16]{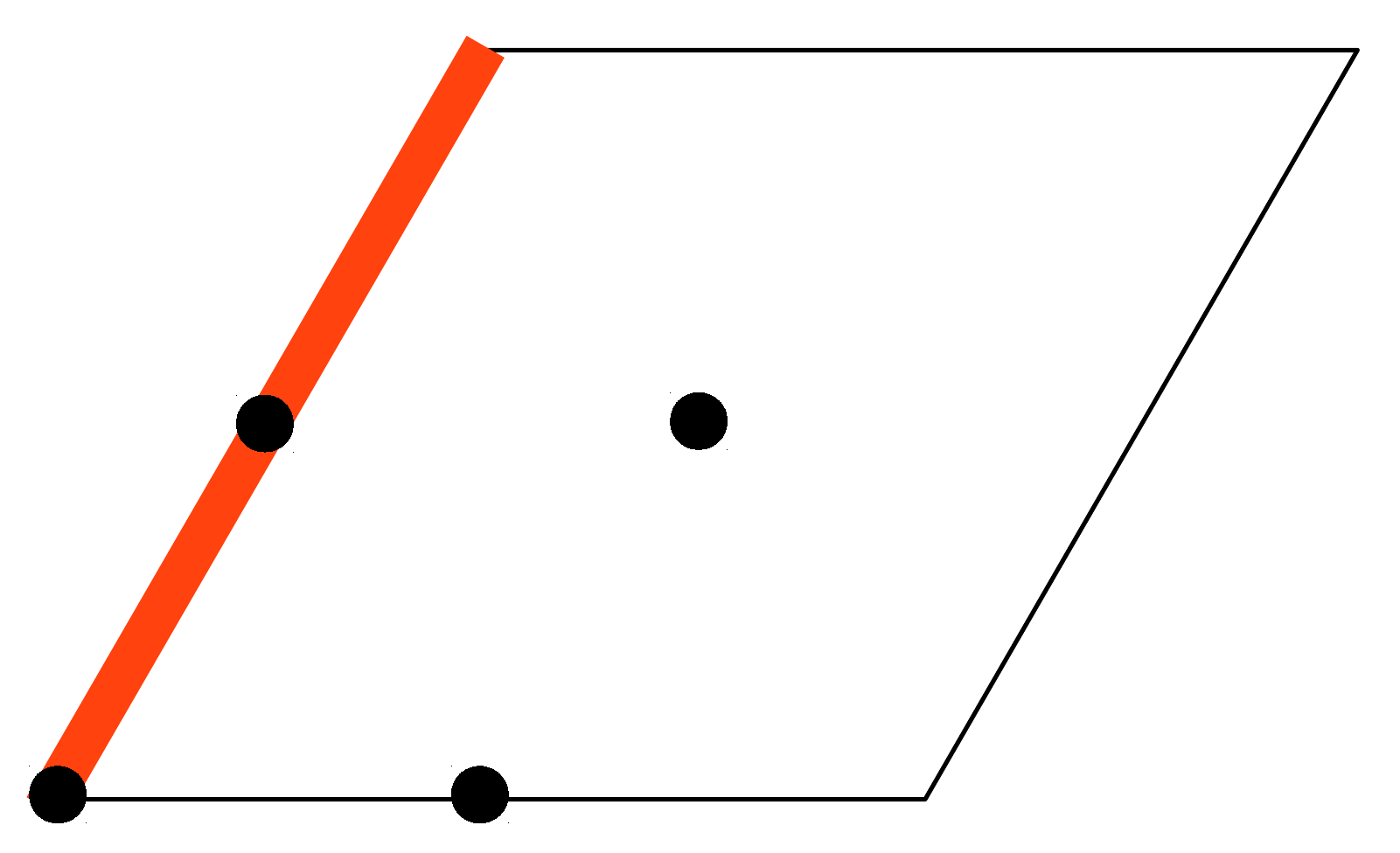}} \\
\cline{2-5} 
 \multirow{-2}{*}{$\pm(0,1)$}
	& \raisebox{-12pt}{1}
	& \raisebox{-12pt}{$| \xi^2 x - 1 |^2 = 3 \,, \Re(\xi^2 x) \le -1/2   $}
   & \raisebox{-12pt}{$\bf bIII^-$} 
   & \raisebox{2ex - .95\height}{\includegraphics[scale=0.16]{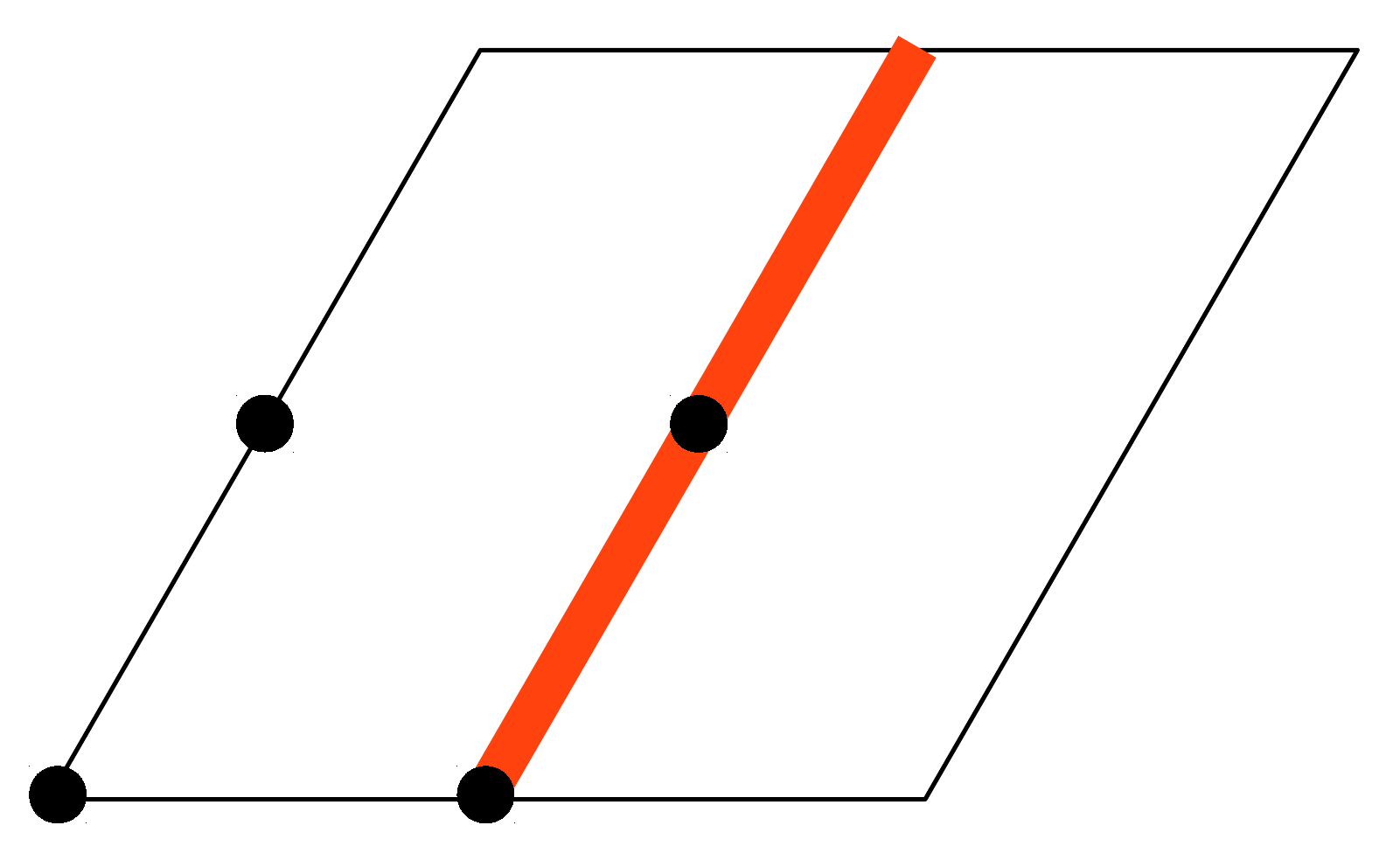}} \\
\cline{1-5} 
	&  \raisebox{-12pt}{$0$}
	& \raisebox{-12pt}{$\xi^2 x \le 1$}
	&  \raisebox{-12pt}{$\bf bII^-$} 
	&  \raisebox{2ex - .95\height}{\includegraphics[scale=0.16]{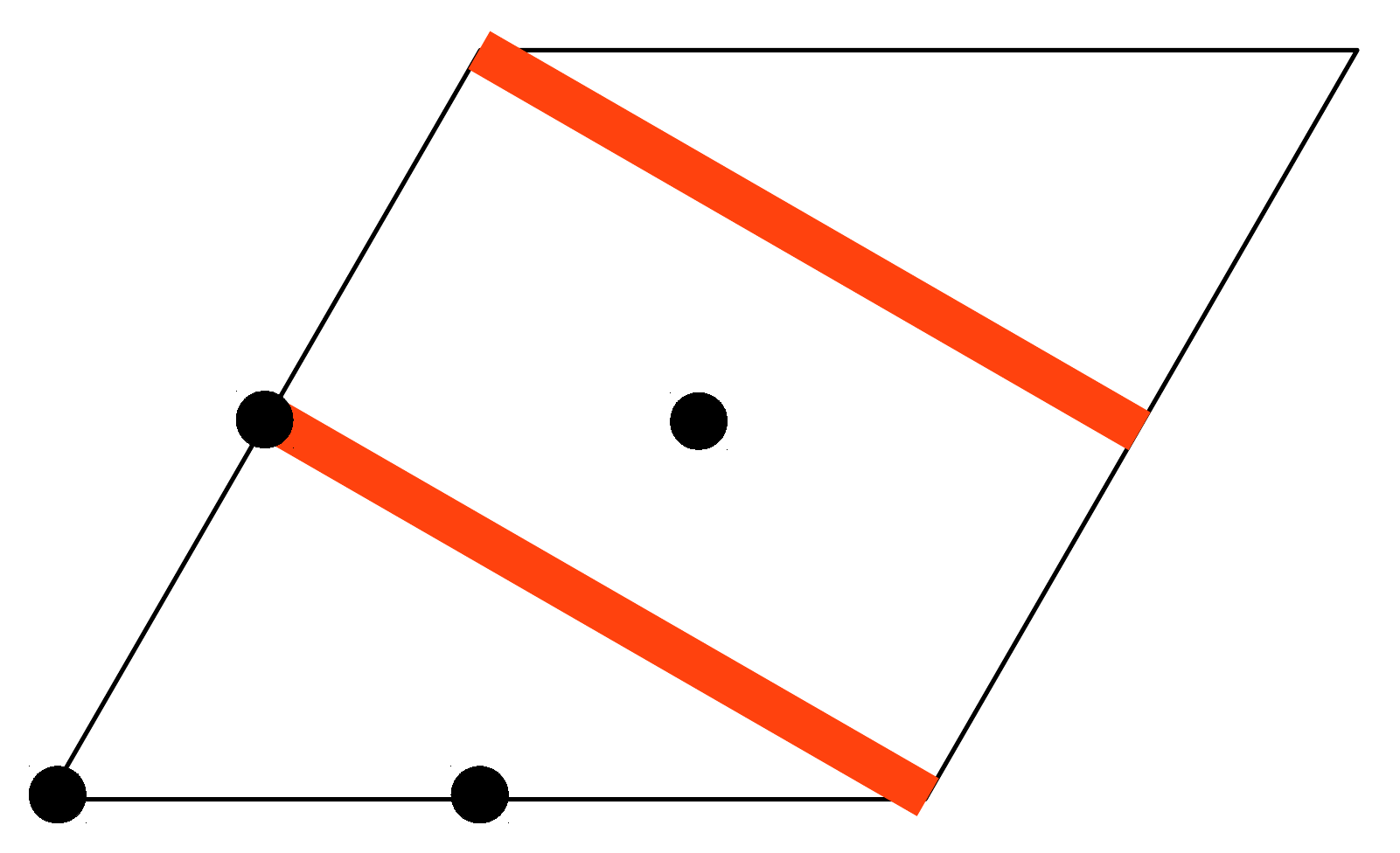}} \\
\cline{2-5} 
\multirow{-2}{*}{$\pm(2,-1)$}
	& \raisebox{-12pt}{$1$}
	& \raisebox{-12pt}{$| \xi^2 x - 1 |^2 = 3 \,, \Re(\xi^2 x) \ge -1/2   $}
 	& \raisebox{-12pt}{$\bf bIV^-$} 
 	& \raisebox{2ex - .95\height}{\includegraphics[scale=0.16]{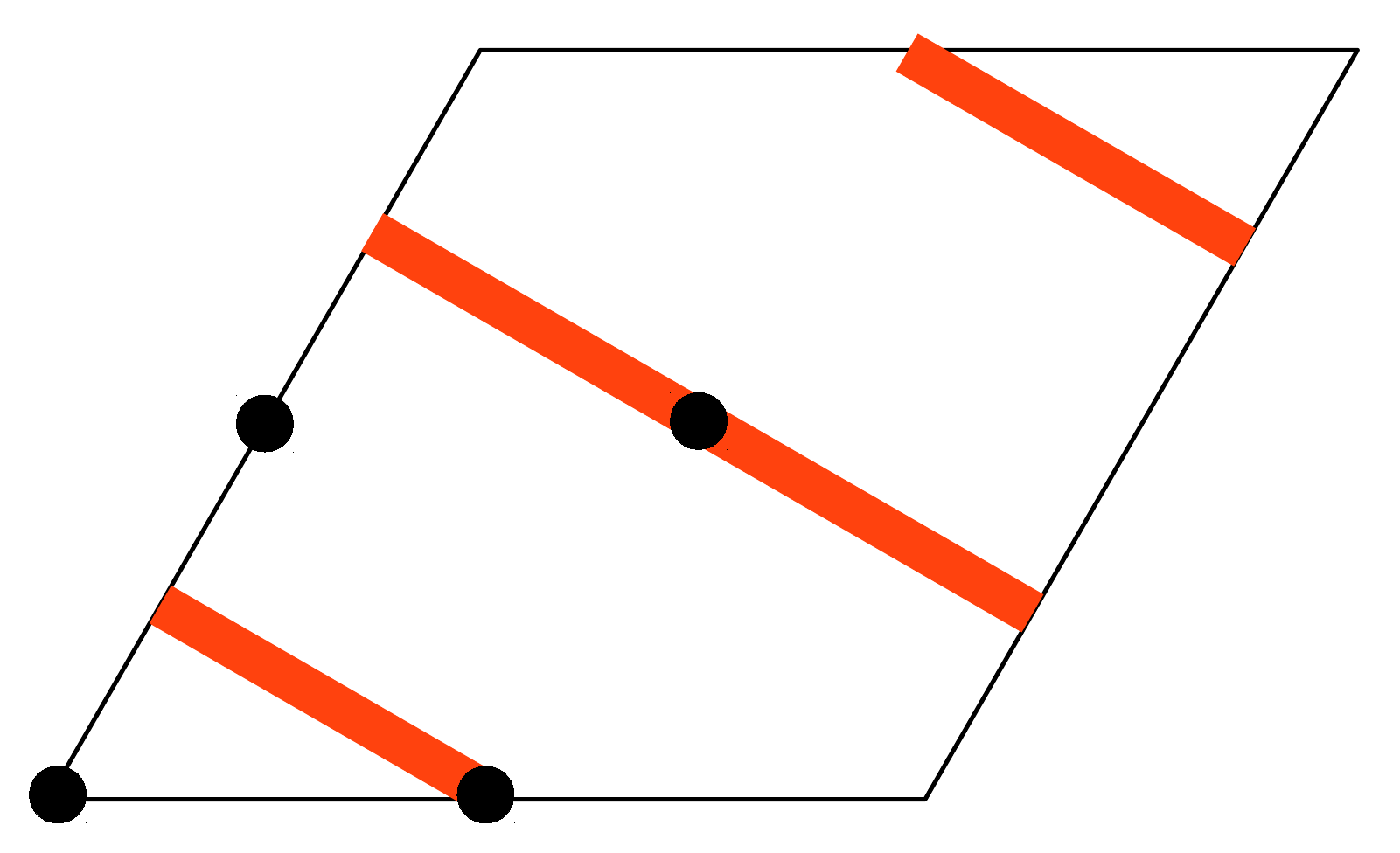}} \\
\hline\hline 
%
 	& \raisebox{-12pt}{0}
	& \raisebox{-12pt}{$\xi x \ge 1$}
	& \raisebox{-12pt}{$\bf bI^+$}
	& \raisebox{2ex - .95\height}{\includegraphics[scale=0.16]{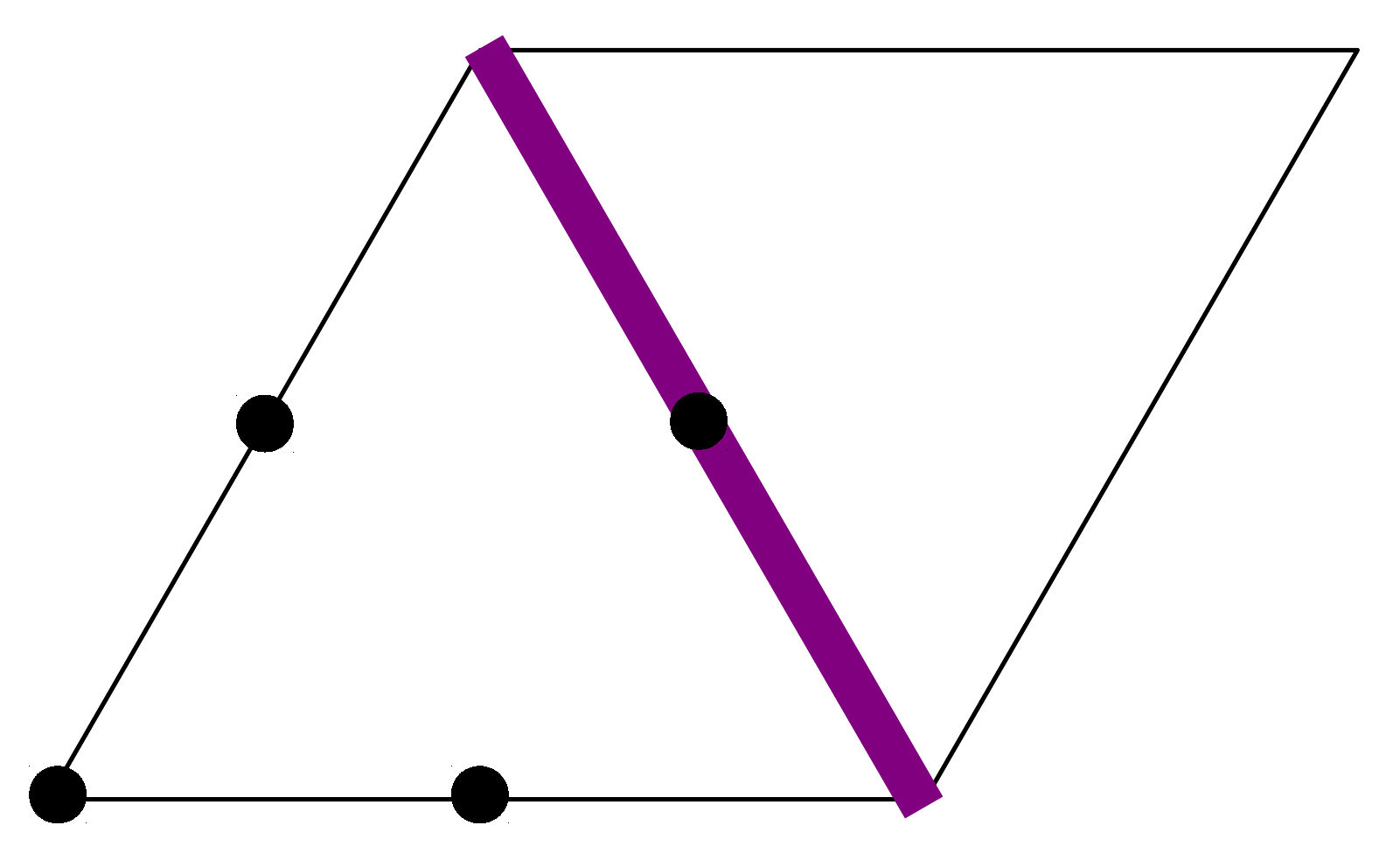}} \\
\cline{2-5} 
 \multirow{-2}{*}{$\pm(1,-1)$}
	& \raisebox{-12pt}{1}
	& \raisebox{-12pt}{$| \xi x - 1 |^2 = 3 \,, \Re(\xi x) \le -1/2   $}
   & \raisebox{-12pt}{$\bf bIII^+$} 
   & \raisebox{2ex - .95\height}{\includegraphics[scale=0.16]{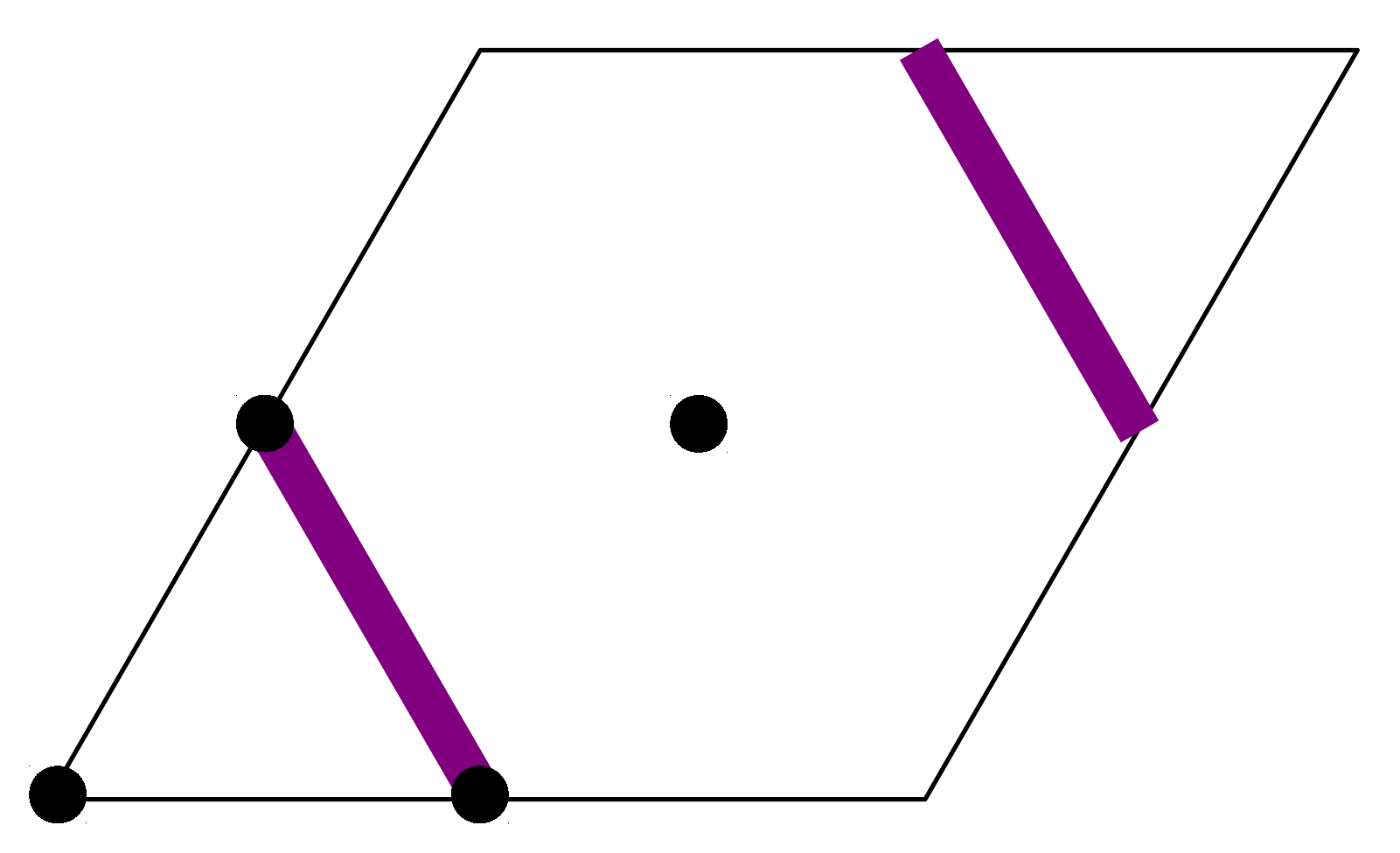}} \\
\cline{1-5} 
	&  \raisebox{-12pt}{$0$}
	& \raisebox{-12pt}{$\xi x \le 1$}
	&  \raisebox{-12pt}{$\bf bII^+$} 
	&  \raisebox{2ex - .95\height}{\includegraphics[scale=0.16]{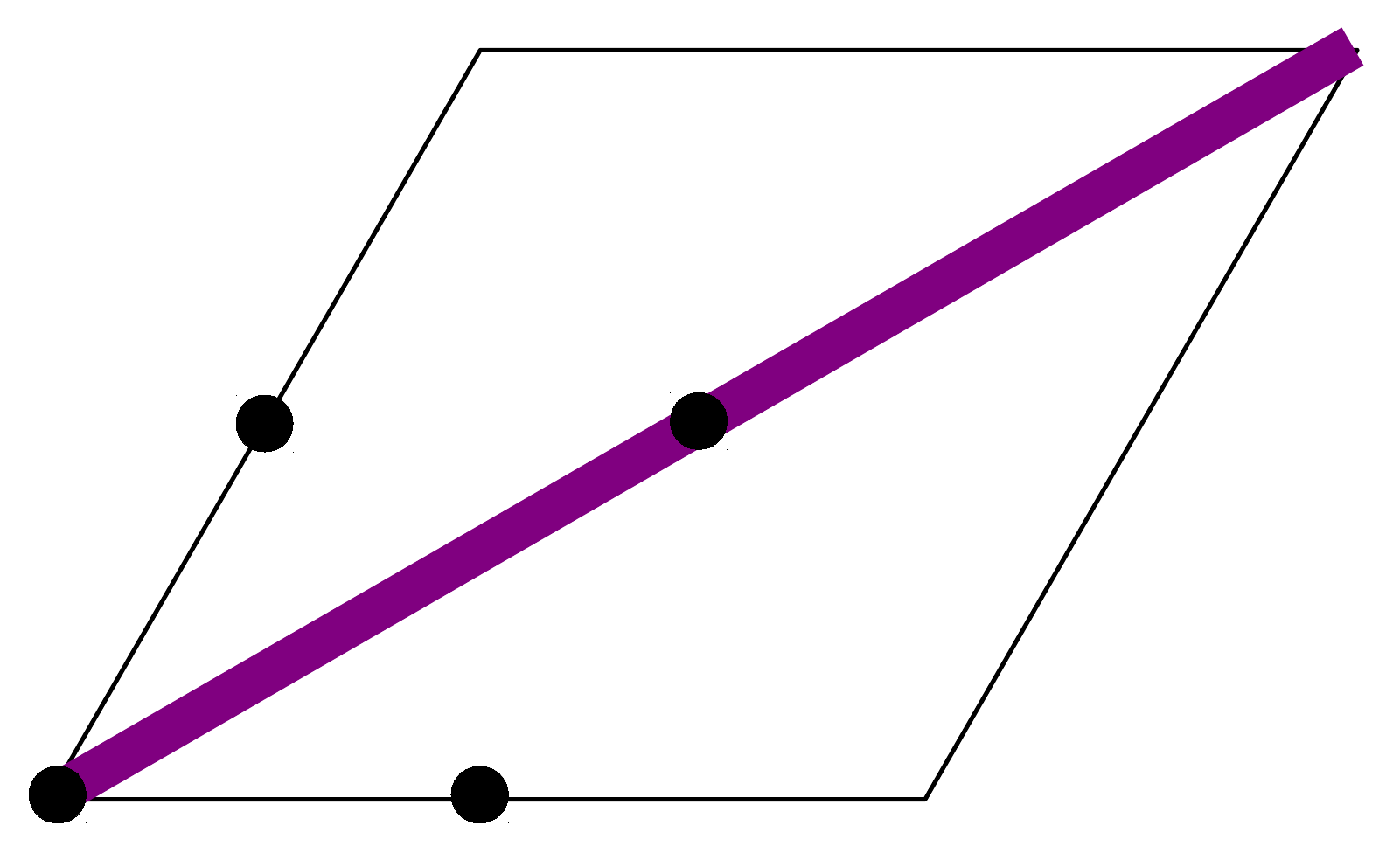}} \\
\cline{2-5} 
\multirow{-2}{*}{$\pm(1,1)$}
	& \raisebox{-12pt}{$1$}
	& \raisebox{-12pt}{$| \xi x - 1 |^2 = 3 \,, \Re(\xi x) \ge -1/2   $}
 	& \raisebox{-12pt}{$\bf bIV^+$} 
 	& \raisebox{2ex - .95\height}{\includegraphics[scale=0.16]{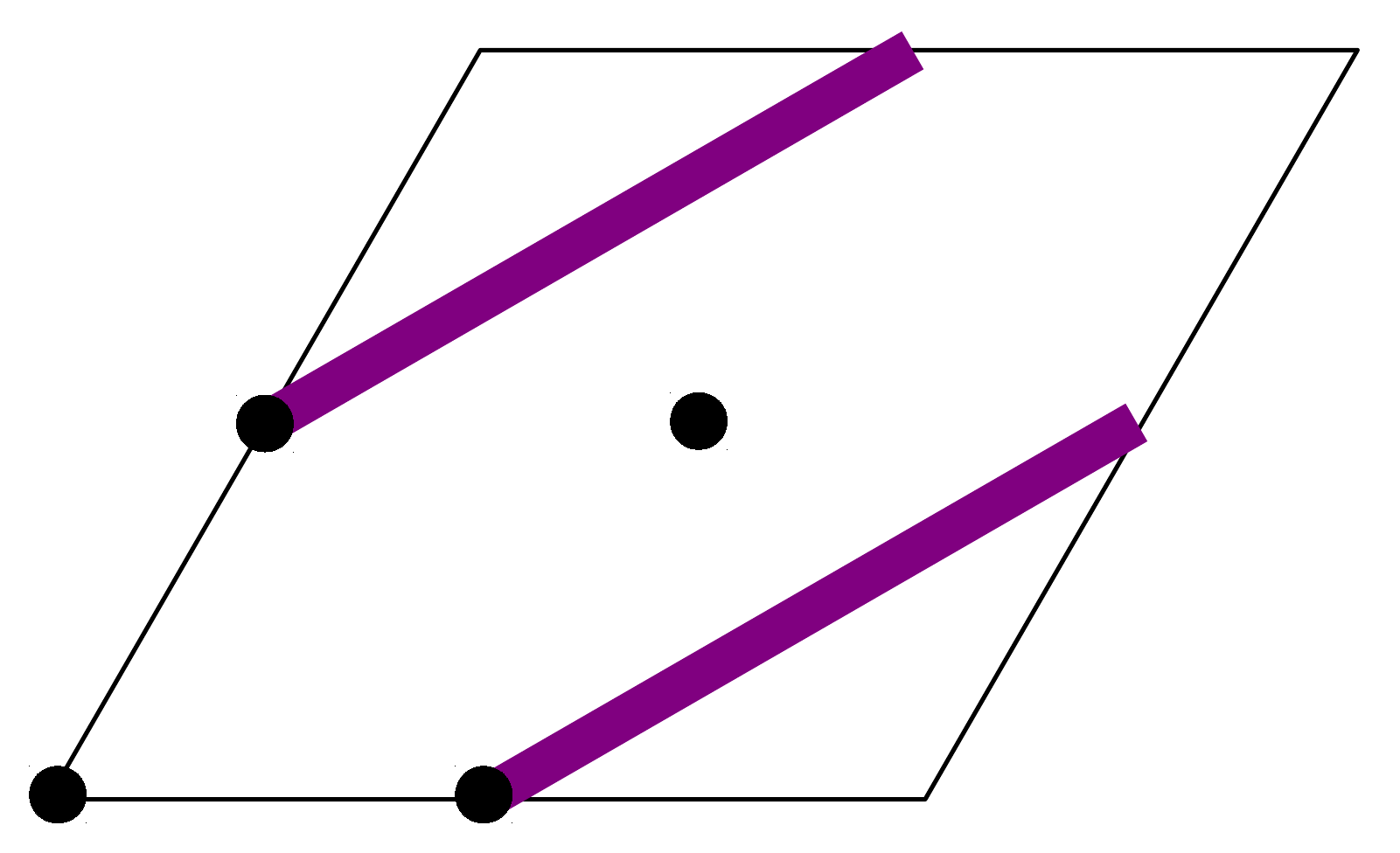}} \\
\hline 
\end{tabular}}
\bCaption{Linearly realised {\it Lag} lines on the hexagonal torus. The constraint equations on the variable $x$ describe straight lines (without displacement) or arcs (with displacement), 
see figures~\ref{fig:Diagram2_Hex_ALL} and~\ref{fig:Diagram2_Hex_0_-}.}
\label{tab:LagLinesHex}
\end{table}

\begin{figure}[ht]
\centering
\includegraphics[width=0.75\textwidth]{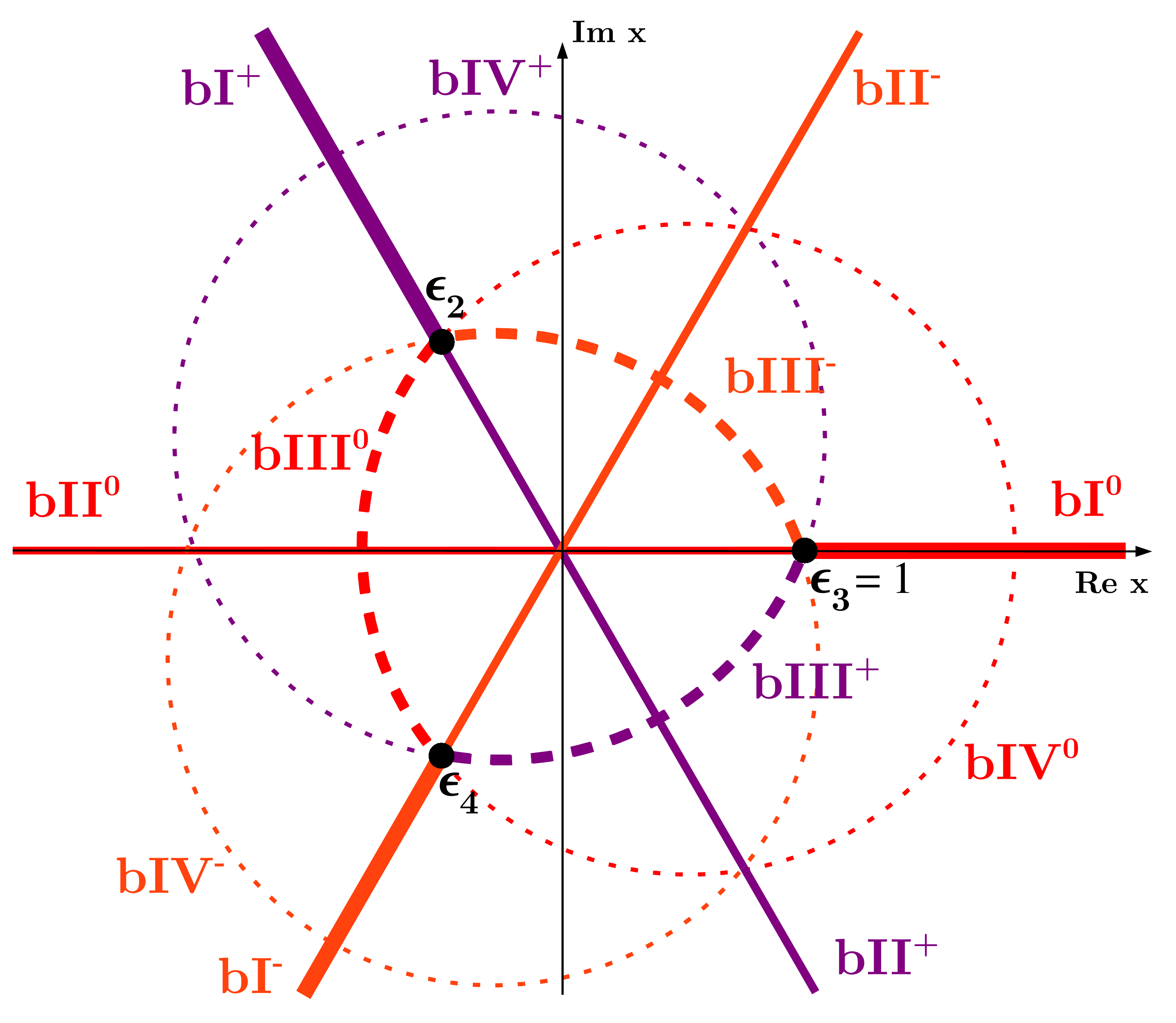}
\caption{All linearly realised {\it Lag} lines of the hexagonal lattice in the complex $x$-plane (with $v \equiv 1$).}
\label{fig:Diagram2_Hex_ALL}
\end{figure}

\begin{figure}[ht]
\centering
\subfloat[{\it Lag} lines with  calibration phase $\boldsymbol{e^{0}}$.]{\label{fig:Diagram2_Hex0} 
\includegraphics[width=0.45\textwidth]{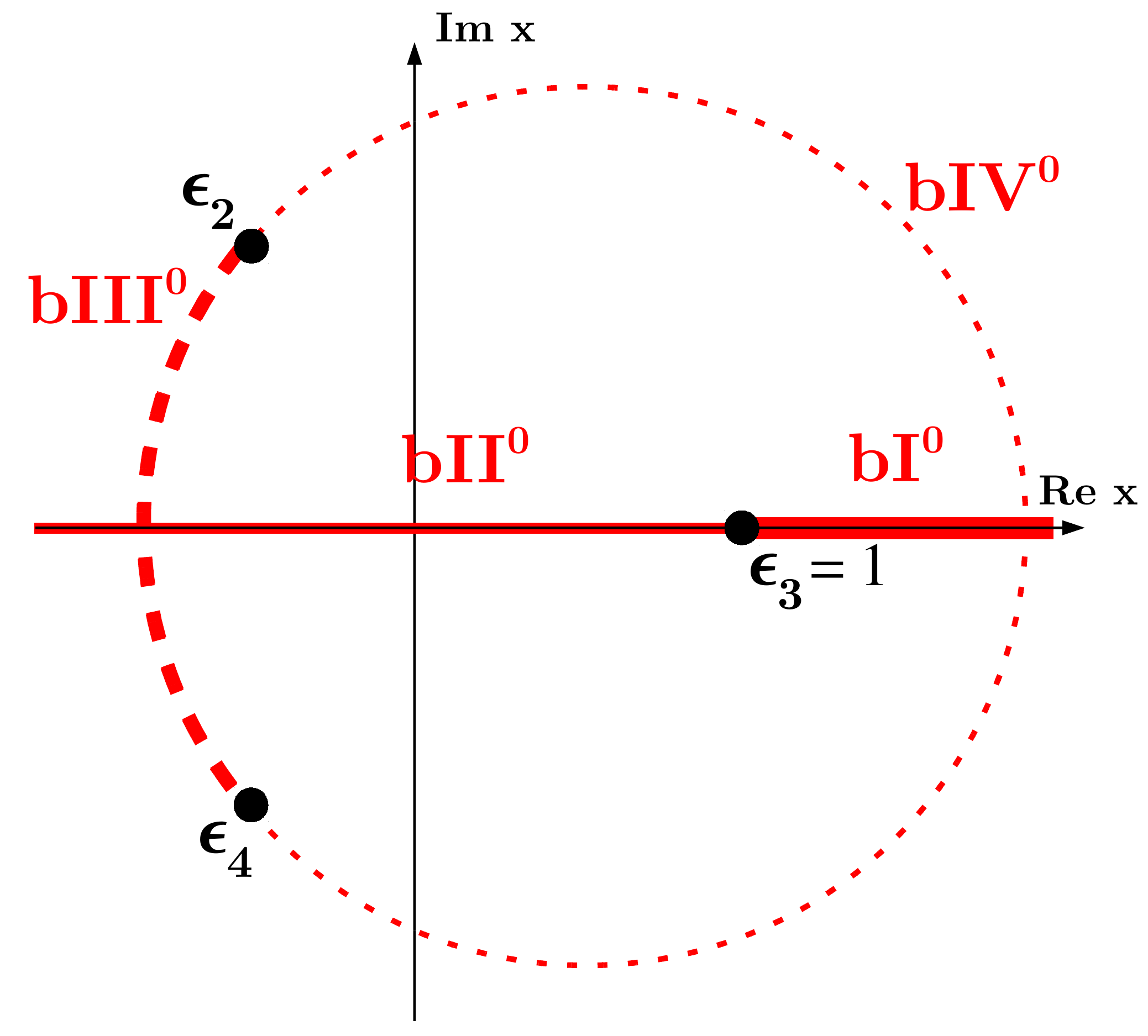}} \quad
\subfloat[{\it Lag} lines with calibration phase $\boldsymbol{e^{-2\pi i / 3}}$.]{\label{fig:Diagram2_Hex-} 
\includegraphics[width=0.45\textwidth]{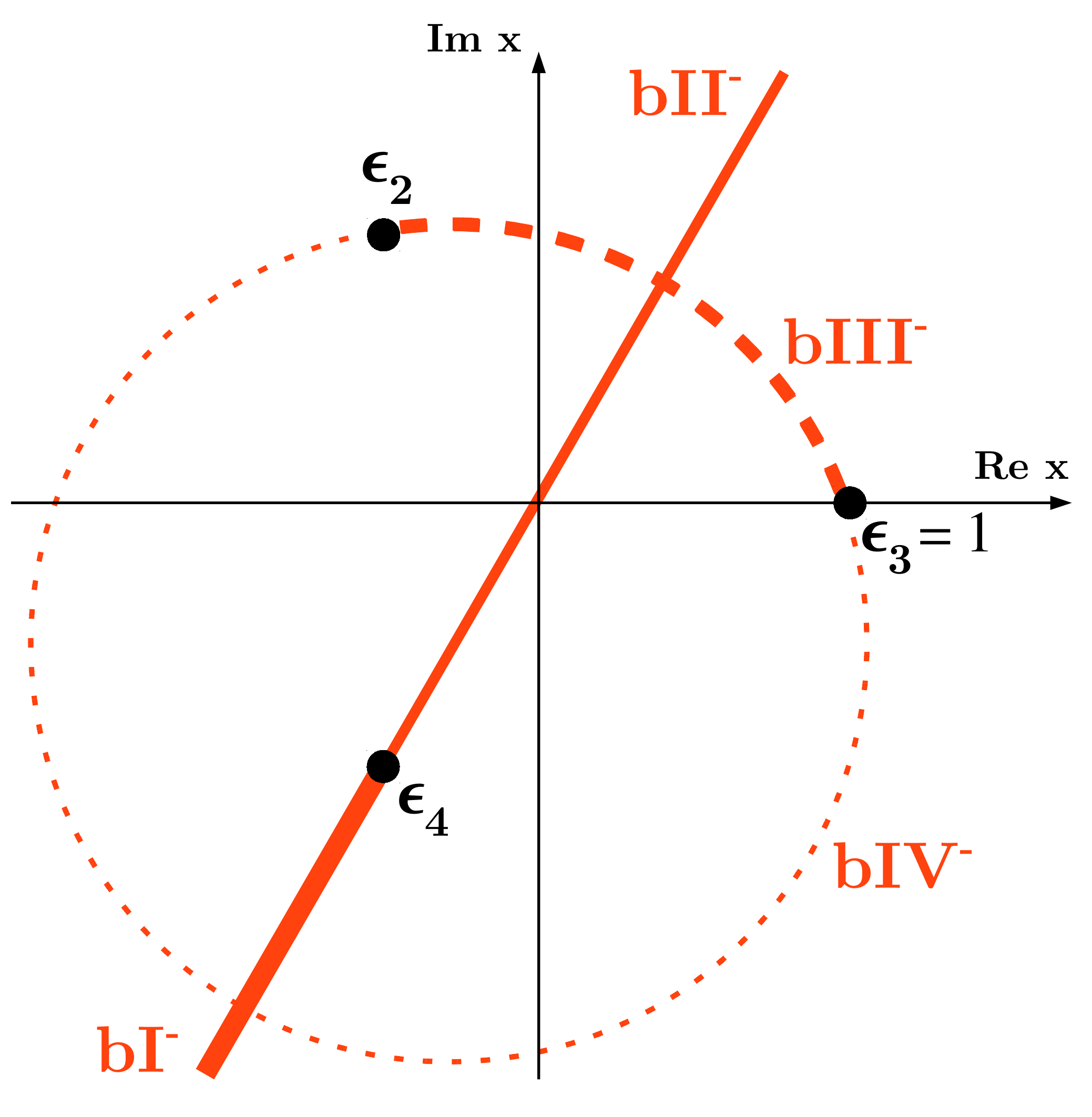}} \quad
\caption{All linearly realised {\it Lag} lines of the hexagonal lattice with a given calibration phase, depicted in the complex $x$-plane (where the illustration for the phase $\boldsymbol{e^{+2\pi i / 3}}$ is omitted). In comparison to figure~\ref{fig:Diagram2_Hex_ALL}, one easily sees that the {\it Lag} lines of figure~\ref{fig:Diagram2_Hex0} and~\ref{fig:Diagram2_Hex-} are by $\pi /3$ rotated versions of each other, as detailed in table~\ref{tab:LagLinesHex}.}
\label{fig:Diagram2_Hex_0_-}
\end{figure}

\begin{figure}[ht]
\includegraphics[width=\textwidth]{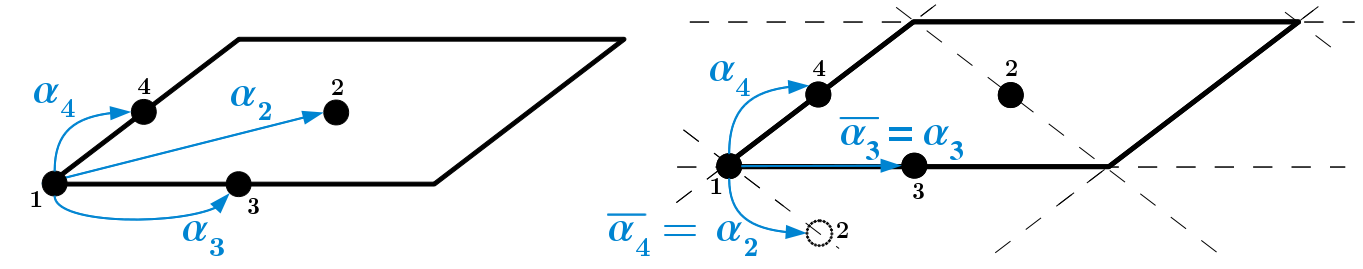}
\caption{Action of half-lattice shifts $\alpha_i$ on $\Z_2$ fixed points on a tilted two-torus (left figure), and complex conjugation of the transformations $\alpha_i$ (right figure).}
\label{fig:CoordTrafo_tilted}
\end{figure}

The complete set of linearly realised cycles on each hexagonal two-torus is displayed in table~\ref{tab:LagLinesHex} as well as depicted in figures~\ref{fig:Diagram2_Hex_ALL} and~\ref{fig:Diagram2_Hex_0_-}. We observe that in the hypersurface language all undisplaced cycles (labelled {\bf bI} and {\bf bII}) are represented by straight lines, whereas the displaced cycles (labelled {\bf bIII} and {\bf bIV}) correspond to arcs of circles centred around a third root of unity. In fact, there are automorphisms which exchange the displaced and undisplaced cycles corresponding to shifts by half lattice vectors on the torus. These maps are linear in $x,v$, so after fixing $v\equiv 1$ they become M\"obius transformations:
\begin{align}
\alpha_i : x \longmapsto \frac{\xi^i x + 2 \xi^{- i }}{ x - \xi^i} \,, \qquad i=2,3,4 \,, \label{Eqn:HalfLatticeShifts}
\end{align}
such that $\alpha_i$ shifts the fixed point 1 to fixed point $i$, see figure~\ref{fig:CoordTrafo_tilted}. 
These transformations will be used to find parametrisations for displaced cycles.

The full set of linearly realised {\it sLag} cycles on the six-torus or its $\Z_2 \times \Z_{2N}$ orbifold is obtained by combining three of those cycles under the condition that their angles\footnote{The angle $\phi$ w.r.t.\ the horizontal axis per two-torus is obtained from the wrapping numbers by \mbox{ tan$\phi = \sqrt{3} m / ( 2n + m ) $} on the hexagonal lattice. } sum up to zero mod $2\pi$. Let us count the total number of such cycles. A priori, there are twelve cycles per two-torus, leading to $12^3 = 1728$ cycles on the six-torus. The $\Z_3\subset \Z_6'$ subgroup identifies them in triplets, leaving $576$ {\it Lag} cycles on the orbifold. The sum of the angles over the three two-tori takes six discrete values $\{0,\pm \frac{\pi}{6}, \pm \frac{\pi}{3}, \frac{\pi}{2}\}$ modulo $\pi$, corresponding to six different calibrations. Thus, for a given calibration, we find $96$ {\it sLag} cycles. These are precisely all the fractional {\it sLag} cycles of minimal and next-to-minimal length.

On the deformed space the O6-planes are still determined by the region in the real $x_1,x_2,x_3$ plane where $y^2 \ge 0$, but in general the boundary\footnote{The O6-planes, like all other {\it sLag} three-cycles, only have boundaries $\{y=0\}$ after projecting to the $x_i$-plane. In the full geometry spanned by $(x_i,v_i,y)$, this boundary becomes a branching surface such that all cycles are closed.} $\{y=0\}$ has a more complicated shape. At the orbifold point, the surface $\{y=0\}$ splits into six separate components, given by $x_i = 1 $ and $x_i = \infty$, whereas after switching on deformations the polynomial $y(x_i)$, defined by equation \eqref{Eqn:T6Z2Z2HypersurfaceEquation}, factorises into fewer parts, and the formerly separate components combine. For a generic deformation, $\{y=0\}$ has only one irreducible component. As a consequence, the regions corresponding to the O6-planes can merge together or get separated. This discussion also applies to other linearly realised {\it sLag} cycles, i.e.\ (partially) displaced ones, via the transformations~\eqref{Eqn:HalfLatticeShifts}. More concretely, the consequences for D6-branes wrapping such cycles are discussed in section~\ref{Sss:U-model-defs}.

\subsubsection{Deformations of $\boldsymbol{\Z_2 \times \Z_6}$}\label{Sss:Z2Z6}

As a cross-check, we repeat the same counting for the other $T^6/(\Z_2 \times \Z_6)$ orbifold with twist vectors $\vec{v}=\frac12(1,-1,0)$ and $\vec{w}=\frac16(0,1,-1)$. Here the additional $\Z_3$ symmetry acts as $\omega^2: (x_1,x_2,x_3) \mapsto (x_1, \xi x_2, \xi^2 x_3)$. Using the same deformation polynomials as for the $T^6/(\Z_2 \times \Z_6')$ orbifold (equation~\eqref{Eq:Def-Polynomials_Z2Z6}), we observe that the deformation polynomials $\delta F_1^{(\alpha)}$ in the first two-torus are invariant, $\delta F_2^{(\alpha)}$ transform according to equation \eqref{Eqn:Z3DeltaF} and 
$\delta F_3^{(\alpha)}$ transform backwards compared to equation \eqref{Eqn:Z3DeltaF}. Hence, we find the following deformations of cycles containing $\Z_2$ fixed points on $T^4_{(k)}$:
\begin{itemize}
\item $\varepsilon^{(1)}_{11}$ is unrestricted,
\item the remaining $\varepsilon^{(1)}_{\alpha\beta}$ get identified in triplets similarly to the other $\Z_2 \times \Z_6'$ orbifold,
\item $\varepsilon^{(2)}_{1\beta}$ are forbidden,
\item $\varepsilon^{(2)}_{2\beta}=\varepsilon^{(2)}_{3\beta}=\varepsilon^{(2)}_{4\beta}$,
\item the restrictions on $\varepsilon^{(3)}_{\alpha\beta}$ are the same as for $\varepsilon^{(2)}_{\alpha\beta}$.
\end{itemize}
To sum up, in the $\Z_2^{(1)}$ sector there are six possible deformations, whereas in each of the $\Z_2^{(2 \text{ or } 3)}$ sectors we find four, in agreement with table~\ref{tab:Hodge-numbers}. Again, the purely geometrical counting of deformations in the hypersurface formalism matches exactly the moduli from the conformal field theory.

The anti-holomorphic orientifold involution reduces here as well the complex deformation parameter to a real one if the $\Z_3$-orbit of $\Z_2$ fixed points is preserved, and it identifies the two complex parameters if two $\Z_3$-orbits are exchanged by the orientifold involution.

\clearpage

\section{Concrete Models}\label{S:ConcreteModels}

In this section, we want to apply the formalism from the previous section to deform the singularities of some concrete models. For this we choose three toy models with D6-branes wrapping orientifold invariant three-cycles, where new models containing SO($2N$) and USp($2N$) gauge groups are described in section~\ref{sec:SOUSp} and a semi-realistic model with Pati--Salam spectrum is shown in section~\ref{sec:PatiSalam}, which was first presented in~\cite{Honecker:2012qr}. In each of these models, the lattice orientation is \textbf{AAA}, and the exotic orientifold-plane is chosen to be $\OR$, i.e.\  $(\eta_{\OR}, \eta_{\OR\Z_2^{(1)}},\eta_{\OR\Z_2^{(2)}},\eta_{\OR\Z_2^{(3)}})=(-1,1,1,1)$. The RR tadpole cancellation conditions are summarised in tables~\ref{tab:Bulk-RR+SUSY-Z2Z6p} and~\ref{tab:twistedRR-Z2Z6p}. Note that the twisted RR tadpole cancellation conditions are homogeneous reflecting the fact that the O6-plane does not wrap exceptional cycles.
\mathtabfix{
\begin{array}{|c||c|c|c|}\hline
\multicolumn{4}{|c|}{\text{\bf Global bulk consistency conditions on } \; \boldsymbol{T^6/(\Z_2 \times \Z_6' \times \OR)} \; \text{\bf with discrete torsion } \boldsymbol{(\eta=-1)}}
\\\hline\hline
\text{Lattice} & \text{\bf Bulk RR tadpole cancellation} & \text{\bf SUSY: necessary}  & \text{\bf SUSY: sufficient} 
\\\hline\hline
{\bf AAA} & \begin{array}{c} 
\sum_a N_a \left(2 \,X_a+Y_a \right)= 4 \left( \eta_{\OR} + 3 \, \sum_{i=1}^3 \eta_{\OR\Z_2^{(i)}} \right) 
\\ \stackrel{\eta_{\OR}=-1}{\Rightarrow} \qquad  \sum_a N_a \left(2 \,X_a+Y_a \right)= 32
\end{array}
& Y_a=0 & 2 \, X_a + Y_a > 0
\\\hline
\end{array}
}{Bulk-RR+SUSY-Z2Z6p}{Bulk RR tadpole cancellation and supersymmetry conditions  on the orientifold $T^6/(\Z_2 \times \Z_6' \times \OR)$ with discrete torsion $(\eta=-1)$. The $\Z_3$ invariant bulk wrapping numbers are defined (see~\cite{Forste:2010gw,Honecker:2012qr}) as 
$X_a \equiv  n^1_a n^2_a n^3_a  - m^1_a m^2_a m^3_a -\sum_{(ijk) \text{ cyclic perm. of} (123)} n^i_a m^j_a m^k_a$ and $Y_a \equiv \!\! \sum_{(ijk)\simeq (123) \text{ cyclic}} \!\! \left(n^i_a n^j_a m^k_a + n^i_a m^j_a m^k_a  \right)$. 
The examples in sections~\protect\ref{sec:SOUSp} and~\protect\ref{sec:PatiSalam} are valid for the choice of exotic O6-plane $(\eta_{\OR},\eta_{\OR\Z_2^{(1)}},\eta_{\OR\Z_2^{(2)}},\eta_{\OR\Z_2^{(3)}})=(-1,1,1,1)$.
}
%
\mathtabfix{
\begin{array}{|c|c||c|c|}\hline
\multicolumn{3}{|c|}{\text{\bf Twisted RR tadpole cancellation conditions on $\boldsymbol{T^6/(\Z_2 \times \Z_6' \times \OR)}$ with discrete torsion $\boldsymbol{(\eta=-1)}$}}
\\\hline\hline
i & \rho & {\bf AAA} 
\\\hline\hline
1,2,3 & 1,2,3 & \sum_a N_a \left[ (1-\eta_{(i)}) \, x_{\rho,a}^{(i)}  -\eta_{(i)} y_{\rho,a}^{(i)} \right]=0 = \sum_a N_a (1+ \eta_{(i)}) \, y_{\rho,a}^{(i)} 
\qquad \stackrel{\eta_{\OR}=-1}{\Rightarrow} \qquad 
\sum_a N_a \left[ 2 \, x_{\rho,a}^{(i)}  + y_{\rho,a}^{(i)} \right]=0 
\\\hline
1,2,3 & 4,5 & \sum_a N_a \left[ \left( 2 \,  x^{(i)}_{4,a} +y^{(i)}_{4,a} \right) -\eta_{(i)} \left( 2 \,  x^{(i)}_{5,a} +  y^{(i)}_{5,a} \right) \right]=0 = \sum_a N_a \left[ y^{(i)}_{4,a} + \eta_{(i)} \,  y^{(i)}_{5,a} \right]
\\
& & \stackrel{\eta_{\OR}=-1}{\Rightarrow} \qquad 
 \sum_a N_a \left[  2 \, (  x^{(i)}_{4,a} + x^{(i)}_{5,a} ) +y^{(i)}_{4,a}  +  y^{(i)}_{5,a}  \right]=0 = \sum_a N_a \left[ y^{(i)}_{4,a} - y^{(i)}_{5,a} \right]
\\\hline
\end{array}
}{twistedRR-Z2Z6p}{Twisted RR tadpole cancellation conditions on $T^6/(\Z_2 \times \Z_6' \times \OR)$ with discrete torsion $(\eta=-1)$.
The choice of exotic O6-plane enters via the sign factor $\eta_{(i)} \equiv \eta_{\OR} \cdot  \eta_{\OR\Z_2^{(i)}} \in \{\pm 1\}$.
The examples in sections~\protect\ref{sec:SOUSp} and~\protect\ref{sec:PatiSalam} have $ \eta_{(1)}=\eta_{(2)}=\eta_{(3)}=-1$.
The five $\Z_3$-orbits per $\Z_2^{(i)}$ twisted sector are labelled by the first lower index $\rho \in \{1 \ldots 5\}$.
}

For later convenience, the orientifold-even bulk and exceptional wrapping numbers can be read off from the RR tadpole cancellation conditions in tables~\ref{tab:Bulk-RR+SUSY-Z2Z6p} and~\ref{tab:twistedRR-Z2Z6p}, i.e.\ for the choice of exotic O6-plane $\eta_{\OR}=-1$ one obtains
\begin{equation}\label{Eq:wrappings-OR-even}
[2X_a+Y_a], \quad
2\, x^{(i)}_{\rho,a} + y^{(i)}_{\rho,a} \text{ for } \rho=1,2,3,
\quad
y^{(i)}_{4,a} - y^{(i)}_{5,a},
\quad
2 (x^{(i)}_{4,a} + x^{(i)}_{5,a}) + (y^{(i)}_{4,a} + y^{(i)}_{5,a})
,
\end{equation}
while the orientifold-odd bulk and exceptional wrapping numbers for $\eta_{\OR}=-1$,
\begin{equation}\label{Eq:wrappings-OR-odd}
Y_a, \quad
y^{(i)}_{\rho,a} \text{ for } \rho=1,2,3,
\quad
y^{(i)}_{4,a} + y^{(i)}_{5,a},
\quad
2 (x^{(i)}_{4,a} - x^{(i)}_{5,a}) + (y^{(i)}_{4,a} - y^{(i)}_{5,a})
,
\end{equation}
can be found in~\cite{Honecker:2013hda,Honecker:2013kda}.
Since supersymmetry breaking by deformations is tied to U(1) gauge groups, which in the effective field theory appears as Fayet-Iliopoulos term as we will argue in detail in section~\ref{Sss:U-model-defs}, we expect that the number of stabilised deformation parameters is set by the number of independent non-vanishing orientifold-odd wrapping numbers. We will now verify this claim in two classes of examples presented in sections~\ref{sec:SOUSp} and~\ref{sec:PatiSalam}, which consist entirely of products of the linearly realised {\it Lag} cycles of table~\ref{tab:LagLinesHex}.

\paragraph{Deformations in the local picture:}
For simplicity, it is best to first discuss each deformation separately in a local patch and focus on the global picture later. To start with, we zoom in on one particular fixed point, say (333), but any other fixed point can as well be accessed by the shift transformations in equation \eqref{Eqn:HalfLatticeShifts} with the only difference being the restrictions on the respective deformation parameters. Locally, the singularity can be described by the simplified hypersurface equation 
\begin{align}
y^2 = x_1 x_2 x_3 - \varepsilon^{(1)} x_1 - \varepsilon^{(2)} x_2 - \varepsilon^{(3)} x_3 + \varepsilon \,,
\end{align}
where $\varepsilon \equiv 2 \sqrt{\varepsilon^{(1)} \varepsilon^{(2)} \varepsilon^{(3)}}$ (and indices $\alpha, \beta$ of $\varepsilon_{\alpha \beta}^{(i)}$ are omitted), which can be achieved by a proper rescaling of the $x_i$ coordinates, setting $v_i \equiv 1$ and expanding around $x_i = 0$. We focus on deformations in one $\Z_2$ twisted sector by setting $\varepsilon^{(1)}=\varepsilon^{(2)}=0$ such that $x_3$ factors out and we can work in the $x_1-x_2$ plane. At the orbifold point, we then recover the fractional cycle $\Pi_{1}$ with wrapping numbers $(1,0;1,0)$ at $x_1,x_2 \ge 0$ (in the {\it local} picture)
 and the fractional cycle $\Pi_{2}$ with wrapping numbers $(-1,2;1,-2)$ at $x_1,x_2 \le 0$. The orientifold symmetry requires the deformation parameter $\varepsilon^{(3)}$ to be real. Then both of these cycles are described by $x_1 \cdot x_2 \ge 0$, however, we find different behaviours for the two sign choices of $\varepsilon^{(3)}$:
\begin{itemize}
\item For $\varepsilon^{(3)} > 0$, the cycles $\Pi_i$ are still two separated cycles with obviously no intersection point in the hypersurface formalism. In \cite{Blaszczyk:2014xla} it was shown that in this case both cycles have a minus sign in front of the exceptional part, which topologically explains their zero intersection. In addition, we recover the exceptional cycle by the equation $x_1 = \overline{x_2}$ and find that it is {\it sLag}. As a consequence, both cycles $\Pi_i$ remain {\it sLag} upon this deformation. For our choice of exotic O-plane ($\eta_{\OR}=-1$), the cycles $\boldsymbol{\varepsilon}^{(i)}_{a=1,2,3}$ are orientifold-even and thus {\it sLag}, therefore we find the restrictions  $\varepsilon^{(i)}_{a=1,2,3} \ge 0$.

Further details are discussed in section~\ref{sec:SOUSp}.
 In particular, the O6-planes can never wrap such a cycle since otherwise they would contain an exceptional contribution.
\item For $\varepsilon^{(3)} < 0$ the situation at first looks similar, but turns out to be completely different. The exceptional cycle is now given by $x_1 = - \overline{x_2}$ and is only {\it Lag}, but not {\it sLag}, i.e.\ it has the wrong calibration. Consequently, the fractional cycles $\Pi_i$ are no longer {\it sLag} since their bulk part and exceptional part are differently calibrated. However, if we take both cycles together such that the exceptional part cancels out, we recover one big merged {\it sLag} cycle $\Pi_1 + \Pi_2$. 
\end{itemize}
When the orientifold involution maps two distinct cycles onto each other, the corresponding deformation parameter may in general take complex values, as happens for fully displaced cycles in tilted tori. If we write $\varepsilon^{(3)} = |\varepsilon^{(3)}| e^{i \phi}$, the exceptional cycle is given by $x_2 = e^{i \phi} \overline x_1$, which, when inserted into the holomorphic two-form, shows that the two-cycle is calibrated with phase $e^{i \phi/2}$. In the $\Z_2 \times \Z_6^{\prime}$ geometry this is the case for the cycles $\boldsymbol{\varepsilon}^{(i)}_{a=4,5}$, which can topologically be decomposed into orientifold-even and -odd parts $\boldsymbol{\varepsilon}^{(i)}_{4} \pm \boldsymbol{\varepsilon}^{(i)}_{5}$.

For the other exceptional cycles, the calibrations are fixed by the choice of the exotic O-plane. Concretely, e.g.\ the cycles $e^{(3)}_{\alpha\beta}\otimes \pi_5$ come in three calibrations
(remember $\xi \equiv e^{2\pi i /3}$):
\begin{itemize}
\item 1: $e_{13}$, $e_{31}$, $e_{33}$,
\item $\xi$: $e_{12}$, $e_{21}$, $e_{22}$,
\item $\xi^2$: $e_{14}$, $e_{41}$, $e_{44}$. 
\end{itemize}
Exceptional three-cycles with a differently oriented one-cycle $(n_3 \pi_5 + m_3 \pi_6)$
then have calibration multiplied by the angle of the one-cycle. Finally, we briefly show how the exceptional cycles can be accessed in the global geometry and what restrictions have to be imposed on the deformation parameters. For convenience we focus on the $\Z_2^{(3)}$ sector and drop the correspondig sector indices, but the other $\Z_2$ sectors work completely analogously.

\paragraph{Exceptional Cycle 3}
 
The easiest cycle to describe is $e_3$ (i.e. $\rho=3$ in table~\ref{tab:Z3Restrictions}), which in two-index notation is represented by $e_{\alpha\alpha}$, $\alpha=2,3,4$. It is described by the equation $x_1 = \overline x_2$ for those deformations which leave the symmetry $x_1 \leftrightarrow \overline x_2$ invariant (cf.\ equation~\eqref{Eqn:Z6HypersurfaceEqn} in appendix~\ref{A:1}). These are given by the restrictions $\varepsilon_1=\varepsilon_2$ and $\varepsilon_{4-5}=0$. In this case, we find a {\it sLag} cycle growing around the fixed point $x_1 = x_2 = 1$ for positive deformation parameter, i.e.\ $\varepsilon_3 > 0$.

\paragraph{Exceptional Cycles 4 and 5}

In each four-torus, the cycles $e_4$ and $e_5$ are orientifold images of one another, which implies that the holomorphic two-form\footnote{Here $\Omega_2$ can be thought of as $\Omega_3$ integrated over a one-cycle on the remaining two-torus.} $\Omega_2$ on $e_4$ is the complex conjugate 
of $\Omega_2$ on $e_5$. To parametrise the cycles $e_4$, $e_5$ we apply the shifts $\alpha_2$ on $x_1$ and $\alpha_4$ on $x_2$, which moves the fixed point $(42)$ to the spot $x_1 = x_2 = 1$. Then the cycle $e_4$ is described by $\alpha_2(x_1) = \alpha_4(\overline  x_2)$. Since $\alpha_{2/4}$ are complex conjugates of one another, this description only holds if the deformation preserves the symmetry $x_1 \leftrightarrow \overline x_2$ as for $e_3$.

\paragraph{Exceptional Cycles 1 and 2}

Directly accessing the cycles $e_1$ and $e_2$ turns out to be difficult. Naively they would be described e.g.\ by the equation $x_1 = \alpha_3(\overline x_2 )$, however, this does not work since it is not the fixed set of an antiholomorphic involution. Luckily, there is another way to compute the volume. For this, consider the fractional two-cycle with wrapping numbers $(1,0;1,0)$, no displacement, and without Wilson lines. It has the decomposition $\Pi_0 = \frac12 \left( \Pi_{13} - e_{13} - e_{31} - e_{33} \right)$. This cycle lies in the real $x_1-x_2$-plane for any type of deformation, and we can always describe it in the hypersurface formalism and in particular compute its volume. Furthermore, we will see in section \ref{Sss:USp+SO-defs} and in figure \ref{fig:IntPath} how to compute the volume of $\Pi_{13}$ for any deformation, thus we can consider the linear combination $\Pi_{13} - 2 \Pi_0$ and obtain the volumes of $e_1$, $e_2$ or $e_3$ if we deform each of the associated $\Z_2$ singularities separately.

\subsection{SO and USp models}
\label{sec:SOUSp}

We present here two classes of toy models whose gauge group contains only SO($2N$) or only USp($2N$) factors. These gauge groups are lacking a central U(1) factor such that the moduli are completely neutral. Therefore, there are no D-terms in the low-energy effective action, which could provide a potential for the deformation moduli. 
We thus expect all deformations to preserve the {\it sLag} property.

\subsubsection{The USp$\boldsymbol{(8)^4}$ and SO$\boldsymbol{(8)^4}$ models at the orbifold point}\label{Sss:USp+SO-orbifold}

In order to realise a SO$(2N)$ or USp$(2N)$ gauge group by some brane configuration, the D6-brane stack  must be either parallel or perpendicular to the O6-planes, which fixes the bulk wrapping numbers. We choose here the most simple configuration of solely orientifold-invariant D6-branes whose twisted RR tadpoles cancel by having an equal number of each type with $\Z_2 \times \Z_2$ eigenvalues in $\{(+,+),(+,-),(-,+),(-,-)\}$.
In this way, the bulk RR tadpole cancellation conditions in table~\ref{tab:Bulk-RR+SUSY-Z2Z6p} dictate a total number of 16 D6-branes which are distributed in four stacks of four branes each. 
In addition, orientifold-invariance of the exceptional parts of the three-cycles poses constraints on the displacement and Wilson line parameters~\cite{Forste:2010gw,Honecker:2012qr}:
\begin{align}
(\eta_{(1)},\eta_{(2)},\eta_{(3)}) \stackrel{!}{=} 
\left( - (-1)^{\sigma^2\tau^2 +\sigma^3\tau^3} \, , \,  - (-1)^{\sigma^1\tau^1 +\sigma^3\tau^3} \, , \,  - (-1)^{\sigma^1\tau^1 +\sigma^2\tau^2}   \right) \,,
\end{align}
where we defined $\eta_{(i)} \equiv \eta_{\OR} \cdot \eta_{\OR\Z_2^{(i)}}$ as before. These conditions depend on the choice of exotic O6-plane, which in our case leads to $\eta_{(i)} = -1$, for $i = 1,2,3$. Thus, the products $\sigma^i \cdot \tau^i$ must be identical for all $i$. Then the case $\sigma^i \cdot \tau^i \equiv 1$ leads to SO-type gauge groups, whereas for $\sigma^i \cdot \tau^i \equiv 0$ one obtains USp-type gauge groups, as can be verified by the method of computing the corresponding beta-function coefficients presented e.g.\ in~\cite{Honecker:2011sm,Honecker:2012qr}. The twisted RR tadpole cancellation conditions in table~\ref{tab:twistedRR-Z2Z6p} require to choose the displacements and Wilson lines to be the same for all 16 branes, after we confined to four stacks with different $\Z_2 \times \Z_2$ eigenvalues. The full configurations of the models 1a, 1b and 1c are shown in table~\ref{tab:Models1a+b+c}.

%

\begin{table}[h]
\centering
\resizebox{\linewidth}{!}{
\begin{tabular}{|c|c|c||c|c|c|c|}\hline
\muc{7}{|c|}{\text{\bf Model 1: D6-brane configuration with USp$\boldsymbol{(8)^4}$ or SO$\boldsymbol{(8)^4}$ gauge group}}
\\\hline\hline
&\bf Wrapping numbers& $\frac{\rm Angle}{\pi}$ &\bf $\boldsymbol{\Z_2^{(i)}}$ eigenvalues  & ($\vec \tau$) & ($\vec \sigma$)& \bf Gauge group\\
\hline \hline
\muc{7}{|c|}{\text{\bf Model 1a}}
\\\hline \hline
 $a_{1 \ldots 4}$&(1,0;1,0;1,0)&$(0,0,0)$&$\begin{array}{c} (+++)\\ (+--)\\(-+-)\\(--+)\end{array}$&$(\tau^1,\tau^2,\tau^3)$ & $(0,0,0)$& USp$(8)^4$
\\\hline \hline
\muc{7}{|c|}{\text{\bf Model 1b}}
\\\hline \hline
 $\tilde{a}_{1 \ldots 4}$&(1,0;1,0;1,0)&$(0,0,0)$&$\begin{array}{c} (+++)\\ (+--)\\(-+-)\\(--+)\end{array}$&$(1,1,1)$ & $(1,1,1)$& SO$(8)^4$
\\\hline \hline
\muc{7}{|c|}{\text{\bf Model 1c}}
\\\hline \hline
 $\hat{a}_{1 \ldots 4}$&(1,0;1,0;1,0)&$(0,0,0)$&$\begin{array}{c} (+++)\\ (+--)\\(-+-)\\(--+)\end{array}$&$(0,0,\tau)$ & $(1,1,0)$& USp$(8)^4$
 \\\hline
\end{tabular}}
\caption{D6-brane configuration with four $\OR$-invariant D-brane stacks yielding the gauge group USp$(8)^4$ or SO$(8)^4$ 
depending on the choice of discrete Wilson lines $(\vec{\tau})$ and displacements $(\vec{\sigma})$.
}
\label{tab:Models1a+b+c}
\end{table}

The orientifold invariance of each stack of D6-branes implies that all orientifold-odd combinations of wrapping numbers in equation~\eqref{Eq:wrappings-OR-odd} vanish, cf.\ table~\ref{tab:Example-1a-XY+x1y1} to table~\ref{tab:Example-1c-x2y2+x3y3}.
Therefore, we expect that all 5+5+5 deformations of $\Z_2$ singularities in these three models constitute flat directions. Said differently, 
we anticipate that D-branes of USp- or SO-type do not contribute to the stabilisation of moduli at the orbifold point.

{\footnotesize
\begin{table}[h]
\begin{equation*}
\begin{array}{|c||c||c|c||c|c|c|c|c||c|}\hline
\muc{10}{|c|}{\text{\bf Bulk and exceptional wrapping numbers of Model 1a, Part I}}
\\\hline\hline
 & N & X & Y & x_{1}^{(1)} & x_{2}^{(1)}  & x_{3}^{(1)}   & x_{4}^{(1)}  & x_{5}^{(1)}   & y_{1,\ldots,5}^{(1)}
\\\hline\hline
a_1 & 4 & 1 & 0 &  (-1)^{\tau^2}& (-1)^{\tau^3}& (-1)^{\tau^2 + \tau^3}& 0& 0&   0
\\
a_2 & 4 & 1 & 0 &  (-1)^{\tau^2}& (-1)^{\tau^3}& (-1)^{\tau^2 + \tau^3}& 0& 0&   0
\\
a_3 & 4 & 1 & 0 &  (-1)^{\tau^2 +1}& (-1)^{\tau^3 +1}& (-1)^{\tau^2 + \tau^3 +1}& 0& 0&   0
\\
a_4 & 4 & 1 & 0 &  (-1)^{\tau^2 +1}& (-1)^{\tau^3 +1}& (-1)^{\tau^2 + \tau^3 +1}& 0& 0&   0
\\\hline
\end{array}
\end{equation*}
\caption{Bulk wrapping numbers and $\Z_2^{(1)}$ exceptional wrapping numbers of Model 1a.}
\label{tab:Example-1a-XY+x1y1}
\end{table}
}
\begin{table}[h]
\bCentering
\resizebox{\linewidth}{!}{
\begin{tabular}{|c||c|c|c|c|c||c||c|c|c|c|c||c|}\hline
\muc{13}{|c|}{{\bf Exceptional wrapping numbers of Model 1a, Part II}}
\\\hline\hline
 & $\!x_{1}^{(2)} \!\!\!$ & $\! x_{2}^{(2)} \!\!\! $ & $\! x_{3}^{(2)} \!\!\!$ & $\! x_{4}^{(2)} \!\!\!$ & $\! x_{5}^{(2)} \!\!\! $ & $\! y_{1,\ldots,5}^{(2)} $
$\!\!\!$ & $\! x_{1}^{(3)} \!\!\!$ & $\! x_{2}^{(3)} \!\!\!$ & $\! x_{3}^{(3)} \!\!\!$ & $\! x_{4}^{(3)} \!\!\!$ & $\! x_{5}^{(3)} \!\!\!$ & $\! y_{1,\ldots,5}^{(3)}$
\\\hline\hline
$a_1$
& $(-1)^{\tau^3}$ & $(-1)^{\tau^1}$ & $(-1)^{\tau^3 + \tau^1}$ & 0& 0&  0
& $(-1)^{\tau^1}$ & $(-1)^{\tau^2}$ & $(-1)^{\tau^1 + \tau^2}$ & 0& 0&  0
\\
$a_2$ 
& $(-1)^{\tau^3 +1}$ & $(-1)^{\tau^1 +1}$ & $(-1)^{\tau^3 + \tau^1 +1}$ & 0& 0&  0
& $(-1)^{\tau^1 +1}$ & $(-1)^{\tau^2 +1}$ & $(-1)^{\tau^1 + \tau^2 +1}$ & 0& 0&  0
\\
$a_3 $
& $(-1)^{\tau^3}$ & $(-1)^{\tau^1}$ & $(-1)^{\tau^3 + \tau^1}$ & 0& 0&  0
& $(-1)^{\tau^1 +1}$ & $(-1)^{\tau^2 +1}$ & $(-1)^{\tau^1 + \tau^2 +1}$ & 0& 0&  0
\\
$a_4 $
& $(-1)^{\tau^3 +1}$ & $(-1)^{\tau^1 +1}$ & $(-1)^{\tau^3 + \tau^1 +1}$ & 0& 0&  0
& $(-1)^{\tau^1}$ & $(-1)^{\tau^2}$ & $(-1)^{\tau^1 + \tau^2}$ & 0& 0&  0
\\\hline
\end{tabular}}
\bCaption{$\Z^{(2)}_2$ and $\Z^{(3)}_2$ exceptional wrapping numbers of Model 1a.}
\label{tab:Example-1a-x2y2+x3y3}
\end{table}
{\footnotesize
\begin{table}[h]
\begin{equation*}
\begin{array}{|c||c||c|c||c|c|c|c|c||c|c|c|c|c|}\hline
\muc{14}{|c|}{\text{\bf Bulk and exceptional wrapping numbers of Model 1b, Part I}}
\\\hline\hline
 & N & X & Y & x_{1}^{(1)} & x_{2}^{(1)}  & x_{3}^{(1)}   & x_{4}^{(1)}  & x_{5}^{(1)}   & y_{1}^{(1)}   & y_{2}^{(1)}  & y_{3}^{(1)}   & y_{4}^{(1)}  & y_{5}^{(1)}  
\\\hline\hline
\tilde{a}_1 & 4 & 1 & 0 &   0& 0& -1& -1& 0&   0& 0& 0&  1& -1
\\
\tilde{a}_2 & 4 & 1 & 0 &   0& 0& -1& -1& 0&   0& 0& 0&  1& -1
\\
\tilde{a}_3 & 4 & 1 & 0 &   0& 0&  1&  1& 0&   0& 0& 0& -1&  1
\\
\tilde{a}_4 & 4 & 1 & 0 &   0& 0&  1&  1& 0&   0& 0& 0& -1&  1
\\\hline
\end{array}
\end{equation*}
\caption{Bulk wrapping numbers and $\Z_2^{(1)}$ exceptional wrapping numbers  of Model 1b.}
\label{tab:Example-1b-XY+x1y1}
\end{table}
}
\begin{table}[h]
\bCentering
\resizebox{\linewidth}{!}{
\begin{tabular}{|c||c|c|c|c|c||c|c|c|c|c||c|c|c|c|c||c|c|c|c|c|}\hline
\muc{21}{|c|}{\text{\bf Exceptional wrapping numbers of Model 1b, Part II}}
\\\hline\hline
 & $\! x_{1}^{(2)} \!\!\!$ & $\!x_{2}^{(2)}  \!\!\!$ & $\!x_{3}^{(2)}   \!\!\!$ & $\!x_{4}^{(2)}  \!\!\!$ & $\!x_{5}^{(2)}   \!\!\!$ & $\!y_{1}^{(2)}  \!\!\!$ & $\!y_{2}^{(2)}  \!\!\!$ & $\!y_{3}^{(2)}   \!\!\!$ & $\!y_{4}^{(2)}  \!\!\! $ & $\!y_{5}^{(2)}  
\!\!\!$ & $\!x_{1}^{(3)} \!\!\!$ & $\!x_{2}^{(3)}  \!\!\!$ & $\!x_{3}^{(3)}   \!\!\!$ & $\!x_{4}^{(3)}  \!\!\!$ & $\!x_{5}^{(3)}   \!\!\!$ & $\!y_{1}^{(3)}  \!\!\!$ & $\!y_{2}^{(3)}  \!\!\!$ & $\!y_{3}^{(3)}   \!\!\!$ & $\!y_{4}^{(3)}  \!\!\!$ & $\!y_{5}^{(3)}  \!\!\!$
\\\hline\hline
$\tilde{a}_1 $
&  0& 0& -1& -1& 0&  0& 0& 0& 1& -1
&  0& 0& -1& -1& 0&  0& 0& 0& 1& -1
\\
$\tilde{a}_2 $
&  0& 0&  1&  1& 0&  0& 0& 0& -1& 1
&  0& 0&  1&  1& 0&  0& 0& 0& -1& 1
\\
$\tilde{a}_3 $
&  0& 0& -1& -1& 0&  0& 0& 0&  1& -1
&  0& 0&  1&  1& 0&  0& 0& 0& -1&  1
\\
$\tilde{a}_4 $
&  0& 0&  1&  1& 0&   0& 0& 0& -1&  1
&  0& 0& -1& -1& 0&   0& 0& 0&  1& -1
\\\hline
\end{tabular}}
\caption{$\Z^{(2)}_2$ and $\Z^{(3)}_2$ exceptional wrapping numbers  of Model 1b.}
\label{tab:Example-1b-x2y2+x3y3}
\end{table}
{\footnotesize
\begin{table}[h]
\begin{equation*}
\begin{array}{|c||c||c|c||c|c|c|c|c||c|c|c|c|c|}\hline
\muc{14}{|c|}{\text{\bf Bulk and exceptional wrapping numbers of Model 1c, Part I}}
\\\hline\hline
 & N & X & Y & x_{1}^{(1)} & x_{2}^{(1)}  & x_{3}^{(1)}   & x_{4}^{(1)}  & x_{5}^{(1)}   & y_{1}^{(1)}   & y_{2}^{(1)}  & y_{3}^{(1)}   & y_{4}^{(1)}  & y_{5}^{(1)}  
\\\hline\hline
\hat{a}_1 & 4 & 1 & 0 &   -1& 0& 0& 0& (-1)^{\tau +1}&   0& 0& 0& (-1)^{\tau +1}& (-1)^{\tau}
\\
\hat{a}_2 & 4 & 1 & 0 &   -1& 0& 0& 0& (-1)^{\tau +1}&   0& 0& 0& (-1)^{\tau +1}& (-1)^{\tau}
\\
\hat{a}_3 & 4 & 1 & 0 &    1& 0& 0& 0& (-1)^{\tau}&      0& 0& 0& (-1)^{\tau}& (-1)^{\tau +1}
\\
\hat{a}_4 & 4 & 1 & 0 &    1& 0& 0& 0& (-1)^{\tau}&      0& 0& 0& (-1)^{\tau}& (-1)^{\tau +1}
\\\hline
\end{array}
\end{equation*}
\caption{Bulk wrapping numbers and $\Z_2^{(1)}$ exceptional wrapping numbers  of Model 1c.}
\label{tab:Example-1c-XY+x1y1}
\end{table}
}
\begin{table}[h]
\bCentering
\resizebox{\linewidth}{!}{
\begin{tabular}{|c||c|c|c|c|c||c||c|c|c|c|c||c|c|c|c|c|}\hline
\muc{17}{|c|}{\text{\bf Exceptional wrapping numbers of Model 1c, Part II}}
\\\hline\hline
 & $\! x_{1}^{(2)} \!\!\!$ & $\!x_{2}^{(2)}  \!\!\!$ & $\!x_{3}^{(2)}   \!\!\!$ & $\!x_{4}^{(2)}  \!\!\!$ & $\!x_{5}^{(2)}   \!\!\!$ & $\!y_{1,\ldots,5}^{(2)}  \!\!\!$ & $\!x_{1}^{(3)} \!\!\!$ & $\!x_{2}^{(3)}  \!\!\! $ & $\!x_{3}^{(3)}   \!\!\!$ & $\!x_{4}^{(3)}  \!\!\!$ & $\!x_{5}^{(3)}   \!\!\!$ & $\!y_{1}^{(3)}  \!\!\!$ & $\!y_{2}^{(3)}  \!\!\!$ & $\!y_{3}^{(3)}   \!\!\!$ & $\!y_{4}^{(3)}  \!\!\!$ & $\!y_{5}^{(3)}  \!\!\!$
\\\hline\hline
$\hat{a}_1 $
&  0& -1& 0& $(-1)^{\tau}$ & $(-1)^{\tau}$ & 0
&  0& 0& -1& -1& 0&  0& 0& 0& 1& -1
\\
$\hat{a}_2 $
&  0& 1& 0& $(-1)^{\tau +1}$ & $(-1)^{\tau +1}$ & 0
&  0& 0& 1& 1& 0&  0& 0& 0& -1& 1
\\
$\hat{a}_3 $
&  0& -1& 0& $(-1)^{\tau}$ & $(-1)^{\tau}$ & 0
&  0&  0& 1& 1& 0&  0& 0& 0& -1& 1
\\
$\hat{a}_4 $
&  0& 1&  0& $(-1)^{\tau +1}$ & $(-1)^{\tau +1}$ & 0
&  0& 0& -1& -1& 0&  0& 0& 0& 1& -1
\\\hline
\end{tabular}}
\caption{$\Z^{(2)}_2$ and $\Z^{(3)}_2$ exceptional wrapping numbers of Model 1c.}
\label{tab:Example-1c-x2y2+x3y3}
\end{table}

\clearpage

\subsubsection{Deformations of the models with USp$\boldsymbol{(8)^4}$ or SO$\boldsymbol{(8)^4}$ gauge group}\label{Sss:USp+SO-defs}

Before we discuss deformations in these models, we first show how to access the relevant cycles. In the SO-model, all D-branes wrap cycles with bulk wrapping numbers $(1,0;1,0;1,0)$ which are displaced. To access them in the hypersurface formalism we would have to apply the shifts $\alpha_2$ or $\alpha_4$, see equation \eqref{Eqn:HalfLatticeShifts} and figure~\ref{fig:CoordTrafo_tilted}, which lead to complex coefficients in the hypersurface equation for any of the deformations. Thus, these cycles are not easily described in the real $x_i$-plane. Instead, we can take the decomposition of these cycles into bulk and exceptional parts and compute their volumes individually. The description of the exceptional parts was already discussed. For the bulk cycle $\Pi_{135}$, one could in principle use any representative, but for technical reasons there is one especially suitable representative, as we will demonstrate now.
\begin{figure}[ht]
\centering
\includegraphics[width=0.9\textwidth]{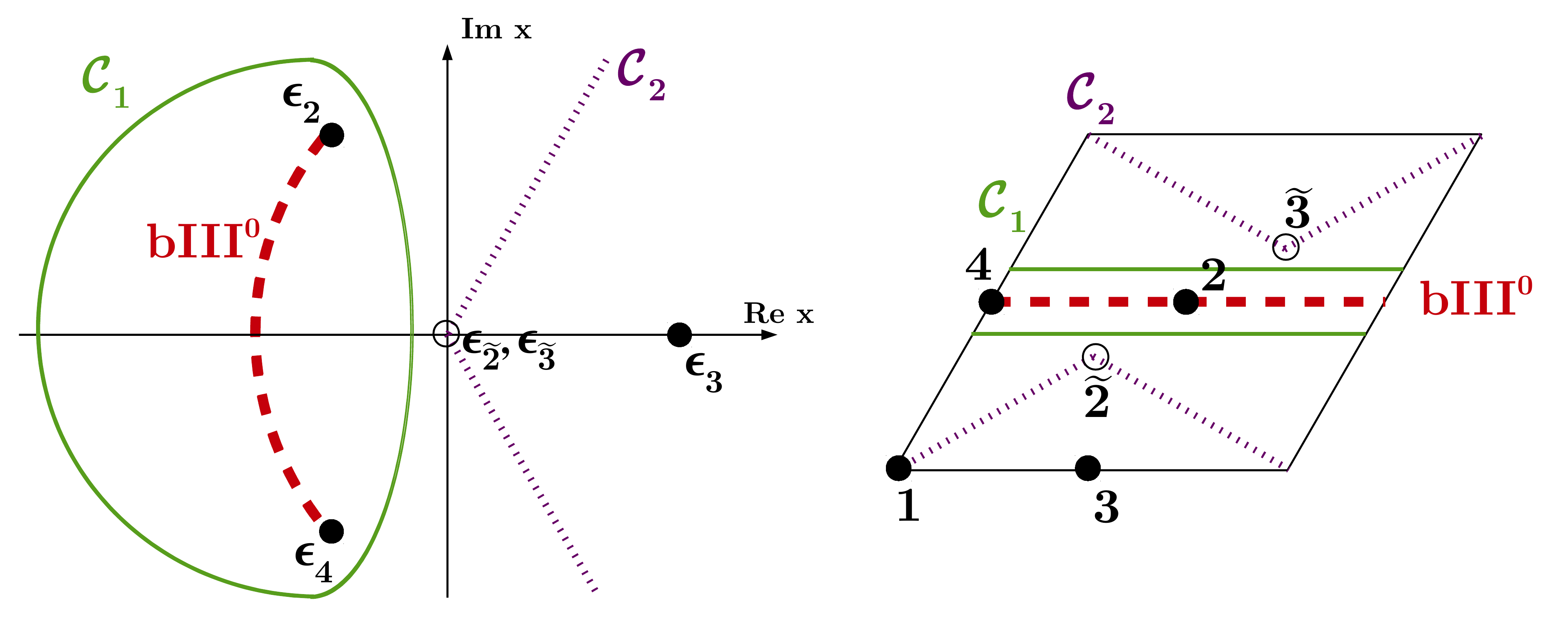}
\caption{The right picture shows the fundamental region of the hexagonal two-torus, and the left picture shows its image via the Weierstrass $\wp$-function into the complex $x$-plane for $v \equiv 1$. In order to describe the bulk part of the displaced fractional cycle {$\bf bIII^0$}, we move it away from the fixed points (cycle $\mathcal{C}_1$). The most useful representative for this cycle 
within the same homology class is $\mathcal{C}_2$, which passes through the fixed point 1 located at $x=\infty$ in the complex $x$-plane (left figure) or at the origin of the fundamental domain of the two-torus (right figure), and through the $\Z_3$ fixed points $\epsilon_{\tilde2 , \tilde3}$ (both in the origin of the $x$-plane, but separated in the fundamental domain). As discussed further in the main text, the sought-after integral is topological, i.e. $\int_{\mathcal{C}_1} \Omega_3 =\int_{\mathcal{C}_2} \Omega_3$.
 }
\label{fig:IntPath}
\end{figure}

The situation is depicted in figure~\ref{fig:IntPath}. A priori, any closed cycle in the real $x$-plane, which encircles the two fixed points $\epsilon_{2,4}$ located at $x=\xi$ and $x=\xi^2$ exactly once while not crossing the line $x \ge 1$, is a representative of the required bulk one-cycle. The bulk two- or three-cycles are simply products of such one-cycles. Now we want to integrate the holomorphic three-form $\Omega_3$ over these cycles, see equation~\eqref{eqn:Omega3}. For technical reasons, it is useful to select a path which avoids the zeros of the real part of $y$. The ideal choice for this are straight lines in the complex $x$-plane through the origin with slope $e^{\pm i \pi /3}$, see path $\mathcal{C}_2$ in figure~\ref{fig:IntPath}. These straight lines pass through the $\Z_6$ fixed point at $x=\infty$ as well as the $\Z_3$ fixed points at $x=0$, which are all not deformable, cf.\ table~\ref{tab:Hodge-numbers}, and thus do not influence the value of the integral. Intuitively, this is the path which has the maximal distance to the deformable $\Z_2$ fixed points and thus is least influenced by their deformations. Although this cycle itself is not {\it sLag}, it is in the same homology class as the ``horizontal" {\it sLag} cycle (called $\mathcal{C}_1$ in figure~\ref{fig:IntPath}) and thus leads to the same value for the integral of the closed holomorphic three-form, i.e.\ $\int_{\mathcal{C}_1} \Omega_3 = \int_{\mathcal{C}_2} \Omega_3$.

Now we are ready to compute the periods of the holomorphic three-form on the cycles wrapped by D6-branes. After normalisation to Vol$(\Pi^{\text{frac}})$ = 1 at the orbifold point, the calculations can be reduced to six cases. The exact formulas for the deformation polynomials and in particular for $\Omega_3$ are given in appendix \ref{A:1}. In figures~\ref{Fig:Int1} we plotted the change of the volume of ``horizontal" fractional cycles with exceptional wrapping number $x_1^{(i)}=1$ (figure~\ref{Fig:Int1+}) and $x_1^{(i)}=-1$ (figure~\ref{Fig:Int1-}), depending on the deformation parameter $\varepsilon_1^{(i)}$. 
\begin{figure}[th]
 \centering
\subfloat[Normalised periods of $\Omega_3$ on fractional {\it sLag} cycles of the form $\Pi_{\rm horizontal} + \boldsymbol{\varepsilon}_1^{(i)}$ plotted against $\varepsilon_1^{(i)}$.]{ \includegraphics[width=194pt]{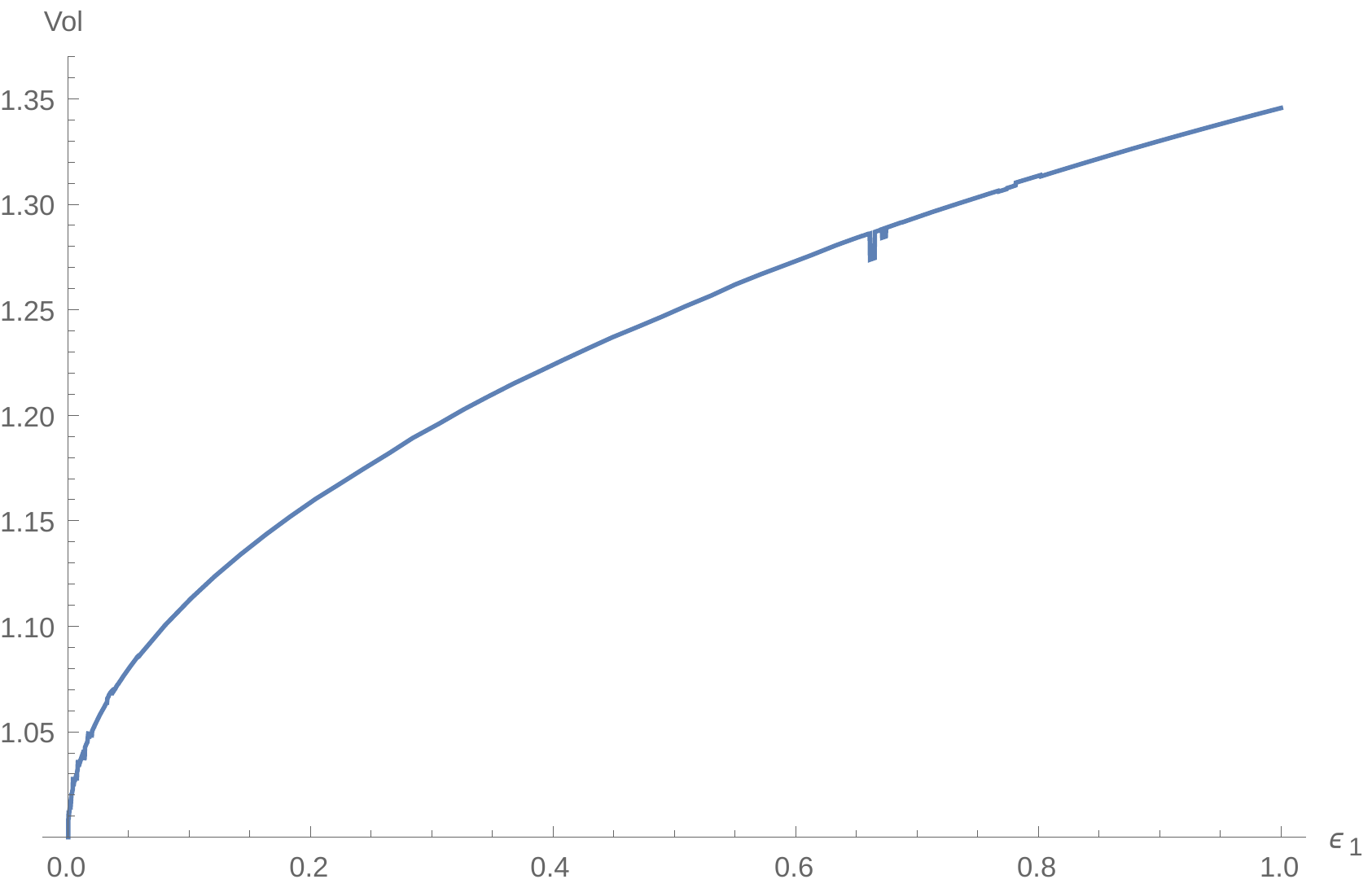}\label{Fig:Int1+}}
\hspace{30pt}
\subfloat[Normalised periods of $\Omega_3$ on fractional {\it sLag} cycles of the form $\Pi_{\rm horizontal} - \boldsymbol{\varepsilon}_1^{(i)}$ plotted against $\varepsilon_1^{(i)}$.]{ \includegraphics[width=194pt]{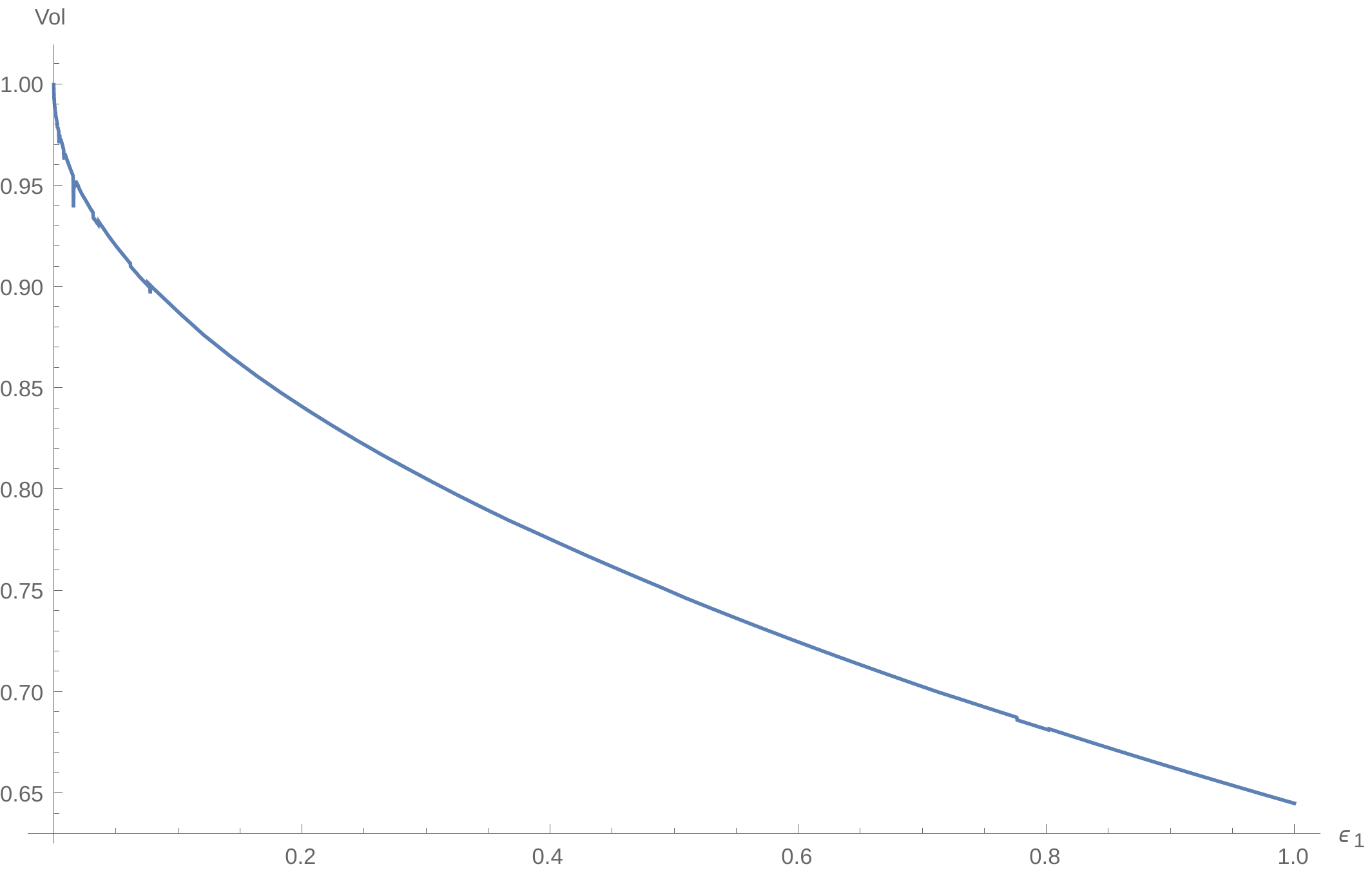}\label{Fig:Int1-}}
\caption{Integrals of $\Omega_3$ on ``horizontal'' fractional  {\it sLag} cycles, depending on $\varepsilon_1^{(i)}$. 
}
\label{Fig:Int1}
\end{figure}
Note that this does not only hold for all $\Z_2^{(i)}$ sectors ($i=1,2,3$), but also for replacing the cycle index \mbox{$\rho =1 \rightarrow \rho=2$} 
due to the permutation symmetry of the two-tori. As was also noticed in~\cite{Blaszczyk:2014xla}, the volume grows~/ shrinks with a $\pm \sqrt{\varepsilon_1^{(i)}}$ behaviour, which continues up to relatively large deformations of $\varepsilon_1^{(i)} \simeq 1$.

\begin{figure}[th]
 \centering
\subfloat[Normalised periods of $\Omega_3$ on fractional {\it sLag} cycles of the form $\Pi_{\rm horizontal} + \boldsymbol{\varepsilon}_3^{(i)}$ plotted against $\varepsilon_3^{(i)}$.]{ \includegraphics[width=194pt]{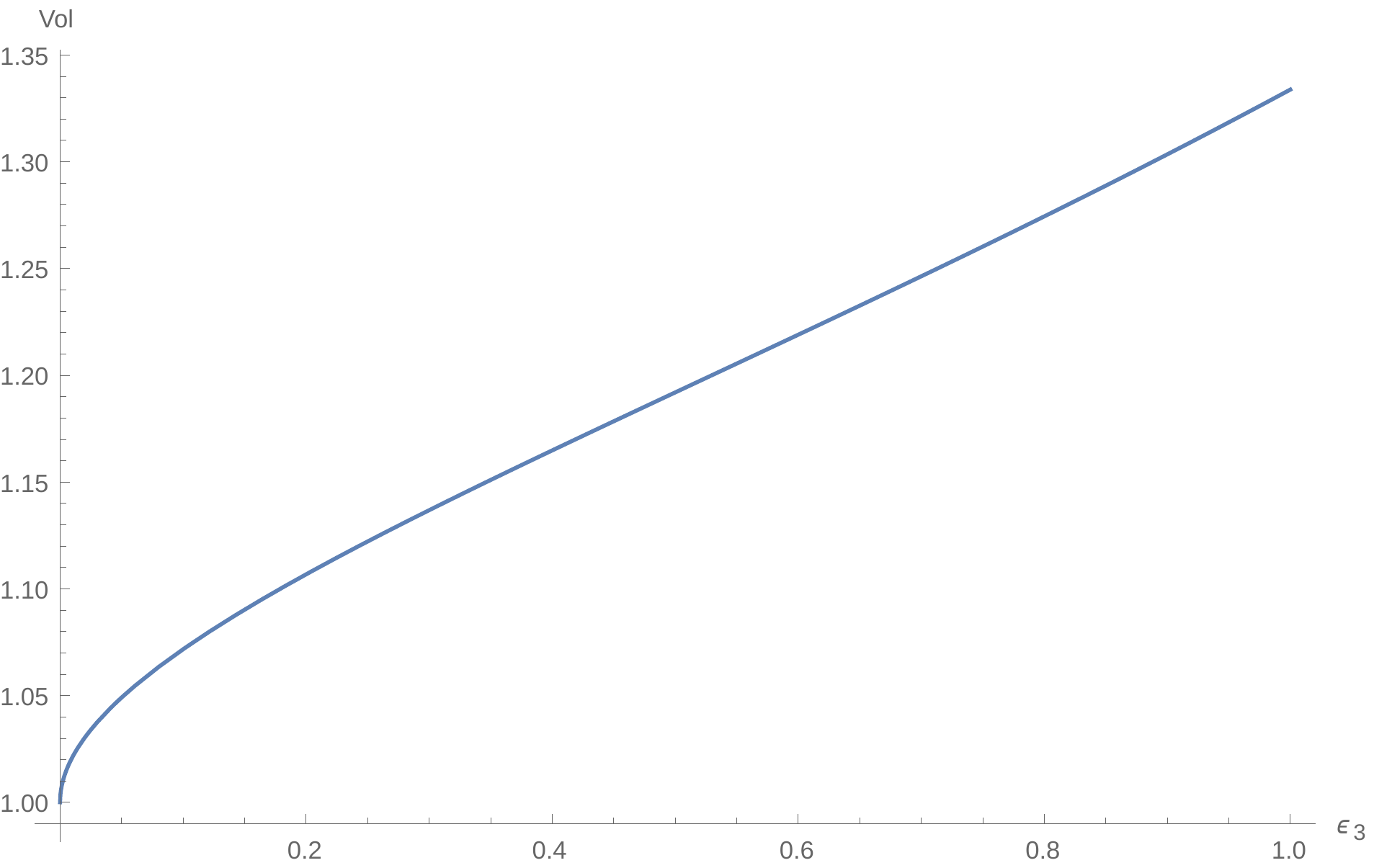}\label{Fig:Int3+}}
\hspace{30pt}
\subfloat[Normalised periods of $\Omega_3$ on fractional {\it sLag} cycles of the form $\Pi_{\rm horizontal} - \boldsymbol{\varepsilon}_3^{(i)}$ plotted against $\varepsilon_3^{(i)}$.]{ \includegraphics[width=194pt]{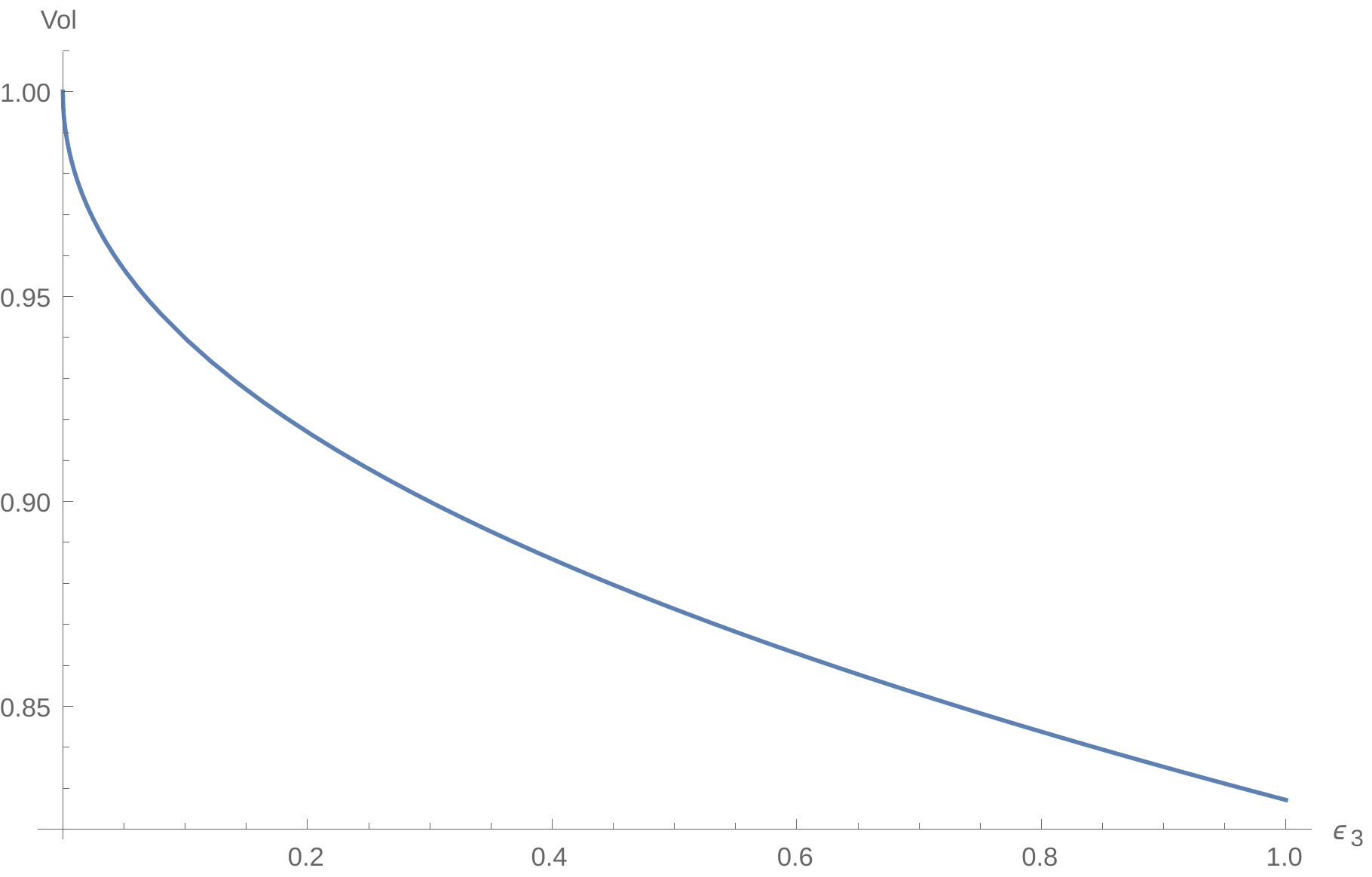}\label{Fig:Int3-}}
\caption{Integrals of $\Omega_3$ on  ``horizontal''  fractional {\it sLag} cycles, depending on $\varepsilon_3^{(i)}$.}
\label{Fig:Int3}
\end{figure}

Similarly, figures~\ref{Fig:Int3} show the variation of the volume of fractional ``horizontal" cycles with exceptional wrapping number $x_3^{(i)}=\pm1$ in dependence of $\varepsilon_3^{(i)}$. 
Again, the growth is of square-root type at the beginning, but now goes over to an almost linear graph when leaving the small-deformation regime.

\begin{figure}[th]
 \centering
\subfloat[Normalised periods of $\Omega_3$ on fractional {\it sLag} cycles of the form $\Pi_{\rm horizontal} + \boldsymbol{\varepsilon}_4^{(i)}$ plotted against $\varepsilon_{4+5}^{(i)}$.]{ \includegraphics[width=194pt]{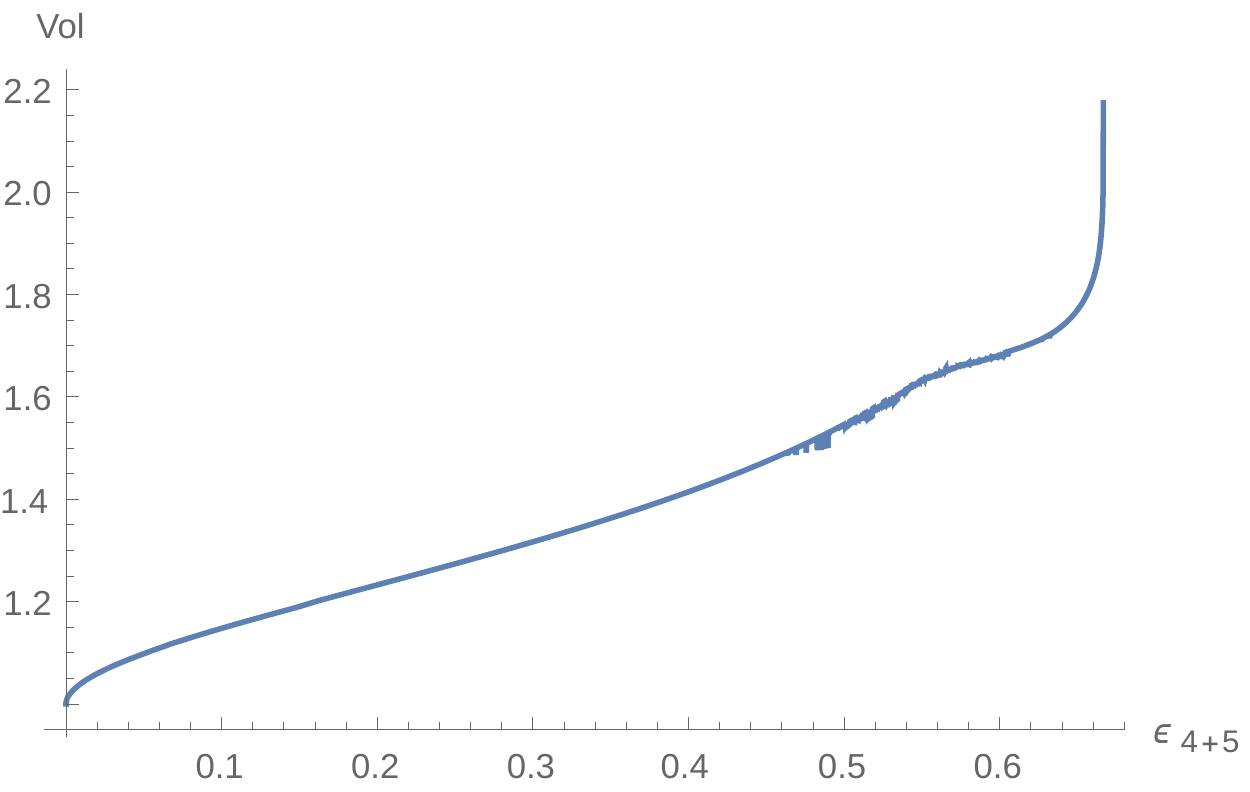}}
\hspace{30pt}
\subfloat[Normalised periods of $\Omega_3$ on fractional {\it sLag} cycles of the form $\Pi_{\rm horizontal} - \boldsymbol{\varepsilon}_4^{(i)}$ plotted against $\varepsilon_{4+5}^{(i)}$.]{ \includegraphics[width=194pt]{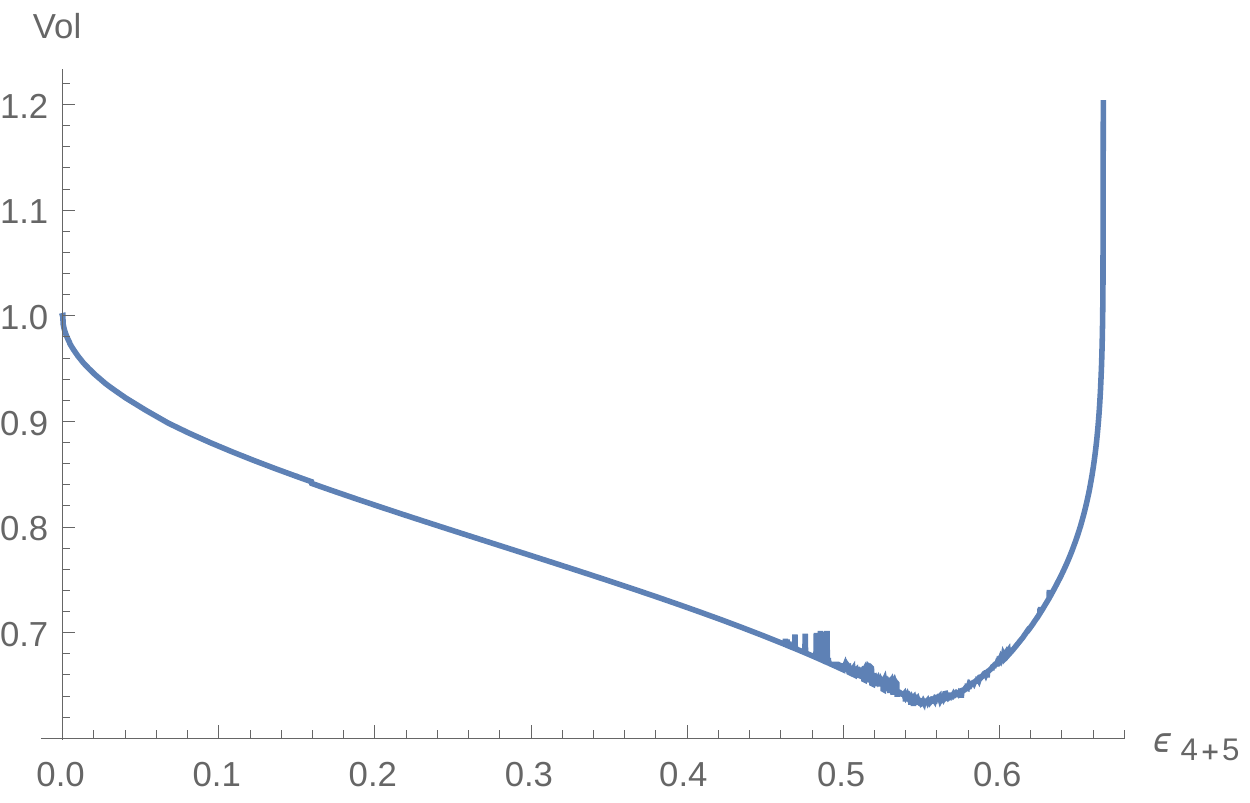}}
\caption{Integrals of $\Omega_3$ on ``horizontal'' fractional  {\it sLag} cycles, depending on $\varepsilon_{4+5}^{(i)}$.}
\label{Fig:Int4}
\end{figure}

Finally, figures~\ref{Fig:Int4} depict the volume of fractional ``horizontal" cycles with exceptional wrapping numbers $x_4^{(i)} = -y_4^{(i)} = y_5^{(i)}=\pm1$, $x_5^{(i)}=0$ or $x_5^{(i)}=y_4^{(i)}=-y_5^{(i)}=\pm1$, $x_4^{(i)}=0$ as a function of $\varepsilon_4^{(i)}=\overline\varepsilon_5^{(i)}$ for $\varepsilon_4^{(i)}\ge 0$. 
Note that although for $\varepsilon_4^{(i)}$ any phase is allowed, we restrict here to just positive values for reasons of simplicity. As before, the curves start with a square root-like behaviour. However, the higher order corrections become important much earlier than in the other cases such that the curves seem to diverge at $\varepsilon_4^{(i)} \simeq 2/3$.
We now comment on each of the three models separately.

In {\bf Model 1a} all D6-branes $\Pi_{a_i}$ are undisplaced, thus one can easily see, even before summing over $\Z_3$ orbits, that each exceptional cycle in the decomposition is {\it sLag} by itself. Therefore, the choice of $\Z_2$ eigenvalues and discrete Wilson lines will only change the signs with which the exceptional cycles contribute to the fractional three-cycles, but not their calibration. More concretely, we find the cycles $e^{(k)}_{\alpha\beta}\otimes \pi_{2k-1}$ with $(\alpha\beta)=(13), (31)$ or $(33)$ plus their $\Z_3$ images. Thus, the deformations $\varepsilon_1,\varepsilon_2,\varepsilon_3$ will change the volume of these cycles while leaving them {\it sLag}. We only have to consider two different kinds of deformations for two possible signs in front of the exceptional cycles, see figures~\ref{Fig:Int1} and \ref{Fig:Int3}. The precise correspondence between the branes and the volume as function of the parameters $\varepsilon_\rho^{(i)}$ can be read off from the wrapping numbers in tables~\ref{tab:Example-1a-XY+x1y1} and \ref{tab:Example-1a-x2y2+x3y3}.

In {\bf Model 1b}, all brane cycles are ``horizontal" and displaced in all three tori, and thus contain only exceptional three-cycles of the form $e_{\alpha\beta}^{(k)}\otimes\pi_{2k-1}$ with $\alpha,\beta=2,4$ plus their $\Z_3$ orbifold images. The Wilson lines are arranged such that, when summing over the $\Z_3$ orbits, only {\it sLag} exceptional cycles remain. For example, we find
\begin{align}
\Pi_{\tilde a_i} \supset \pm e^{(k)}_{24} \otimes \left[  (-\pi_{2k-1} - \pi_{2k} ) + \pi_{2k}\right] =  \mp e^{(k)}_{24} \otimes \pi_{2k-1}  \label{eqn:e33inIb} \,,
\end{align}
i.e.\ the non-{\it sLag} parts cancel out, and we are allowed to switch on the deformation parameters $\varepsilon_
{4 \pm 5}^{(k)}$. A similar discussion can be applied to the exceptional cycles containing the divisor $e_{33}^{(k)}$, thus allowing for deformations $\varepsilon_{3}^{(k)}$.  Therefore, we expect a direct dependence of the volume of $\Pi_{\tilde a_i}$ only on the two parameters $\varepsilon_3^{(k)}$ and $\varepsilon_{4\pm 5}^{(k)}$. Since the situation is the same for permutations of the two-tori, we are essentially left with four cases to compute, see figures~\ref{Fig:Int3} and \ref{Fig:Int4}, where the precise assignment can be read off in tables~\ref{tab:Example-1b-XY+x1y1} and \ref{tab:Example-1b-x2y2+x3y3}.

Finally, in {\bf Model 1c}, we have D6-branes which are displaced on two two-tori and undisplaced in the remaining two-torus. Thus, depending on the $\Z_2$ sector which we are deforming, we obtain different exceptional contributions. However, again the relative signs are arranged such that all exceptional parts with non-{\it sLag} calibration cancel out after summing over $\Z_3$ orbits, thus again all deformations are allowed. The shapes of the periods are, depending on the wrapping numbers in tables~\ref{tab:Example-1c-XY+x1y1} and \ref{tab:Example-1c-x2y2+x3y3}, again found in figures~\ref{Fig:Int1}, \ref{Fig:Int3} or  \ref{Fig:Int4}.


\subsection{A Pati--Salam model}
\label{sec:PatiSalam}

In this second example we present a Pati-Salam-like model where each stack of $N$ D6-branes supports a gauge factor of U$(N)$-type. The crucial difference to the toy models of the previous section is that this gauge group develops a non-trivial D-term upon deformation away from the orbifold point.
This statement holds true provided that the fractional D6-brane has wrappings along the corresponding exceptional cycle.

\subsubsection{The five-stack U$\boldsymbol{(4) \times }$U$\boldsymbol{(2)^4}$ model at the orbifold point}\label{Sss:PS-orbifold}

The global three-generation Pati-Salam model that we present here is taken from~\cite{Honecker:2012qr}. It has five stacks of D6-branes whose configuration is shown in table~\ref{Tab:decentPatiSalam1}. In particular we list their wrapping numbers $(n_i,m_i)$, $\Z_2$-eigenvalues, discrete Wilson lines $(\vec\tau)$ and displacements $(\vec\sigma)$. A first check shows that all twisted and untwisted RR tadpoles are cancelled, all supersymmetry conditions are fulfilled and the K-theory constraints are satisfied trivially, i.e.\ the model is globally consistent. Each brane stack wraps a cycle which is not equal to its orientifold image, thus the gauge group is composed of U($N$) factors only. Furthermore, in \cite{Honecker:2012qr} it was shown that all diagonal U$(1)\subset$U$(N)$ factors are anomalous and  become massive via the St\"uckelberg mechanism. Therefore the D-terms of these five diagonal U(1)s lead to D-term potentials in the effective action, which can potentially stabilise some of the deformation moduli dynamically. 

\begin{table}[h]
\Centering
\resizebox{\linewidth}{!}{
\begin{tabular}{|c||c|c||c|c|c||c|}\hline 
\muc{7}{|c|}{\bf Model 2: D6-brane configuration of a global Pati-Salam model}
\\\hline \hline
&\bf Wrapping numbers &$\frac{\rm Angle}{\pi}$&\bf $\boldsymbol{\Z_2^{(i)}}$ eigenvalues  & ($\vec \tau$) & ($\vec \sigma$)&\bf Gauge group\\
\hline \hline
 $a$&(0,1;1,0;1,-1)&$(\frac{1}{3},0,-\frac{1}{3})$&$(+++)$&$(0,0,1)$ & $(1,1,1)$& $U(4)$\\
 $b$&(0,1;1,0;1,-1)&$(\frac{1}{3},0,-\frac{1}{3})$&$(--+)$&$(0,1,1)$ & $(1,1,1)$&$U(2)_L$\\
 $c$&(0,1;1,0;1,-1)&$(\frac{1}{3},0,-\frac{1}{3})$&$(-+-)$&$(1,0,1)$ & $(1,1,1)$&$U(2)_R$\\
 \hline $d$ & (-1,2;2,-1;1,-1) &$(\frac{1}{2},-\frac{1}{6},-\frac{1}{3})$ & $(--+)$ &$(0,0,1)$&$(1,1,1)$&$U(2)_d$\\
 $e$ & (1,0;1,0;1,0)& $(0,0,0)$  &$(+--)$&$(1,1,1)$& $(1,1,0)$&$U(2)_e$\\
 \hline
\end{tabular}}
\caption{D6-brane configuration with five stacks of D6-branes yielding a globally defined Pati-Salam model with gauge group 
SU$(4)_a\times $SU$(2)_b\times $SU$(2)_c \times $SU$(2)_d \times $SU$(2)_e\times $U$(1)^5_{\text{massive}}$.\label{Tab:decentPatiSalam1}}
\end{table}

The corresponding exceptional wrapping numbers $(x^{(i)}_{\rho},y^{(i)}_{\rho})$ were amongst others used in~\cite{Honecker:2013hda,Honecker:2013kda} to determine the conditions on the existence of discrete $\Z_n$ gauge symmetries at low energies. 
It turned out that the model in table~\ref{Tab:decentPatiSalam1} contains a generation-dependent $\Z_2$ symmetry, which dictates the section rules on perturbative and non-perturbative matter couplings.

For the five stack model at hand, the exceptional wrapping numbers given in tables~\ref{tab:Example-1-XY+x1y1} and~\ref{tab:Example-1-x2y2+x3y3} show that none of the fractional D6-branes couples to the $2^{\rm nd}$ $\Z_2^{(i)}$ fixed point orbit for all $i=1,2,3$, and also all entries in the tables are zero for the $1^{\rm st}$ orbit in the $\Z_2^{(3)}$ sector.

\begin{table}[h]
\bCentering
\resizebox{\linewidth}{!}{
\begin{tabular}{|c||c||c|c||c|c|c|c|c||c|c|c|c|c|}\hline
\muc{14}{|c|}{\text{\bf Bulk and exceptional wrapping numbers of the U$\boldsymbol{(4) \times }$U$\boldsymbol{(2)^4}$ example, Part I}}
\\\hline\hline
 & N & X & Y & $x_{1}^{(1)}$ & $x_{2}^{(1)}$  & $x_{3}^{(1)}$   & $x_{4}^{(1)}$  & $x_{5}^{(1)} $  & $y_{1}^{(1)} $ &$ y_{2}^{(1)}$  &$ y_{3}^{(1)} $  & $y_{4}^{(1)}$  & $y_{5}^{(1)}  $
\\\hline\hline
a & 4 & 1 & 0 & 0& 0& 1& 1& -2& 0& 0& 0& -1& 1
\\
b & 2 & 1 & 0 & 0& 0& -1& 1& 0& 0& 0& 0& -1& 1
\\
c & 2 & 1 & 0 & 0& 0& -1& -1& 2& 0& 0& 0& 1& -1
\\
d & 2 & 3 & 0 &0& 0& 1& -3& 2& 0& 0& -2& 3& -1
\\ 
e & 2 & 1 & 0 &-1& 0& 0& 0& 1& 2& 0& 0& -1& -1 
\\\hline
\end{tabular}}
\caption{Bulk wrapping numbers and exceptional wrapping numbers from the $\Z_2^{(1)}$ twisted sector.
None of the stacks couples to the $\Z_3$-orbit of $\Z_2^{(1)}$ fixed points labelled by the lower index $\rho=2$.}
\label{tab:Example-1-XY+x1y1}
\end{table} 
\begin{table}[h]
\bCentering
\resizebox{\linewidth}{!}{
\begin{tabular}{|c||c|c|c|c|c||c|c|c|c|c||c|c|c|c|c||c|c|c|c|c|}\hline
\muc{21}{|c|}{\text{\bf Exceptional wrapping numbers of the U$\boldsymbol{(4) \times }$U$\boldsymbol{(2)^4}$  example, Part II}}
\\\hline\hline
 & $\! x_{1}^{(2)} \!\!\!$ & $\!x_{2}^{(2)}  \!\!\!$ & $\!x_{3}^{(2)}   \!\!\!$ & $\!x_{4}^{(2)}  \!\!\!$ & $\!x_{5}^{(2)}   \!\!\!$ & $\!y_{1}^{(2)}  \!\!\!$ & $\!y_{2}^{(2)}  \!\!\!$ & $\!y_{3}^{(2)}   \!\!\!$ & $\!y_{4}^{(2)}  \!\!\!$ & $\!y_{5}^{(2)}  
\!\!\!$ & $\!x_{1}^{(3)} \!\!\!$ & $\!x_{2}^{(3)}  \!\!\!$ & $\!x_{3}^{(3)}   \!\!\!$ & $\!x_{4}^{(3)}  \!\!\! $& $ \!x_{5}^{(3)}   \!\!\!$ & $\!y_{1}^{(3)}  \!\!\!$ & $\!y_{2}^{(3)}  \!\!\!$ & $\!y_{3}^{(3)}   \!\!\!$ & $\!y_{4}^{(3)}  \!\!\!$ & $\!y_{5}^{(3)}  \!\!\!$
\\\hline\hline
a 
&0&0& 1& -1& 0& 0& 0& 0& -1& 1
& 0& 0& -1& 1& 0& 0& 0& 0& -1& -1
\\
b 
& 0& 0& -1& 1& 0& 0& 0& 0& 1& -1
& 0& 0& 1& 1& -2& 0& 0& 0& -1& 1
\\
c 
& 0& 0& -1& 1& 0& 0& 0& 0& -1& 1
& 0& 0& 1& 1& 2& 0& 0& 0& -1& -1
\\
d 
& 0& 0& -1& 1& 0& 0& 0& 2& 1& -3
& 0& 0& -1& -1& 0& 0& 0& 0& 1& 1
\\ 
e 
& 1& 0& 0& 0& -1& -2& 0& 0& 1& 1
& 0& 0& 1& -1& 0& 0& 0& 0& 1& -1 
\\\hline
\end{tabular}}
\caption{Exceptional wrapping numbers from the $\Z_2^{(2)}$ and $\Z_2^{(3)}$ twisted sectors.
None of the stacks couples to $\rho=2$ in both $\Z_2^{(i)}$ sectors, or $\rho=1$ in the $\Z_2^{(3)}$ sector.
Also $\rho=3$ in the $\Z_2^{(3)}$ sector has only vanishing orientifold-odd combinations defined in equation~\protect\eqref{Eq:wrappings-OR-odd}.
}
\label{tab:Example-1-x2y2+x3y3}
\end{table}

In addition to the four  $\Z_2$ fixed point orbits already discussed above, none of the brane stacks in the example couples to the orientifold-odd part of the $3^{\rm rd}$ orbit of the $\Z_2^{(3)}$ sector.
Our expectation is thus that 4+4+2 out of the 5+5+5 deformation parameters from the three $\Z_2$ twisted sectors are stabilised at the supersymmetric orbifold point, and any deformation along one of these directions will break supersymmetry and provide a larger value for the scalar potential. On the other hand, the 1+1+3 deformations without D6-brane contributions 
to orientifold-odd combinations of  exceptional wrapping numbers are expected to constitute flat directions.

\subsubsection{Deformations of the model with U$\boldsymbol{(4) \times }$U$\boldsymbol{(2)^4}$ gauge symmetry}
\label{Sss:U-model-defs}

\paragraph{D-branes and effective action:}
The restrictions of the O6-planes on the deformations could be deduced from purely geometric considerations. For the restrictions from the D-branes, we must also take into account that they are dynamical objects, in particular because we first have to specify the multiplicity of D-branes on each cycle. Each stack of D-branes comes with an ${\cal N}=1$ supersymmetric gauge theory and thus with D-terms, which serve as a potential for the deformation moduli~\cite{Blumenhagen:2006ci},
\begin{equation}
\begin{aligned}
{\cal V}_{\text{scalar}}^{\rm NS-NS} & \propto \left( \sum_x N_x \bigl[ \text{Vol}(\Pi_x) +  \text{Vol}(\Pi_x') \bigr] -  \text{Vol} (\Pi_{O6})
\right)
&
\begin{cases}
=0 \qquad \text{if all D$6_x$-branes are {\it sLag}}
\\
> 0 \qquad \text{else}
\end{cases} \,.
\end{aligned}
\end{equation}

For a gauge group to be U($N$), the stack of $N$ identical D-branes must wrap a cycle which is not orientifold invariant. These fractional cycles always contain orientifold-odd exceptional contributions, although their bulk part may be orientifold-invariant (e.g.\ $\Pi_e$ in the Pati-Salam model) or not (e.g.\ $\Pi_{a,b,c,d}$). For most of these exceptional cycles this implies that they are not {\it sLag}, but only {\it Lag}, for non-vanishing volume. 
Thus, the model is {\it a priori}  only supersymmetric when such cycles remain singular.

Giving a vacuum expectation value ({\it vev}) to the deformation modulus corresponds to generating a Fayet-Iliopoulos term, resulting in a non-zero D-term of the U(1)$\subset$U($N$) and in breaking supersymmetry. The coefficient is proportional to the orientifold-odd wrapping numbers, see e.g.\ equation \eqref{Eq:wrappings-OR-odd}. Thus, in the general situation, D-branes with U($N$) gauge factors do stabilise the twisted moduli of the fixed points which they pass through at zero {\it vev}, i.e.\ at the orbifold point. 

However, in certain situations the D-term can be cancelled by assigning {\it vev}s to charged open string states. Consider an ${\cal N}=1$ supersymmetric SU($N$) gauge theory with $M$ chiral multiplets in the fundamental representation. Denote their scalar components by $\phi_i^a$ with SU($N$) index $a$ and flavour index $i$. Then the D-terms of the SU($N$) factor allow for a flat direction only if $M \ge N$ by assigning a {\it vev} of the form $\langle \phi_i^a \rangle \propto \delta_i^a$ up to gauge and flavour symmetry transformations. If we now have $N_i$ D-branes on two cycles $\Pi_i$, $i=1,2$, then the massless spectrum contains a chiral multiplet in the $\left( N_1, \overline{N_2} \right) $ representation precisely if $\Pi_1$ and $\Pi_2$ intersect. 
We find that this is for example the case if the bulk part of $\Pi_1$ and $\Pi_2$ is orthogonal in two of the three two-tori and parallel in the third one, and their exceptional part differs in a global sign.
Then a flat direction for both SU$(N_i)$ gauge factors can only be found if $N_1 = N_2$. The {\it vev} of the fundamental representations also appears in the D-terms of the U(1) factors and there indeed cancels the {\it vev} of the deformation modulus. 

Geometrically, the following happens: the exceptional contribution of the D-branes wrapped on $\Pi_i$ cancels only if $N_1 = N_2$, thus only in this case the sum of D-branes stays {\it sLag} when performing the deformation. The former two stacks of D-branes merge to only one stack of $N_1 = N_2$ D-branes resulting in just a U($N_1$) gauge theory. This agrees with the field theoretical picture since here the diagonal {\it vev} in the bifundamental representation is responsible for the symmetry breaking SU($N$)$\times$SU($N$)$\rightarrow$SU($N$).

\paragraph{Allowed deformations in the U(4)$\boldsymbol{\times}$U$\boldsymbol{(2)^4}$ model:}

\begin{figure}[th]
 \centering
\subfloat[Normalised periods of $\Im(\Omega_3)$ on fractional {\it sLag} cycles of the form $\Pi_{\rm horizontal} + 2 \boldsymbol{\tilde\varepsilon}_1$ plotted against $\varepsilon_{1}^{(i)}$. The real part $\Re(\Omega_3)$ is identical to figure \ref{Fig:Int1}. ]{ \includegraphics[width=194pt]{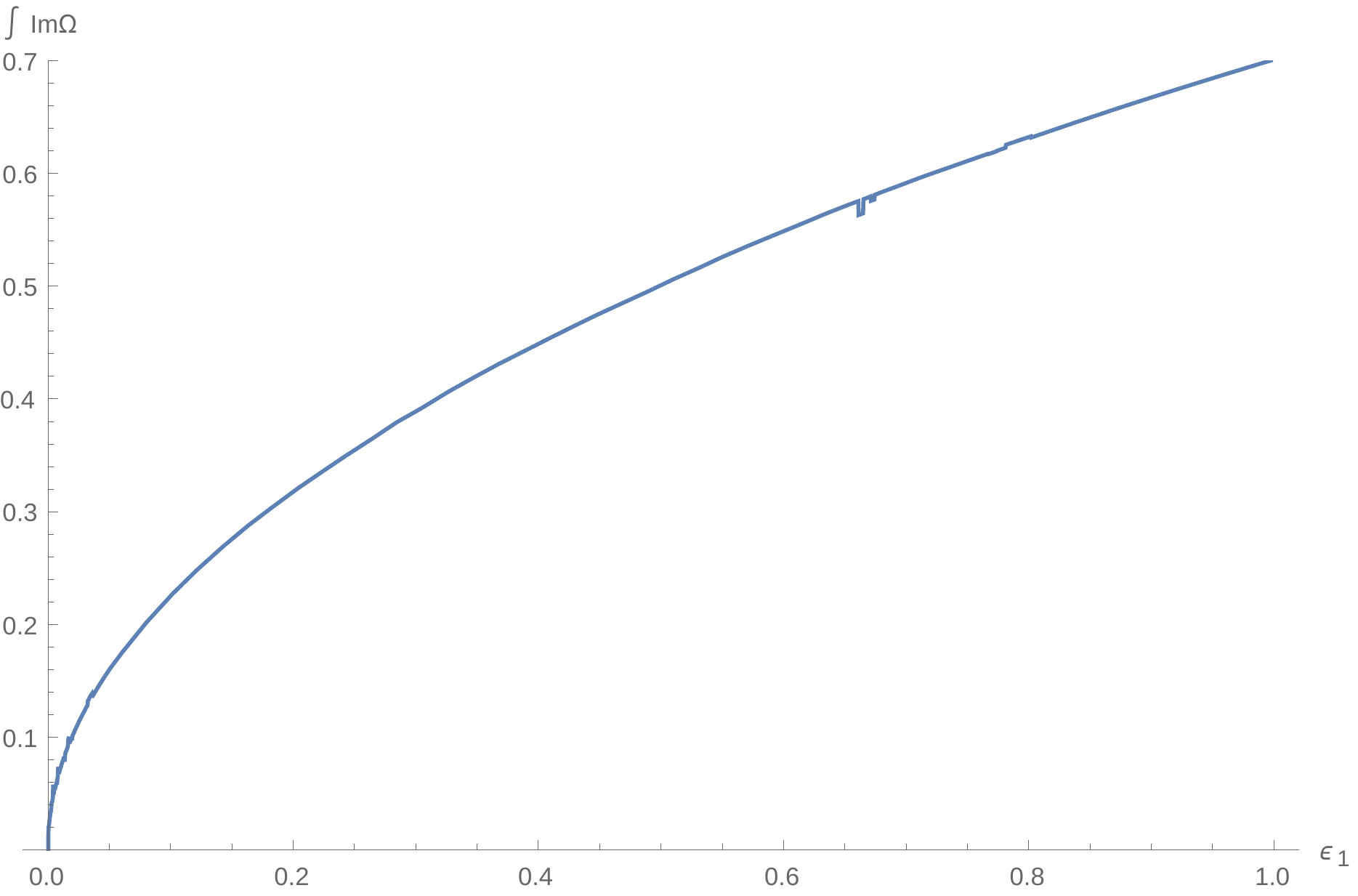}\label{Int1Imaginary}}
\hspace{30pt}
\subfloat[Normalised periods of $\Omega_3$ on fractional {\it sLag} cycles not containing $\boldsymbol{\varepsilon}_1$ plotted against $\varepsilon_{1}^{(i)}$. Notice the absence of the $\sqrt{\varepsilon_{1}^{(i)}}$ behaviour but only a higher order dependence.]{ \includegraphics[width=194pt]{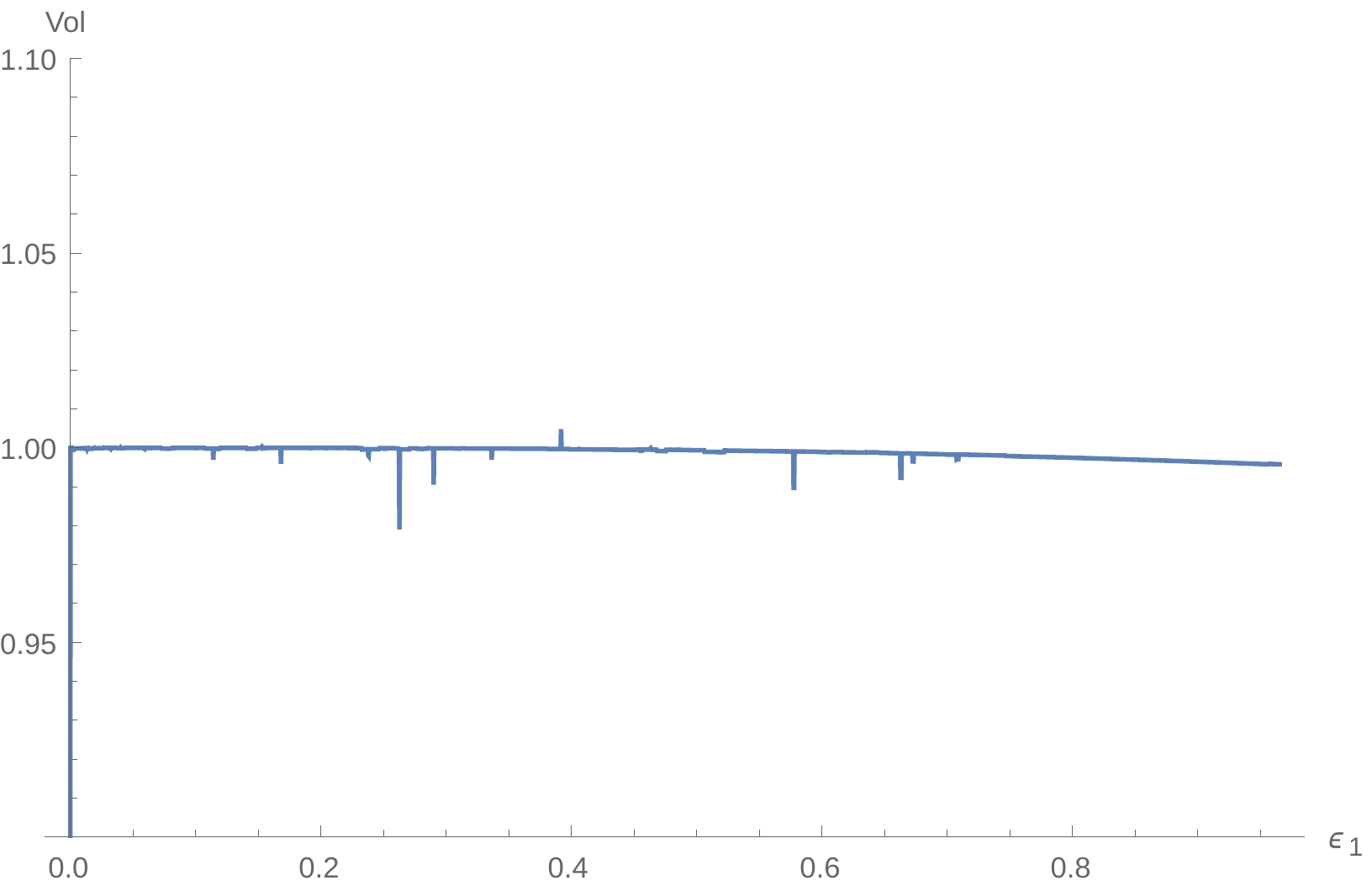}\label{Fig:Int1bulk}}
\caption{Integrals of $\Omega_3$ in the five-stack Pati-Salam model, depending on $\varepsilon_{1}^{(i)}$.}
\label{Fig:IntNonsLag}
\end{figure}

We will first focus on the five brane stacks and check, in a meaningful order, which of the deformations are forbidden. Then we will consider the remaining deformations and discuss their influence on the model.

\paragraph{The Pati--Salam branes $\boldsymbol{\Pi_{a,b,c}}$:} 
These three stacks of branes wrap the same bulk cycle and differ only in their $\Z_2$ eigenvalues and discrete Wilson lines, i.e.\ the signs of their exceptional cycles. As we can read off from tables \ref{tab:Example-1-XY+x1y1} and \ref{tab:Example-1-x2y2+x3y3}, they wrap only four distinct orientifold-odd cycles. For example, when summing over $\Z_3$ images, 
\begin{align}
\Pi_{ a} \supset  e^{(2)}_{24} \otimes \left[  (\pi_{4} - \pi_{3} ) + \pi_{4}\right] =   e^{(2)}_{24} \otimes \left( 2 \pi_{4} - \pi_3 \right)  \label{eqn:e24inPS} \,,
\end{align}
whose calibration depends on the phase of $\varepsilon_4^{(2)}$. In particular, these branes lead to the restrictions
\begin{align}
\varepsilon_4^{(1)} \ge 0 \,, \qquad \varepsilon_4^{(2)} \ge 0 \,, \qquad \varepsilon_4^{(3)} = 0 \,.
\end{align}

\paragraph{The hidden branes $\boldsymbol{\Pi_{d,e}}$:}
Once we imposed the restrictions from the visible branes, we turn to the two remaining brane stacks. Put together, they wrap six additional orientifold-odd cycles. Note that e.g.\ the $e$-branes differ from the $\hat{a}_2$-branes in model 1c 
of section~\ref{sec:SOUSp} only by the choice of discrete Wilson lines. This, however, leads to the non-cancellation of their orientifold-odd parts, e.g.\
\begin{align}
\Pi_{e} \supset  e^{(1)}_{13} \otimes \left[  (\pi_{1} - \pi_{2} ) - \pi_{2}\right] =   e^{(1)}_{13} \otimes \left(  \pi_{1} - 2 \pi_2 \right)  \label{eqn:e12inPS} \,.
\end{align}
For the deformation $\varepsilon_1^{(1)} > 0$, allowed by our choice of exotic O-plane,  this cycle is calibrated with $i \Omega_3$ and thus not {\it sLag}. Therefore, a supersymmetric vacuum requires $\varepsilon_1^{(1)} = 0$. Altogether, we find the following restrictions, in addition to the ones imposed by the three stacks of Pati-Salam-branes:
\begin{align}
\varepsilon_1^{(i)} =  \varepsilon_3^{(i)} =  \varepsilon_4^{(i)} = 0\,, \quad \text{ for } i=1,2 \,, \qquad \varepsilon_4^{(3)} = 0\,. 
\end{align}
For illustrational purposes we computed the integral of $\Im (\Omega_3)$ on the cycle $\Pi_d$, as shown in figure~\ref{Int1Imaginary}, where 
 $\Im (\Omega_3) \neq  0$ for $\varepsilon^{(i)}_1 >0$ corresponds to broken supersymmetry.

To sum up, the only allowed deformations in this model are
\begin{align}
\varepsilon_2^{(i)} \ge 0\,, \quad \text{ for } i=1,2,3\,, \qquad \varepsilon_1^{(3)} \ge 0\,, \qquad \varepsilon_3^{(3)} \ge 0\ , 
\end{align}
in perfect agreement with our expectation from section~\ref{Sss:PS-orbifold}.

We will now discuss the effects of these deformations.

\paragraph{Deformation  $\boldsymbol{\varepsilon_3^{(3)}}$:}
The deformation $\varepsilon_3^{(3)} > 0$ is of particular interest. This is because the visible branes contain only an orientifold-even cycle $\boldsymbol{\varepsilon}_3^{(3)}$, whose coefficient $x_3^{(3)}$ can be read off from table \ref{tab:Example-1-x2y2+x3y3}. Therefore, switching on this deformation will reduce the volume of the U$(4)$ branes and enhance the volume of the U$(2)_{\rm L/R}$ branes. Qualitatively, this effect is shown in figures \ref{Fig:Int3}, where figure \ref{Fig:Int3+} corresponds to the left/right branes and figure \ref{Fig:Int3-} to the U(4) brane. This can be phenomenologically interesting since in this way we can make the gauge coupling of the QCD SU(3) $\subset$ U(4) stronger compared to the weak interactions SU$(2)_{\rm L}$. 

\paragraph{Other deformations:} 

Finally, we discuss the effect of the four remaining deformations. First we observe that none of the branes wraps an orientifold-even cycle of this type, i.e. $(x^{(i)}_2 , y^{(i)}_2)^{i=1,2,3}=
(x^{(3)}_1,  y^{(3)}_1)=(0,0)$ in tables~\ref{tab:Example-1-XY+x1y1} and~\ref{tab:Example-1-x2y2+x3y3} for all five stacks. 
Therefore, we only expect a higher order dependence of the volumes on these parameters. For example, we plotted the normalised volume of $\Pi_{a,b,c,e}$ against the parameter $\varepsilon_1^{(3)}$ in figure \ref{Fig:Int1bulk}. Indeed, we find just a negligibly small change in the volume. Qualitatively the effect of the three other parameters $\varepsilon_2^{(i)}$ is the same, thus their value is not of great importance for the physics of the low-energy gauge theory in this model.


\section{Discussion and Conclusions}\label{S:Conclusions}

In this work we have analysed the behaviour of D6-brane models on toroidal orbifolds with discrete torsion under deformations of the $\Z_2$ fixed points. We focussed here on tilted tori, and most notably on tori which exhibit an additional $\Z_3$ symmetry, as they are particularly appealing for phenomenological models with three quark/lepton generations. In contrast to the deformations on untilted tori, as studied before in \cite{Blaszczyk:2014xla}, we observed some differences which required additional investigations. 

First, here we were interested in orbifolds with point group $\Z_2 \times \Z_6^\prime$, of which there exists no direct description as a hypersurface in an ambient toric space. Thus we started with a fully deformable description of $T^6/(\Z_2 \times \Z_2)$ and modded out the remaining $\Z_3$ symmetry by hand. This led to restrictions on the deformation parameters which are in agreement with the CFT expectations. Furthermore, this formalism is sufficient for our purposes, since the $\Z_3$ and $\Z_6$ sectors, whose singularities are not resolvable here, do not contain exceptional three-cycles which D6-branes could wrap.

In addition, each brane stack appeared as an orbit under the remaining $\Z_3$ symmetry, thus it may happen that a brane intersects its $\Z_3$ images at the fixed points. In such a case, the contributions from the corresponding exceptional cycles add up between the various $\Z_3$ images, whereas in $\Z_2 \times \Z_2$ models each brane has at most one exceptional cycle at each fixed point. This could result either in a {\it sLag} or a non-{\it sLag} cycle, depending on the value of the associated discrete Wilson lines. 

Another technical difficulty arose from the fact that the deformation parameters had to be switched on in a $\Z_3$-invariant way. As a consequence, we observed that some fixed points with different indices than the non-zero deformation parameters get deformed at higher order. Keeping them singular requires the introduction of counter terms in the hypersurface polynomial so that the map between the deformation parameters and the moduli {\it vev}s becomes more involved compared to the $\Z_2 \times \Z_2$ orbifold, see appendix \ref{A:1}.

For simplicity we focussed on just one choice of orientifold involution, $\sigma_\mathcal{R}$, corresponding to tilted tori with an {\bf AAA}-type lattice. Although this requires the parameters of the hypersurface polynomial to be real, many fractional cycles of interest cannot be described by restricting the homogeneous coordinates to be real, in contrast to the analogous case on untilted tori. Therefore, the parametrisation of these fractional cycles is difficult when leaving the orbifold point, and it turned out to be more feasible to access them indirectly by other representatives in their homology class. In this way, the periods of the holomorphic three-form can be computed for the relevant cycles under many deformations.

Moreover, in contrast to the untilted tori studied in \cite{Blaszczyk:2014xla}, on tilted tori not every exceptional cycle is (plus/minus) its own orientifold image, e.g.\ $\sigma_\mathcal{R} : \boldsymbol{\varepsilon}_4^{(i)} \leftrightarrow \boldsymbol{\varepsilon}_5^{(i)}$. These cycles appear in complex conjugate pairs, and their deformation parameter is allowed to take {\it a priori}  arbitrary complex values, leading to exceptional cycles of arbitrary calibration phase. However, in the global geometry, these cycles can only be nicely parametrised for a single calibration, corresponding to real deformation parameters.

After discussing the technical setup, we focussed on two types of D6-brane models, those with only USp($2N$) or only SO($2N$) gauge groups, and a model which contained only U($N$) gauge groups. From the effective action point of view, the difference is that the latter class of models contains U(1) gauge factors whose D-term potential can lead to the stabilisation of deformation moduli.

In the first class of models, we found that the D6-branes only wrap orientifold-even cycles, i.e.\ they remain {\it sLag} for any type of deformation. We confirmed this by computing the integrals of the holomorphic three-form on these cycles and found that the integrals indeed stay real, but can change in their absolute values. More precisely, if a D6-brane wraps a certain exceptional cycle, its volume shows a square root-like dependence on the corresponding deformation parameter with the sign depending on the associated $\Z_2$ eigenvalues and Wilson lines as shown in figures~\ref{Fig:Int1},~\ref{Fig:Int3} and~\ref{Fig:Int4}. 
Thus, in such models we are able to vary the gauge coupling constants along flat directions in the complex structure moduli space.

Finally, we looked at a Pati-Salam model with five U($N$) gauge factors, whose massless spectrum, generation-dependent discrete $\Z_2$ symmetry and Yukawa interactions
had been studied in \cite{Honecker:2012qr,Honecker:2013hda,Honecker:2015ela}. This model leads to five U(1) subgroups with five D-term potentials, but it turns out that indeed $4+4+2=10$ of the 15 deformation moduli can be stabilised at the orbifold point. The explanation is that each of these deformations locally leads to non-{\it sLag} contributions of exceptional three-cycles
for some of the D6-branes, independently of what happens at the other orbifold fixed points. We discussed the possibility of cancelling the Fayet-Iliopoulos terms with chiral states in the bifundamental representation in general, but the basic conditions to do so are not fulfilled for the present model. As a result, we found five D-flat directions among the deformation moduli for this model. One of these directions turns out to be able to change the SU(4) coupling constant against the SU$(2)_{\rm R/L}$ couplings, whereas the other four flat directions have no influence on the low energy-effective gauge theory. 
Analogously, we expect from the counting of non-vanishing orientifold-odd wrapping numbers for the global six-stack Pati-Salam model with spectrum and $\Z_n$ symmetries considered in~\cite{Honecker:2012qr,Honecker:2013hda,Honecker:2013kda}
that  4+3+4 out of 5+5+5 twisted deformation moduli are stabilised at the orbifold point, and only four flat directions in the complex structure moduli space exist.

Our findings are of importance also for phenomenologically appealing models on the $T^6/(\Z_2 \times \Z_6)$ orbifold with a different $\Z_6$ action~\cite{Ecker:2014hma}, in particular for the new global MSSM-like and left-right symmetric models~\cite{Ecker:2015vea}. In these models, one rectangular two-torus and two hexagonal two-tori respect the $\Z_2 \times \Z_6 \times \OR$ action.

The work in this article focussed on complex structure deformations in Type IIA orbifolds/orientifolds. For the $T^6/(\Z_2 \times \Z_2)$ orbifold on untilted tori, our findings can straightforwardly be T-dualised 
to blow-ups (by K\"ahler moduli) in the Type IIB/$\Omega$ language, see e.g.~\cite{Angelantonj:2002ct,Larosa:2003mz,Angelantonj:2009yj,Camara:2010zm,Angelantonj:2011hs}. T-duality for toroidal backgrounds results in identifying tilted tori in Type IIA/$\OR$ with a non-vanishing $B$-field in Type IIB/$\Omega$~\cite{Blumenhagen:2000ea,Larosa:2003mz,Pradisi:2003ct}, but to our best knowledge, the mapping of twisted sectors in the kinds of models considered in the present article has not been worked out, but is expected to differ from the rectangular torus due to the non-trivial permutation of some orbifold singularities under the orientifold involution.
If also the symmetric $\Z_3$ action of the article at hand is included, T-duality will further translate it to an asymmetric $\Z_3$ action in Type IIB/$\Omega$ orientifolds~\cite{Pradisi:1999ii}. The advantage of the Type IIA/$\OR$ language thus clearly lies in the geometric description.

Last but not least, we showed how to stabilise the majority of complex structure moduli in the presence of D6-branes. The dilaton and remaining geometric moduli, in particular the K\"ahler moduli, will have to be stabilised by non-perturbative effects and possibly a small number of RR-fluxes, see e.g.~\cite{Kerstan:2011dy,Grimm:2011dx} for the generic shape,  with assumed small back-reaction on the three-cycle geometry. 
We expect that low-energy parameters like the  gauge and Yukawa  couplings will not depend on the twisted K\"ahler moduli (blow-up modes) but rather only on the areas of the two-tori, reinforcing our focus on stabilising complex structure moduli.  Moreover, as recently argued in~\cite{Marchesano:2014iea}, the interplay between closed and open string moduli is expected to also stabilise the latter. The mixing of their axionic open and closed string partners has also been recently discussed, see e.g.~\cite{Honecker:2013mya,Honecker:2015ela}.
A detailed analysis of a {\it complete} closed and open string moduli stabilisation scenario in phenomenologically appealing models and their impact on cosmology, however, goes well beyond the scope of this article and will be investigated in the future.

\noindent
{\bf Acknowledgements:} 
This work is partially supported by the {\it Cluster of Excellence `Precision Physics, Fundamental Interactions and Structure of Matter' (PRISMA)} DGF no. EXC 1098,
the DFG research grant HO 4166/2-1, and the DFG Research Training Group {\it `Symmetry Breaking in Fundamental Interactions'} GRK 1581.


\appendix
\section[Collection of Formulas]{Collection of Formulas}\label{A:1}

In this appendix, we collect the precise expressions for the hypersurface polynomial in equation~\eqref{Eqn:T6Z2Z2HypersurfaceEquation} for finite deformations. First we apply the restrictions from the $\Z_3$ symmetry and the orientifold projection as in table~\ref{tab:Z3Restrictions} and rewrite the terms in a more compact form. We work only in the $\Z_2^{(3)}$ sector since the other ones are equivalent, and we will drop the dependence of the coordinates $x_3,v_3$. Furthermore, we relabel the deformation parameters $\varepsilon^{(3)}_{\alpha\beta}$ as $\varepsilon_{\rho}$, where $\rho=1,2,3$ represent the two-cycles $e_{\rho=1,2,3}$ following the convention in section~\ref{Ss:Orbs-with-Z3}, whereas $\varepsilon_{4\pm5}$ simultaneously deform $e_{4/5}$. Then the hypersurface polynomial reads:
 \begin{align}
 \begin{split}
  f = -y^2 + \left( v_3 x_3^3 - v_3^4 \right) \cdot \Big\{ &  \left( v_1 x_1^3 - v_1^4 \right) \left( v_2 x_2^3 - v_2^4 \right)  \\
  -& \varepsilon_1 x_1^2 v_1^2 \left( x_2^4 - x_2 v_2^3 \right) \\
  -& \varepsilon_2 \left( x_1^4 - x_1 v_1^3\right) x_2^2 v_2^2  \\
  -& \varepsilon_3 \left( v_1^4 v_2^4 + v_1^2 x_1^2 v_2^3 x_2 + v_1^3 x_1 v_2^2 x_2^2 \right) \\
  -& \varepsilon_{4+5}\left( 2 v_1^4 v_2^4 - v_1^2 x_1^2 v_2^3 x_2 - v_1^3 x_1 v_2^2 x_2^2 \right) \\
  -& \varepsilon_{4-5} \left( v_1^2 x_1^2 v_2^3 x_2 - v_1^3 x_1 v_2^2 x_2^2 \right) \Big\} \,.
 \end{split}
 \label{Eqn:Z6HypersurfaceEqn}
 \end{align}
For simplicity we will work in the patch $v_i \equiv 1$.

\paragraph{Deforming Fixed Points 4 and 5}
 If we switch on $\varepsilon_{4+5}$ we observe how the exceptional cycles $e_4$ and $e_5$ grow out of the singularities. However, also fixed point $3$ gets deformed in this process, but to higher order in $\varepsilon_4$. Therefore, we must also vary $\varepsilon_3$ as a function of $\varepsilon_4$ to keep $e_3$ singular, and we find that this is fulfilled if
 \begin{align}
  \varepsilon_3 = - \frac{3}{2} -  \varepsilon_4 + \sqrt{\frac94 - 3 \varepsilon_4} = \frac{1}{3}\varepsilon_4^2 - \frac{2}{9}\varepsilon_4^3 + \mathcal{O}(\varepsilon_4^4) \,.  \label{eqn:3StaysSingular}
 \end{align}
In this case the fixed point $(33)$ is located at $x_1 = x_2 = \frac{-1 + \sqrt{9 + 12 \varepsilon_4} }2$. Furthermore, we find that this is only valid for $\varepsilon_4 \ge -3/4$, which seems to be the boundary of the range of small deformations. The hypersurface equation for this particular deformation becomes
\begin{align}
 y^2 = \left( x_1^3 - 1 \right) \left( x_2^3 - 1 \right) - \varepsilon_4 \left( 2 - x_1 x_2^2 - x_1^2 x_2 \right) + \left( \frac{3}{2} +  \varepsilon_4 - \sqrt{\frac94 - 3 \varepsilon_4} \right) \left( 1 + x_1 x_2^2 + x_1^2 x_2 \right) \,. \label{Eqn:HypDef45Full}
\end{align}


\paragraph{Deforming Fixed Point 3}
In the next step we want to switch the roles of the fixed points studied above, i.e.\ we want to deform the fixed point (33) while leaving fixed point (24) singular.  
To do so, we give $\varepsilon_3$ a finite value and tune $\varepsilon_4$ accordingly. Unfortunately, we cannot find a nice analytic expression as in equation \eqref{eqn:3StaysSingular}, but we are able to write down a Taylor series expansion:
\begin{align}
 \varepsilon_4 = - \frac19 \varepsilon_3^2 - \frac1{81}\varepsilon_3^3 + \mathcal{O}(\varepsilon_3^5) \,.
\end{align}
Note that the fourth order vanishes so that we can stop the expansion here. Using this, and after shifting the $x_1$ with $\alpha_2$ and $x_2$ with $\alpha_4$, see equation \eqref{Eqn:HalfLatticeShifts}, we find the fixed point (24) at $x_1 = \overline x_2 = 1 + i b$ where $b$ is the real root of the equation
\begin{align}
\varepsilon_3 b^6 - ( 9 + 3 \varepsilon_3 ) b^5 + ( 9 + 6 \varepsilon_3 ) b^4 - (27 + 7 \varepsilon_3 ) b^3 + ( 9 + 6 \varepsilon_3 ) b^2 - (9 + 3 \varepsilon_3 ) b + \varepsilon_3 = 0 \,.
\end{align}
In order to find the exceptional cycle (33), we go back to the unshifted hypersurface equation
\begin{align}
  y^2 = \left( x_1^3 - 1 \right) \left( x_2^3 - 1 \right)  - \varepsilon_3 \left( 1 + x_1 x_2^2 + x_1^2 x_2 \right)+ \left( \frac19 \varepsilon_3^2 + \frac1{81}\varepsilon_3^3 + ... \right)  \left( 2 - x_1 x_2^2 - x_1^2 x_2 \right) \,. \label{Eqn:HypDef3Full}
\end{align}
Again, using the symmetry $x_1 \leftrightarrow \overline x_2$, we recover the exceptional cycles at $x_2 = \overline x_1$ and $y \in \R$.


\paragraph{Deforming Fixed Points 1 and 2}
Since in each four-torus the exceptional two-cycles $e_1$ and $e_2$ are equivalent via the permutation symmetry of the two-tori, we discuss their deformation simultaneously. Here it is useful to focus on two different directions of deformations:
\begin{itemize}
 \item Deforming only one of those fixed points, e.g.\ $\varepsilon_1 > 0, \varepsilon_2=0$: In this set-up one can study the volume and {\it sLag} property of the resulting exceptional cycles. Switching on only one such deformation leaves all other fixed points singular and thus leads to a rather tractable hypersurface equation \eqref{Eqn:Z6HypersurfaceEqn}.
 \item Simultaneously deforming both fixed points, $\varepsilon:= \varepsilon_1=\varepsilon_2 > 0$: Here the advantage is that the symmetry between $x_1$ and $x_2$ is preserved and we thus have easy access to the cycles $e_{3,4,5}$. However, in this case we observe a deformation of all other fixed points at higher order so we need to switch on $\varepsilon_3$ and $\varepsilon_{4+5}$ to keep them singular. For $\varepsilon_3$ we indeed find an analytic espression, 
 \begin{align}
   \varepsilon_3 = -3/2 + 2 \varepsilon + \sqrt{9/4 - 6 \varepsilon + 3 \varepsilon^2} = - \frac{\varepsilon^2}3 - \frac{4 \varepsilon^3}9 + \mathcal{O}(\varepsilon^4) \,,
 \end{align}
 whereas $\varepsilon_{4+5}$  can only be given approximately,  $\varepsilon_{4+5} = - \frac{\varepsilon^2}3 + \frac{\varepsilon^3}3 - \frac{\varepsilon^4}2 + \mathcal{O}(\varepsilon^5)$. To sum up, the full deformed hypersurface equation becomes
 \begin{align}
   y^2 &= \left( x_1^3 - 1 \right) \left( x_2^3 - 1 \right) - \varepsilon \left( x_1^4 x_1^2 + x_1^2 x_2^4 - x_1 x_2^2 - x_1^2 x_2 \right) \\
\nonumber   &- \left( -3/2 + 2 \varepsilon + \sqrt{9/4 - 6 \varepsilon + 3 \varepsilon^2} \right)  \left( 1 + x_1 x_2^2 + x_1^2 x_2 \right) - \left( - \frac{\varepsilon^2}3 + \frac{\varepsilon^3}3 - \frac{\varepsilon^4}2 \right) \left( 2 - x_1 x_2^2 - x_1^2 x_2 \right)\,. \label{Eqn:HypDef12}
   \end{align}

\end{itemize}

\section{Untilted, Tilted and Square Torus Lattice}\label{A:2}

In the previous discussion we focussed on the hexagonal torus, but for completeness we also want to describe the {\it Lag} lines that appear in the general case of an untilted ({\bf a}-type) or tilted ({\bf b}-type) torus, or in the special case of a square torus (either {\bf a}- or {\bf b}-type). Especially for the square torus there appear some interesting symmetries among the {\it Lag} lines.

\begin{figure}[ht]
\centering
\subfloat[Untilted ({\bf a}-type) lattice. The four {\it Lag} lines {\bf aX} (blue) on the real axis are the ones intersecting the singularities ($\epsilon_2, \epsilon_3, \epsilon_4, \infty$). The two additional circles (green) correspond to {\it Lag} lines not intersecting the singularities, i.e.\ pure bulk cycles, and were discussed in detail in~\cite{Blaszczyk:2014xla}, where they were called III (left circle) and IV (right circle).]{\label{fig:Diagram_Untilted_PLUS4} \raisebox{1cm}{\includegraphics[width=0.45\textwidth]{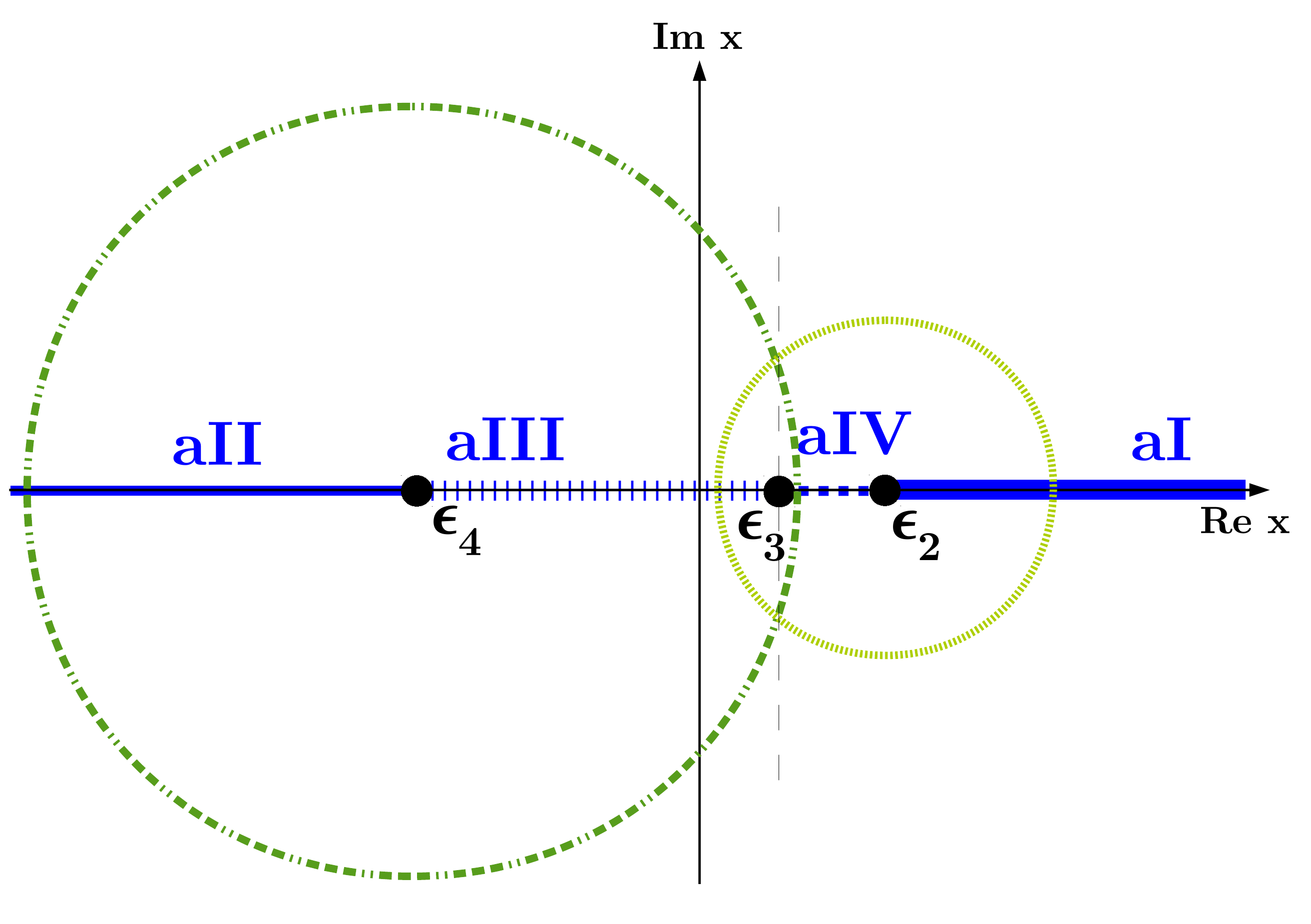}}} \quad
\subfloat[Tilted ({\bf b}-type) lattice. The {\it Lag} lines on the real axis, called {\bf bI} and {\bf bII}, correspond to undisplaced cycles ($\sigma = 0$), while the two arcs of the circle named {\bf bIII} and {\bf bIV} represent displaced ones ($\sigma = 1$). Each cycle passes through two of the singularities $\epsilon_2, \epsilon_3, \epsilon_4, \infty$. The notation in the image is $\epsilon_\text{R} \equiv \Re(\epsilon)$ and $\epsilon_\text{I} \equiv \Im(\epsilon)$.]{\label{fig:Diagram_Tilted} \includegraphics[width=0.45\textwidth]{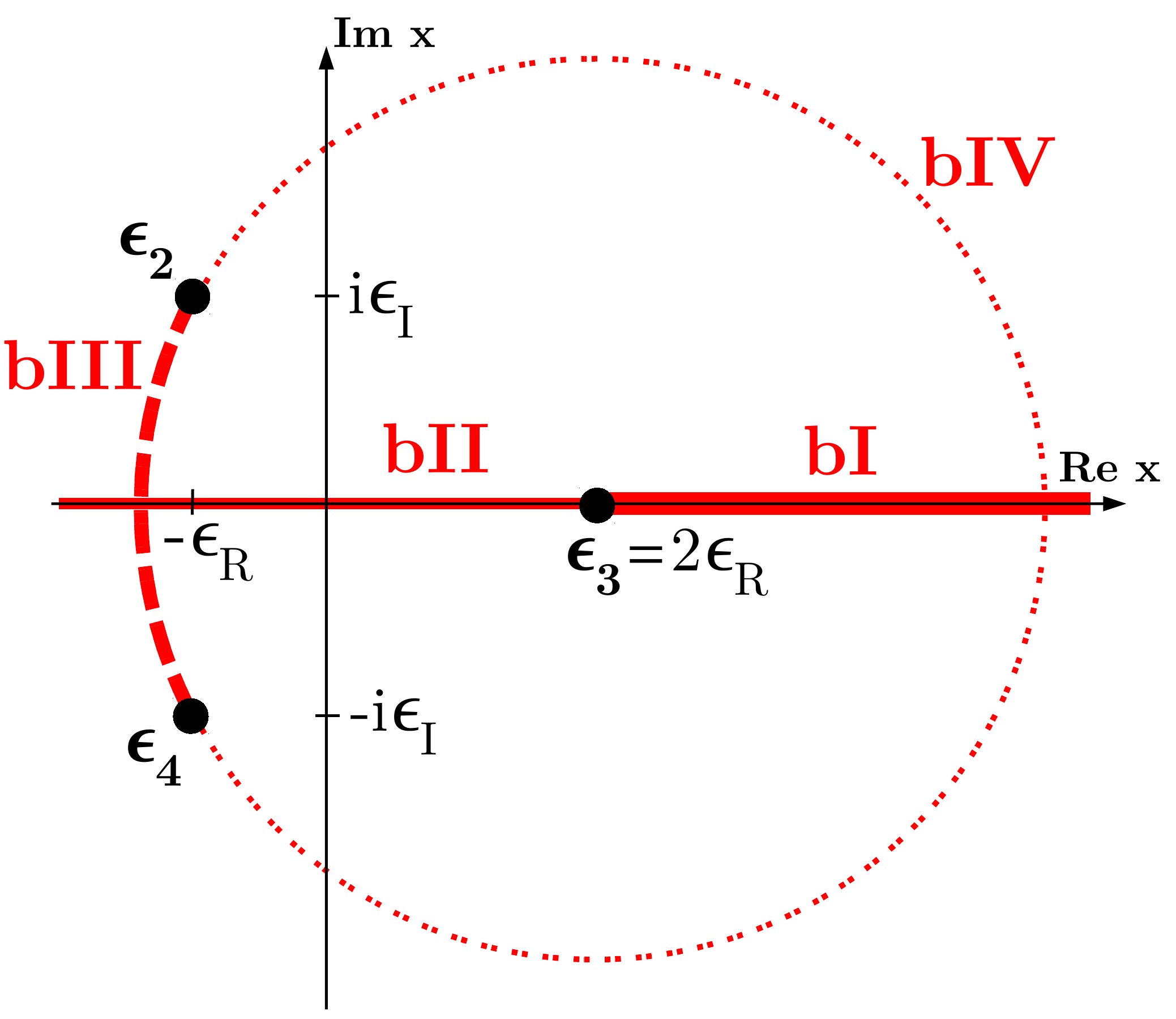}}
\caption{{\it Lag} lines in the complex $x$-plane (with $v \equiv 1$), cf.\ also formulas of  table~\ref{tab:T2LagLines_Untilted_Tilted}.}
\label{fig:Diagram_Untilted_Tilted}
\end{figure}

The properties of the {\it Lag} lines for the untilted and tilted lattice can be found in table~\ref{tab:T2LagLines_Untilted_Tilted}. Here one can read off that all {\it Lag} lines of the general {\bf a}-type lattice which intersect singularities lie on the real $x$-axis (if $v \equiv 1$), as one can also see in figure~\ref{fig:Diagram_Untilted_PLUS4}. Therefore it is easy to work with orbifolds of the {\bf a}-type lattice, especially since the deformation parameters take also only real values (at least for all deformations that are interesting for our discussion). This makes it possible to visualise the {\it Lag} lines in the hypersurface formalism in a two- or three-dimensional picture, as was done in~\cite{Blaszczyk:2014xla}, where also deformations can be studied in a qualitative way. The two additional circles depicted in figure~\ref{fig:Diagram_Untilted_PLUS4} corresponding to pure bulk cycles were omitted in the previous discussion because they are uninteresting in the context of deformations. 
However, their existence allows to use an analogous trick on the rectangular torus to that used in figure~\ref{fig:IntPath} for the hexagonal one to compute the integral over another bulk cycle in the same homology class.
More details about the {\it Lag} lines on the rectangular {\bf a}-type torus can be found in~\cite{Blaszczyk:2014xla}. For the untilted torus, one has two parameters which can be varied freely, which leads to another shape and size of the torus, e.g. vary $\epsilon_2$ and $\epsilon_4$, while $\epsilon_3 = -\epsilon_2 - \epsilon_4$.

\begin{figure}[ht]
\centering
\subfloat[Square lattice of {\bf a}-type. The blue lines, called {\bf aX} (all lying on the horizontal axis), are the same as in the general untilted torus (depicted in figure~\ref{fig:Diagram_Untilted_PLUS4}), while the red curves, labelled {\bf bX}, are additional {\it Lag} lines appearing on the square lattice due to the enhanced symmetry.]{\label{fig:Diagram_SquareA} \includegraphics[width=0.5\textwidth]{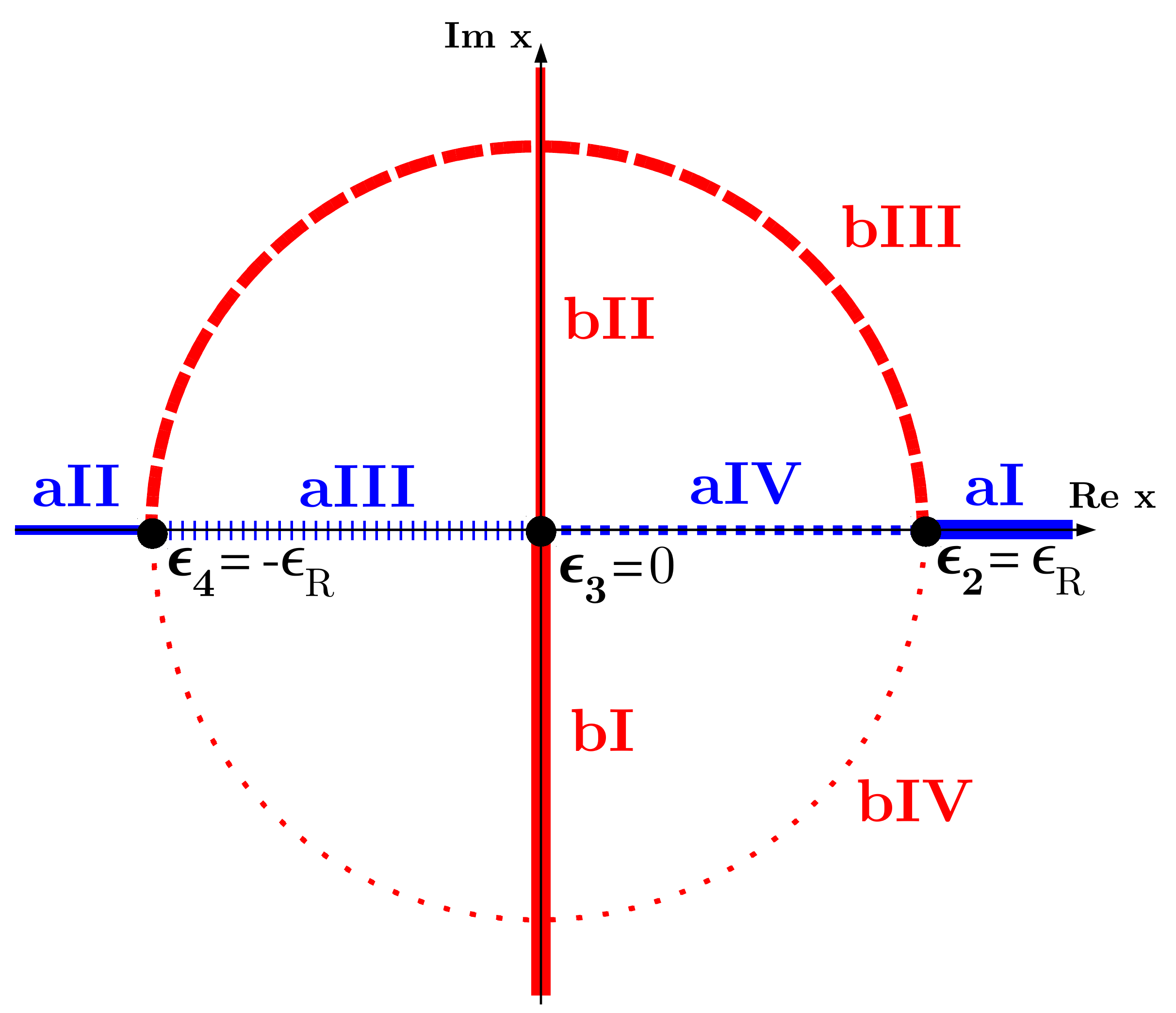}} \quad
\subfloat[Square lattice of {\bf b}-type. Here the red curves (again called {\bf bX}) correspond to the general tilted lattice illustrated in figure~\ref{fig:Diagram_Tilted}, while here the blue lines {\bf aX} (now all lying on the vertical axis) are the additional {\it Lag} lines appearing on the square lattice.]{\label{fig:Diagram_SquareB} 
\includegraphics[width=0.4\textwidth]{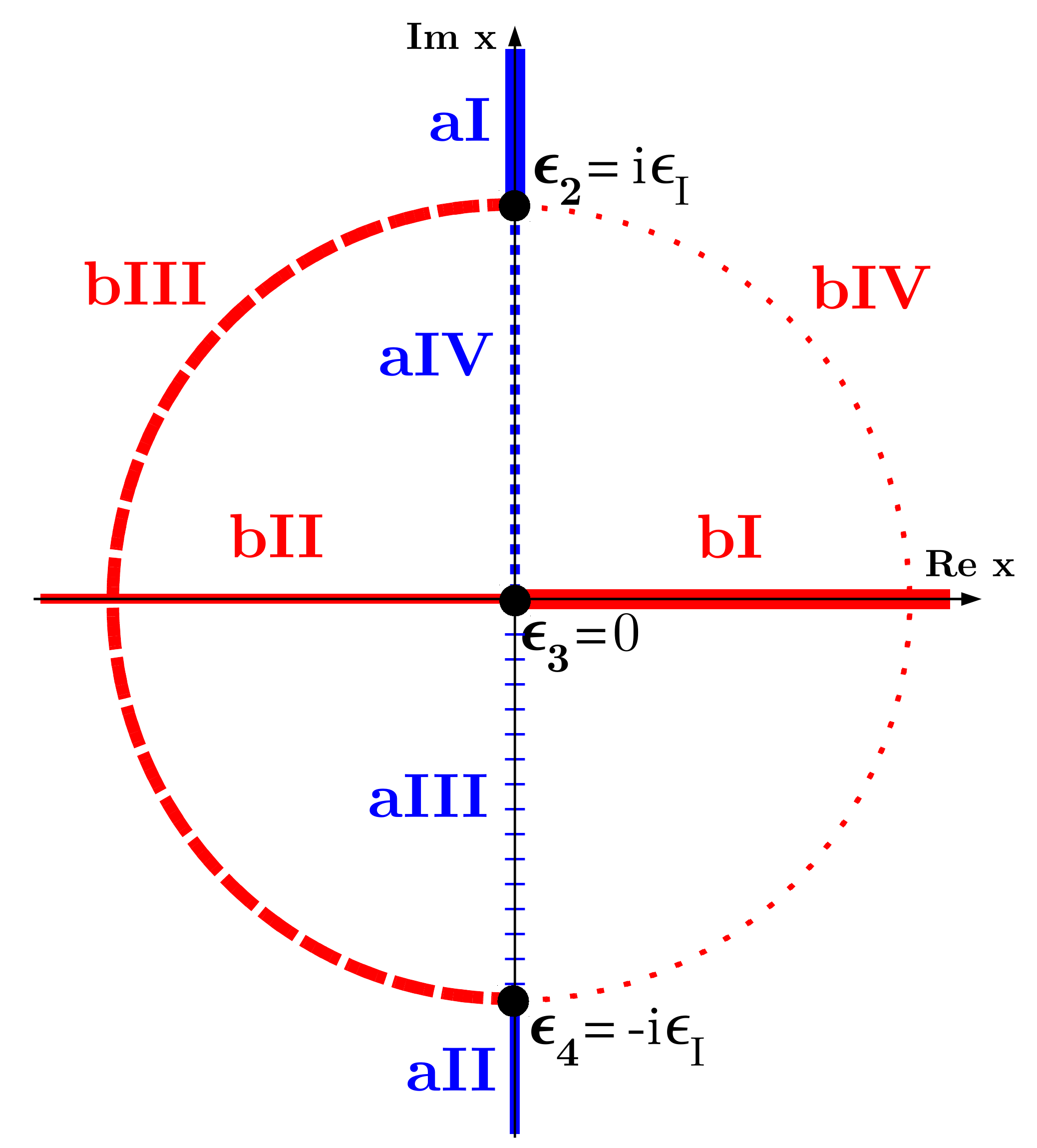}}
\caption{{\it Lag} lines for square tori in the complex $x$-plane (with $v \equiv 1$), cf.\ also formulas of  table~\ref{tab:T2LagLines_Untilted_Tilted} with $\epsilon_3 = 0$ ({\bf a}-type) or $\Re(\epsilon) = 0$ ({\bf b}-type).}
\label{fig:Diagram_SquareAB}
\end{figure}

The {\it Lag} lines of the tilted ({\bf b}-type) torus, depicted in figure~\ref{fig:Diagram_Tilted}, are more complicated to handle because the displaced {\it Lag} lines {\bf bIII} and {\bf bIV} take now complex values in the $x$-coordinate (for $v$ as well if we do not work in the patch $v \equiv 1$). This is similar to the special case of a hexagonal torus described in section~\ref{Ss:Orbs-with-Z3}, but for the generic tilted torus there exist only the four {\it Lag} lines given in figure~\ref{fig:Diagram_Tilted} and table~\ref{tab:T2LagLines_Untilted_Tilted}, i.e.\ no additional cycles appear. The singularities of the tilted torus in the hypersurface formalism are given by $\epsilon_3 = 2 \Re(\epsilon)$ and $\epsilon_2 = -\Re(\epsilon) + i \Im(\epsilon) = \overline{\epsilon_4}$, so here one has, as for the untilted case, two parameters which determine shape and size of the tilted torus and which can be varied freely. The circle depicted in figure~\ref{fig:Diagram_Tilted} has radius $\sqrt{(3 \Re(\epsilon))^2 + \Im(\epsilon)^2}$ and is centred around the singularity $\epsilon_3$. In the previous chapters the parameter $\Re(\epsilon)$ was sometimes chosen as $\Re(\epsilon) = \frac{1}{2}$ to simplify the equations.

\begin{figure}[ht]
\centering
\includegraphics[width=0.7\textwidth]{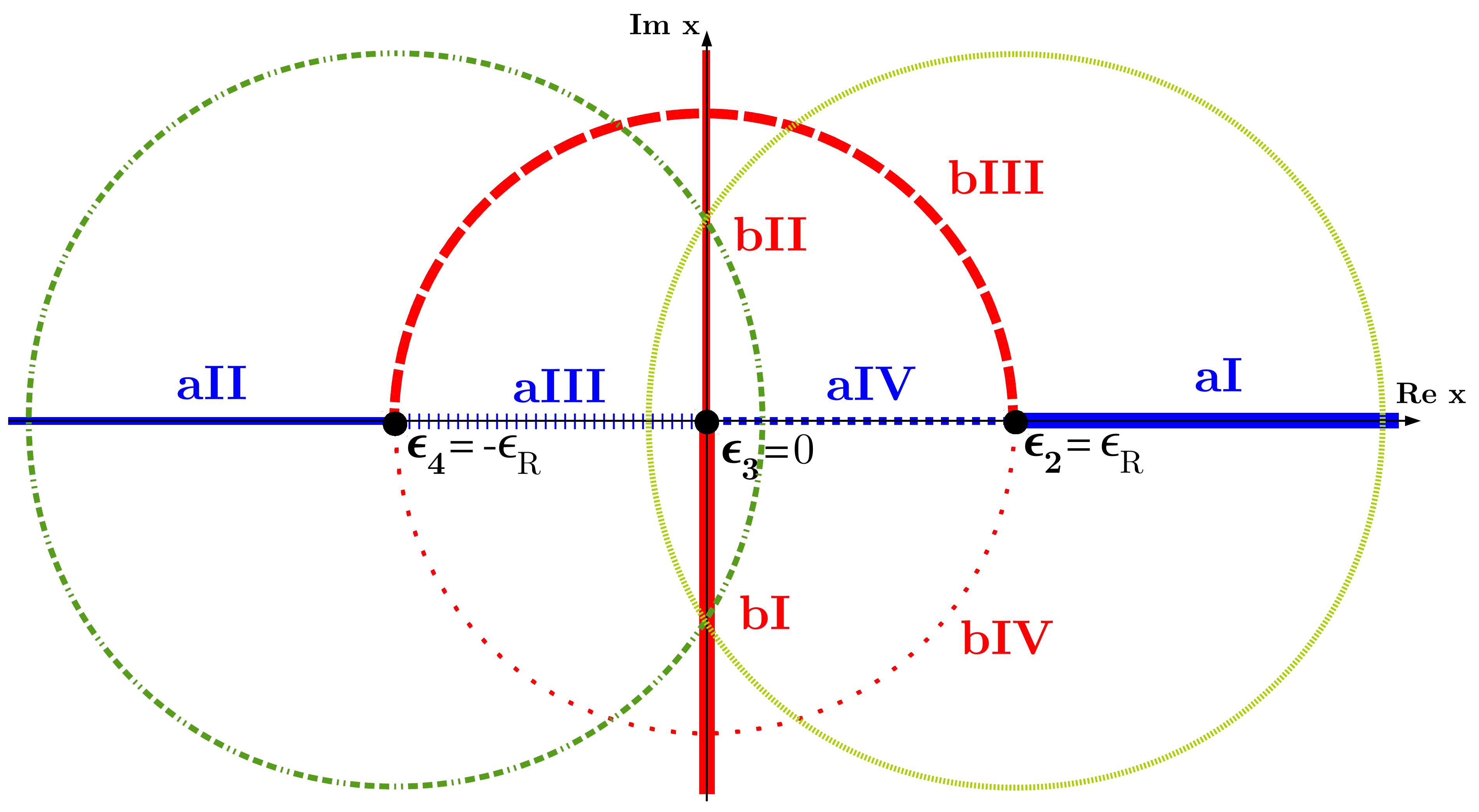}
\caption{All {\it Lag} lines for the square {\bf a}-type torus in the complex $x$-plane (with $v \equiv 1$).}
\label{fig:Diagram_SquareA_PLUS4}
\end{figure}

As already mentioned before, the square torus lattice has the special property that it can exist on an underlying {\bf a}- or {\bf b}-type lattice, cf.\ figure~\ref{fig:Square_tilted_lattice}, and fixed point 3 always has the value $\epsilon_3 = 0$ implying $\epsilon_2 \equiv \epsilon_\text{R} = -\epsilon_4$. Therefore, only one parameter remains that can be varied freely, changing the size of the torus lattice. As for the hexagonal torus lattice (see table \ref{tab:LagLinesHex} for the {\it Lag} lines), the number of {\it Lag} lines on the square lattice increases by the enhanced symmetry such that one counts eight {\it Lag} lines (running through the fixed points) for both lattice types {\bf a} or {\bf b}. This was already discussed in detail in~\cite{Blaszczyk:2014xla}. One nicely sees the symmetry in figures~\ref{fig:Diagram_SquareA} and~\ref{fig:Diagram_SquareB}, where the right picture is just a by $\frac{\pi}{2}$ rotated version of the left picture. Diagram~\ref{fig:Diagram_SquareB} represents the special case of the tilted lattice where $\epsilon_2$ and $\epsilon_4$ have only values in $i \R$.

In figure~\ref{fig:Diagram_SquareA_PLUS4} one finds all {\it Lag} lines of the {\bf a}-type square torus, including the pure bulk cycles that already appeared in figure~\ref{fig:Diagram_Untilted_PLUS4}, 
and one could analogously draw a diagram for the {\bf b}-type lattice (i.e.\ rotated by $\frac{\pi}{2}$).
But, while these additional {\it Lag} lines naturally also appear in the rotated {\bf b}-type version of the square lattice, they are not apparent in the set of {\it Lag} lines of the general tilted torus, as already discussed above.

\vspace{10mm}

\clearpage


\addcontentsline{toc}{section}{References}
\bibliographystyle{ieeetr}
\bibliography{refs_Deformations}

\end{document}